\documentclass[apj,twocolumn]{openjournal}

\usepackage{newtxtext,newtxmath,enumitem,multirow,ctable}
\usepackage[T1]{fontenc}
\usepackage[breaklinks,colorlinks,citecolor=blue,urlcolor=blue]{hyperref}

\newcommand{\orcidauthor}[3]{\author{\href{http://orcid.org/#1}{#2$^{#3}$}}}

\newcommand{\Alf}{{Alfv\'en}}

\newcommand{\gizmourl}{\href{http://www.tapir.caltech.edu/~phopkins/Site/GIZMO.html}{\url{http://www.tapir.caltech.edu/~phopkins/Site/GIZMO.html}}}

\newcommand{\papertwo}{Paper {\small II}}
\newcommand{\paperthree}{Paper {\small III}}

\shorttitle{FORGE'd in FIRE I}
\shortauthors{Hopkins et al.}

\begin{document}

\title{\vspace{-0.8cm}FORGE'd in FIRE: Resolving the End of Star Formation and Structure of AGN Accretion Disks from Cosmological Initial Conditions\vspace{-1.5cm}}

\orcidauthor{0000-0003-3729-1684}{Philip F. Hopkins}{1,*}
\orcidauthor{0000-0002-1655-5604}{Michael Y. Grudi{\'c}}{2,\dagger}
\orcidauthor{0000-0003-1598-0083}{Kung-Yi Su}{3}
\orcidauthor{0000-0002-3977-2724}{Sarah Wellons}{4}
\orcidauthor{0000-0001-5769-4945}{Daniel Angl{\'e}s-Alc{\'a}zar}{5,6}
\orcidauthor{0000-0001-8867-5026}{Ulrich P. Steinwandel}{6}
\orcidauthor{0000-0001-5541-3150}{D\'avid Guszejnov}{7,\dagger}
\orcidauthor{0000-0002-8659-3729}{Norman Murray}{8}
\orcidauthor{0000-0002-4900-6628}{Claude-Andr\'{e} Faucher-Gigu\`{e}re}{9}
\orcidauthor{0000-0001-9185-5044}{Eliot Quataert}{10}
\orcidauthor{0000-0002-1666-7067}{Du\v{s}an Kere\v{s}}{11}

\affiliation{$^{1}$TAPIR, Mailcode 350-17, California Institute of Technology, Pasadena, CA 91125, USA}
\affiliation{$^{2}$Carnegie Observatories, 813 Santa Barbara St, Pasadena, CA 91101, USA}
\affiliation{$^{3}$Black Hole Initiative, Harvard University, 20 Garden St, Cambridge, MA 02138, USA}
\affiliation{$^{4}$Department of Astronomy, Van Vleck Observatory, Wesleyan University, 96 Foss Hill Drive, Middletown, CT 06459, USA}
\affiliation{$^{5}$Department of Physics, University of Connecticut, 196 Auditorium Road, U-3046, Storrs, CT 06269, USA}
\affiliation{$^{6}$Center for Computational Astrophysics, Flatiron Institute, 162 5th Ave., New York, NY 10010 USA}
\affiliation{$^{7}$Department of Astronomy, University of Texas at Austin, TX 78712, USA}
\affiliation{$^{8}$Canadian Institute for Theoretical Astrophysics, University of Toronto, Toronto, ON M5S 3H8, Canada}
\affiliation{$^{9}$CIERA and Department of Physics and Astronomy, Northwestern University, 1800 Sherman Ave, Evanston, IL 60201, USA}
\affiliation{$^{10}$Department of Astrophysical Sciences, Princeton University, Princeton, NJ 08544, USA}
\affiliation{$^{11}$Department of Physics, Center for Astrophysics and Space Sciences, University of California San Diego, 9500 Gilman Drive, La Jolla, CA 92093, USA}

\thanks{$^*$E-mail: \href{mailto:phopkins@caltech.edu}{phopkins@caltech.edu}},
\thanks{$\dagger$NASA Hubble Fellow}

\begin{abstract}
It has recently become possible to ``zoom-in'' from cosmological to sub-pc scales in galaxy simulations to follow accretion onto supermassive black holes (SMBHs). However, at some point the approximations used on ISM scales (e.g.\ optically-thin cooling and stellar-population-integrated star formation [SF] and feedback [FB]) break down. We therefore present the first cosmological radiation-magnetohydrodynamic (RMHD) simulation which self-consistently combines the FIRE physics (relevant on galactic/ISM scales where SF/FB are ensemble-averaged) and STARFORGE physics (relevant on small scales where we track {\em individual} (proto)stellar formation and evolution), together with explicit RMHD (including non-ideal MHD and multi-band M1-RHD) which self-consistently treats both optically-thick and thin regimes. This allows us to span scales from $\sim 100\,$Mpc down to $< 100\,$au ($\sim 300$ Schwarzschild radii) around a SMBH at a time where it accretes as a bright quasar, in a {\em single simulation}. We show that accretion rates up to $\sim 10-100\,{\rm M_{\odot}\,yr^{-1}}$ can be sustained into the accretion disk at $\ll 10^{3}\,R_{\rm schw}$, with gravitational torques between stars and gas dominating on sub-kpc scales until star formation is shut down on sub-pc scales by a combination of optical depth to cooling and strong magnetic fields. There is an intermediate-scale, flux-frozen disk which is gravitoturbulent and stabilized by magnetic pressure sustaining strong turbulence and inflow with persistent spiral modes. In this paper we focus on how gas gets into the small-scale disk, and how star formation is efficiently suppressed.
\end{abstract}


\maketitle

\section{Introduction}
\label{sec:intro}

\begin{deluxetable*}{ll}
\tabletypesize{\footnotesize}
\tablecaption{Summary of physics included in our default simulation.\label{tbl:physics}}
\tablehead{Cosmology & Fully-cosmological baryons+dark matter simulation from $z\sim100$, with a $\sim (6\,{\rm cMpc})^{3}$ scale zoom-in volume in a $\sim (100\,{\rm cMpc})^{3}$ box. \\
Gravity & Full self-gravity, 5th-order Hermite integrator, adaptive softening for gas, consistent softenings for collisionless particles.}
\startdata
Hydrodynamics & Fluid dynamics with 2nd-order finite-volume MFM solver, refinement to $< 0.01\,M_{\odot}$, non-ideal (ion+atomic+molecular) EOS. \\
Magnetic Fields & Integrated with constrained-gradient MHD solver, trace seed cosmological fields amplified self-consistently. \\
Non-Ideal MHD & Kinetic terms: anisotropic Spitzer-Braginskii conduction \&\ viscosity, plus ambipolar diffusion, (optional) Hall MHD, Ohmic resistivity. \\
\hline
Thermo-chemistry & Detailed processes for $1-10^{10}$\,K. Non-equilibrium H \&\ He ions, H$_{2}$ formation/destruction, dust destruction. Fully coupled to RHD.\\
Radiation & M1 solver. Photo-ionizing, Lyman-Werner, photo-electric, NUV, optical \&\ near-IR, and adaptive (multi-wavelength grey-body) FIR followed. \\
Opacities & Dust, molecular, metal, atomic, ion, H$^{-}$, free $e^{-}$ with Kramers or bound-bound, bound-free, free-free, Compton, Thompson, Rayleigh.\\ 
Cosmic Rays & Dynamically-evolved with LEBRON approximation, coupled to chemistry, sourced from fast shocks from SNe and stellar mass-loss.\\
\hline
SSP Particles & FIRE: Formation in self-gravitating, Jeans-unstable gas, sampling IMF, when cell resolution $> 1\,M_{\odot}$.\\
SSP Feedback & FIRE: Main-sequence IMF-sampled tracks: radiation, stellar mass-loss (O/B \&\ AGB), supernovae (Types I \&\ II), cosmic rays.\\ 
Star Particles & STARFORGE: Formation in self-gravitating, isolated, resolved Larson cores inside own Hill sphere, accrete bound mass, resolution $<1\,M_{\odot}$.\\
Star Feedback & STARFORGE: Protostellar \&\ main-sequence single-star tracks with accretion, radiation, jets, surface mass-loss, end-of-life explosions.\\
\hline
Supermassive BH & Live sink particles formed dynamically. Refinement centers on $\sim 1.3\times10^{7}\,M_{\odot}$ BH, accretion at $<300\,R_{\rm schw}$. 
\enddata
\end{deluxetable*}

The origins and growth of super-massive black holes (SMBHs) represents one of the most important open problems in extragalactic astrophysics. Most sufficiently-massive galaxies host SMBHs whose masses correlate with various host galaxy bulge properties and reach masses as large as $\sim 10^{10}\,M_{\odot}$ (\citealt{magorrian,FM00,Gebhardt00,hopkins:bhfp.theory,hopkins:bhfp.obs,aller:mbh.esph,kormendy:2011.bh.nodisk.corr}; for a review see \citealt{kormendy:2013.review.smbh.host.correlations}). Many constraints indicate that most of this BH mass is assembled via accretion of gas in a few bright quasar phases \citep{Soltan82,salucci:bhmf,yutremaine:bhmf,hopkins:old.age}, giving rise to a picture of ``co-evolution'' between galaxies and active galactic nuclei (AGN) or quasars \citep{merloni:synthesis.model}. Understanding this ``co-evolution'' has crucial consequences far beyond the BHs themselves, for example in the form of AGN ``feedback'' launching galactic winds  \citep{silkrees:msigma,king:msigma.superfb.1,dimatteo:msigma,murray:momentum.winds,hopkins:lifetimes.methods,hopkins:lifetimes.obscuration,debuhr:momentum.feedback,fgq2012,torrey:2020.agn.wind.bal.gal.fx.fire}, regulating galaxy masses \citep{croton:sam,hopkins:qso.all,hopkins:red.galaxies,hopkins:groups.ell}, and changing the structure of the circum-galactic or inter-galactic medium (CGM or IGM) around galaxies \citep{ciottiostriker:cooling.flow.selfreg.1,cox:xray.gas,best:radio.loudness,Voit_2017}. 

Essential to understanding this, of course, is to understand how gas is transported from the cosmic web on $\gtrsim$\,Mpc scales down to scales of order the innermost stable circular orbit (ISCO) or event horizon at $\sim R_{\rm s} \sim 2\,R_{\rm g} \sim 2\,G\,M_{\rm BH}/c^{2} \sim {\rm au}\,(M_{\rm BH}/5\times10^{7}\,M_{\odot})$. Not only must the specific angular momentum of accreted gas decrease by factors of $\sim 10^{7}$, but it must do this sufficiently-rapidly to avoid being turned into stars or ejected from the galaxy via stellar feedback processes along the way. This challenge is far more serious for the most luminous quasars, which must sustain gas inflow rates of up to $\gtrsim 10\,M_{\odot}\,{\rm yr^{-1}}$ -- which would naively imply that the outer accretion disk is gravitationally unstable and present a unique ``last parsec problem'' \citep{goodman:qso.disk.selfgrav}. Local magnetic or Reynolds-type stresses (let alone micro-physical viscosity) as assumed to dominate angular momentum transport in a classical \citet{shakurasunyaev73}-like ``accretion disk'' \citep{balbus.hawley.review.1998} are inefficient at larger scales $\gg 0.01\,$pc \citep{shlosman:inefficient.viscosities,goodman:qso.disk.selfgrav,thompson:rad.pressure}, as is random accretion of individual gas clumps or molecular clouds \citep{hopkins:seyferts,kawakatu:disk.bhar.model,nayakshin:forced.stochastic.accretion.model}\footnote{As those authors and e.g.\ \citet{meece:2017.lowlum.agn.fueling.summary,aird:2018.xray.agn.constraints,yesuf:2020.seyfert.acc.corr.gas.fractions,2021ApJ...919..129L,guo:2022.superzoom.riaf.in.m87.nosf.nomhd.etc} more recently note, those processes could be important for much lower-accretion rate AGN, e.g.\ systems like M87 today accreting several orders of magnitude below their Eddington limit, but they cannot sustain quasar-level accretion rates.} However, the last couple of decades have seen considerable progress on this front at scales $\gtrsim 0.1-1\,$pc. Initial analytic arguments \citep{shlosman:bars.within.bars}, followed by semi-idealized numerical simulations of different ``levels'' of the scale hierarchy \citep{escala:bh.mgr.idealized,escala:nuclear.gas.transport.to.msigma,mayer:bh.binary.sph.zoom.sim, wise2007:protogalaxy.collapse,levine2008:nuclear.zoom,hopkins:zoom.sims,costa:2022.sims.qso.outflows.lya.blobs}, and then simulations using ``super-Lagrangian'' or ``hyper-refinement'' techniques to probe small scales \citep{curtis:2015.refinement.bondi.radius,prieto:2016.zoomin.sims.to.fewpc.hydro.cosmo.highz,prieto:2017.zoomin.sims.agn.fueling.sne.fb,bourne:2017.agn.jet.models.cosmo.sims,su:2018.stellar.fb.fails.to.solve.cooling.flow,su:turb.crs.quench,su:2021.agn.jet.params.vs.quenching,franchini:2022.binary.bh.hyperzoom.dynamics,talbot:2022.jet.sims.refinement.near.jet,sivasankaran:2022.iso.galaxy.sims.refined.center.bhgrowth.smuggle} within larger boxes eventually reaching up to cosmological scales \citep{beckmann:cosmological.sims.super.refinement.for.time.variability,2019MNRAS.490..343B,2021MNRAS.506..488B,daa:20.hyperrefinement.bh.growth} have led to a robust emergent picture wherein on large scales, gravitational torques between non-axisymmetric structures (including e.g.\ mergers, bars, large clumps, lopsided/warped disks; see \citealt{hopkins:zoom.sims}), especially {\em between collisionless and collisional components of the galaxy} (e.g.\ torques from stars on gas not only driving angular momentum exchange but inducing shocks which can orders-of-magnitude enhance inflow rates over classical single-component disk models; see \citealt{hopkins:inflow.analytics}) can produce inflows of gas on timescales of order the dynamical time, ensuring some can reach sub-pc radii without turning into stars (references above and e.g.\ \citealt{levine2008:nuclear.zoom,levine:sim.mdot.pwrspectrum,hopkins:cusp.slopes,hopkins:torus,hopkins:qso.stellar.fb.together}). 

While these represent an enormous progress, there are still many open questions and key issues unresolved by these simulations. In particular, it has not yet been possible to ``bridge the gap'' between these ($\gtrsim$\,pc) scales and the traditional ($Q \gg 1$) accretion disk. This is not just a question of dynamic range, but of physics: the physics believed to drive accretion on small scales -- physics like the magneto-rotational instability (MRI) -- are qualitatively different from the physics of gravitational torques on larger scales. And it is by no means clear what physically occurs when the different physics most relevant on different scales intersect. On large scales $\gtrsim 10-100$\,pc, simulations of high-redshift quasar fueling require cosmological dynamics following optically-thin cooling from dusty ionized, atomic, and molecular gas, self-gravity, and detailed models of star formation and stellar feedback which model the formation and collective effects of entire stellar populations (spanning the range of the entire stellar initial mass function [IMF]), including their radiation, acceleration of cosmic rays, mass-loss, and supernovae. On smaller scales $\lesssim 10\,$pc, simulations of star formation need to follow individual stars and protostars as they form, accrete, and grow, while injecting feedback in the form of jets, winds, radiation, and (eventually) supernovae, all in a dusty medium which spans both optically-thin and optically-thick cooling. At even smaller scales around a SMBH ($\lesssim 10^{4}\,R_{g}$, where $R_{g} = G\,M_{\rm BH}/c^{2}$) traditional ``accretion disk'' simulations must be able to accurately evolve radiation-magneto-hydrodynamics, with global simulations that can accurately follow the growth of the magneto-rotational instability even in warped or irregular disks, radiation-pressure dominated fluids with explicit radiation-dynamics (accounting for finite-speed-of-light effects), with opacities dominated by partially-ionized (largely dust-free) gas, and gravity integrators which must follow huge numbers of orbits accurately. 

As a result, there have not been simulations that can span all three of these regimes simultaneously and self-consistently. Even today, very few codes include all of the physics listed for even just one of the three scale regimes described above, let alone two or all three. So simulations using super-Lagrangian hyper-refinement have generally either (a) had to stop at some radius or resolution where the physics prescriptions simply cease to make sense (e.g.\ at $\sim$\,pc scales, for simulations with traditional ``galaxy-scale'' cooling, star formation and feedback prescriptions as in e.g.\ \citealt{daa:20.hyperrefinement.bh.growth}); or (b) consider only restricted special cases like accretion onto low-redshift SMBHs at extremely low accretion rates ($\lesssim 10^{-4}$ times the Eddington limit) in gas-poor ellipticals in ``hot halos'' (where star formation and many other physical processes above can be neglected relatively ``safely''; as in \citealt{guo:2022.superzoom.riaf.in.m87.nosf.nomhd.etc}), or (c) simply neglect most of the physics above even on scales where it could be important.

In this paper, we present the first simulation to span all three of these regimes including all of the physics above. The key to this is to leverage a suite of physics that has been developed and extensively studied in the {\small GIZMO} code \citep{hopkins:gizmo,hopkins:gizmo.public.release} over the last several years. On large scales, all of the physics above (and more) has been developed into a physics suite as part of the Feedback In Realistic Environments (FIRE) \citep{hopkins:2013.fire,hopkins:fire2.methods,hopkins:fire3.methods} project, designed fundamentally for simulations of galaxies on scales where stars can be treated as ensemble stellar populations, so star formation occurs in environments which should fragment to form stars and produces ``stellar population particles'' which represent many stars that can act on the environment in the form of radiation, cosmic rays, mass-loss, and supernovae. In parallel, we have also developed a suite of physics as part of the STARFORGE project \citep{grudic:starforge.methods,guszejnov:2020.starforge.jets}, designed for simulations which resolve {\em individual} star formation, where sink particles form representing individual (proto)stars which then follow individual (proto)stellar evolution tracks as they grow, accrete, evolve, ending up on the main sequence, and eventually ending their main sequence lives as remnants or SNe, explicitly modeling jets, radiation, mass-loss, and supernovae. As a part of this, we have developed gravity and radiation-magnetohydrodynamics solvers which have been applied to high accuracy to evolving e.g.\ the MRI in global disk simulations \citep{gaburov:2012.public.moving.mesh.code,hopkins:mhd.gizmo,hopkins:cg.mhd.gizmo,hopkins:gizmo.diffusion,deng:2019.mri.turb.sims.gizmo.methods,deng:2020.global.magnetized.protoplanetary.disk.sims.gravito.turb.leads.to.large.B.saturation.vs.mri,deng:2022.warped.disk.dynamics.need.mfm.mfv}, dynamics of strongly radiation-pressure dominated fluids \citep{hopkins:2019.grudic.photon.momentum.rad.pressure.coupling,hopkins:radiation.methods,hopkins:2021.dusty.winds.gmcs.rdis,williamson:2020.rad.hydro.agn.sims.with.gizmo,williamson:2022.gizmo.rhd.psph.sims.binary.smbh.torii.radiation.reduces.grav.torques,lupi:2022.qso.feedback.gizmo.radiation.and.others.z6,braspenning:2022.mfm.mfv.sph.cloud.crushing.problem.comparisons,soliman:agn.variability.from.dust.instabilities} including radiation-pressure-dominated AGN accretion disks, and accurately evolving individual gravitational orbits (allowing for ``hard'' N-body dynamics) for up to millions of orbits \citep{grudic:2020.tidal.timestep.criterion,grudic:starforge.methods,guszejnov:starforge.environment.multiplicity,hopkins:tidal.softening}. Crucially, the physics of all of the above are built in a modular fashion in the code, allowing for cross-compatibility -- this allows us to evolve all of the relevant physics simultaneously for the first time. 

These physics and numerical methods allow us to ``zoom in'' from truly cosmological initial conditions down to $<300\,R_{\rm s}$ around a super-massive BH during an extremely high-accretion-rate quasar episode, and to see the formation of the true accretion disk and cessation of star formation on sufficiently small scales in a self-consistent manner. In \S~\ref{sec:methods}, we summarize the numerical methods and physics included (\S~\ref{sec:methods:overview}) including both the FIRE (\S~\ref{sec:methods:fire.mode}) and STARFORGE (\S~\ref{sec:methods:starforge.mode}) regimes, and initial conditions (\S~\ref{sec:methods:ics}) and architecture (\S~\ref{sec:methods:particles}) of the fiducial simulation studied here. In \S~\ref{sec:results} we study the results of the simulation (including some variants with different physics): we describe the qualitatively different behaviors over the vast hierarchy of scales (\S~\ref{sec:scales}), our effective resolution (\S~\ref{sec:resolution}), the (gas/stellar/dark matter) mass density and accretion rate profiles (\S~\ref{sec:mdot}), the plasma and thermodynamic properties on these scales (\S~\ref{sec:plasma}), dynamics of fragmentation and star formation and its suppression at small radii (\S~\ref{sec:sf}), and the torques driving inflow (\S~\ref{sec:torques}). In \S~\ref{sec:no.mhd} we contrast a simulation that ignores magnetic fields entirely, and in \S~\ref{sec:scales.vs.physics} we summarize the scales where different physics ``ingredients'' play a crucial role. In \S~\ref{sec:previous} we compare to previous work in different regimes from galactic (\S~\ref{sec:previous:galactic}) through accretion disk (\S~\ref{sec:previous:disk}) scales. We summarize our conclusions in \S~\ref{sec:conclusions}.

\begin{figure*}
	\centering\includegraphics[width=0.95\textwidth]{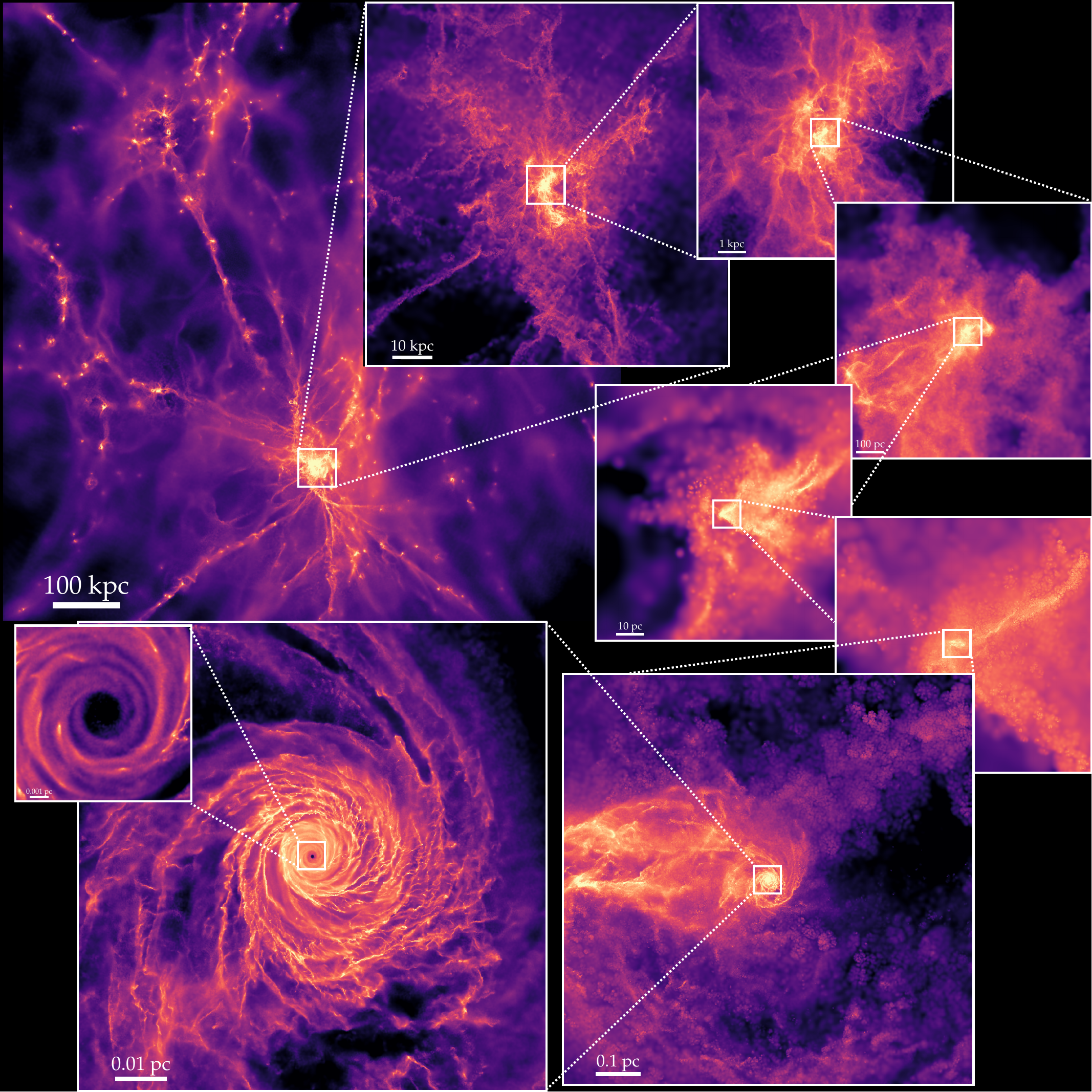} \\
	\caption{Series of images of the projected gas density in our simulation (\S~\ref{sec:methods:ics}) at one moment in time at redshift $z\approx 4$ typical of when we analyze it. Color encodes surface density increasing black-to-white on a logarithmic scale (each panel rescaled owing to the different dynamic range) -- a median pixel in the largest-scale panel ({\em top-left}) has column $N_{H} \sim 10^{19}\,{\rm cm^{-2}}$ (density $n_{H} \sim 10^{-5}\,{\rm cm^{-3}}$), while in the smallest-scale panel $N_{H} \sim 10^{27}\,{\rm cm^{-2}}$ ($n_{H} \sim 10^{12}\,{\rm cm^{-3}}$). We see structure on all scales, with a chaotic, cold, disordered morphology on most scales until an ordered disk forms from capture of gas from a passage of a giant molecular cloud complex (itself triggered by an ongoing galaxy merger in the rapidly-accreting proto-galaxy), forming the accretion disk at $\lesssim 0.1\,$pc.
	\label{fig:images.faceon.stylized}}
\end{figure*} 

\begin{figure*}
	\centering\includegraphics[width=0.95\textwidth]{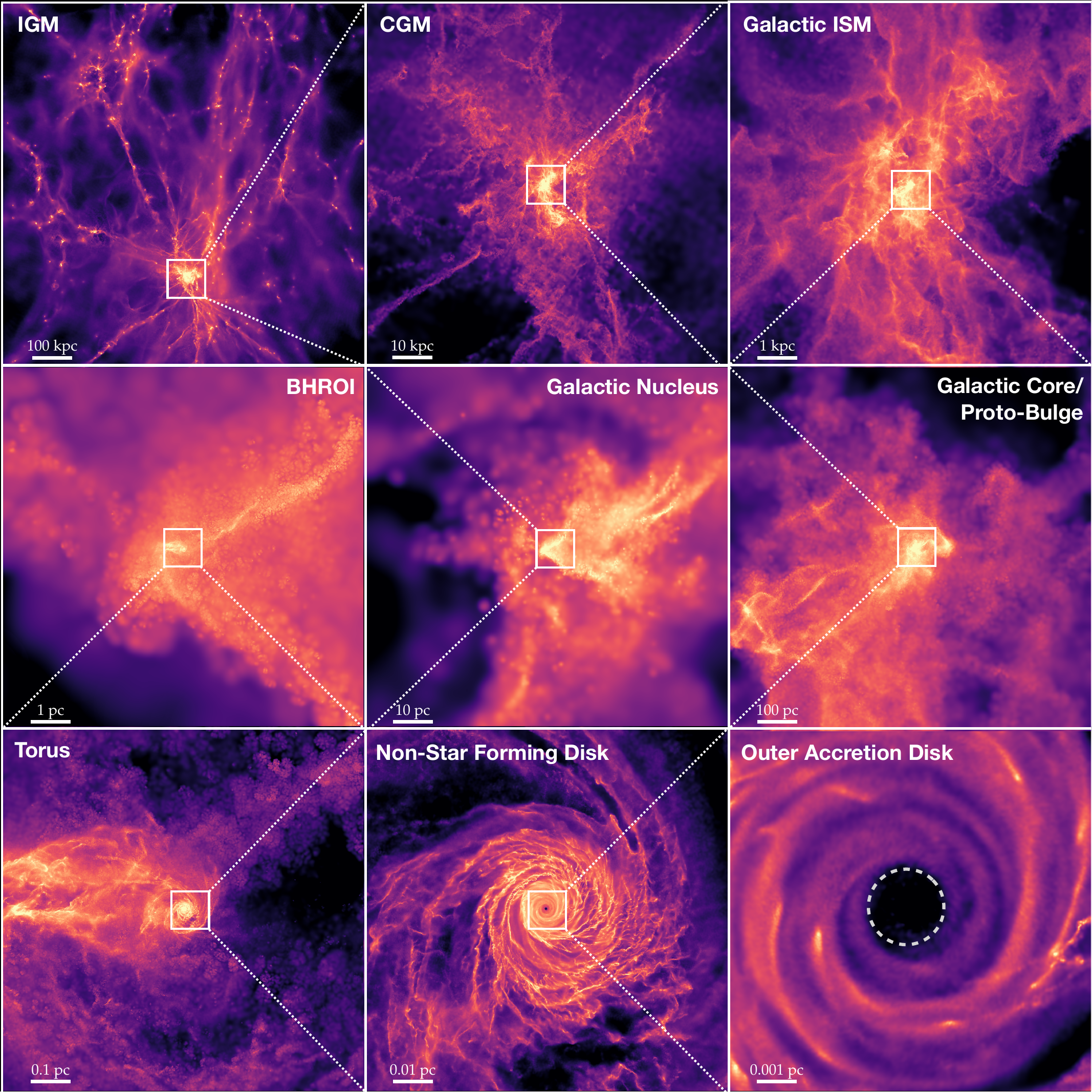} \\
	\caption{As Fig.~\ref{fig:images.faceon.stylized}, but tiling the images so more structure can be seen and identifying each with the heuristic label appropriate to the range of scales shown, per \S~\ref{sec:scales}. In order, each image zooms in by a factor of $10$ around the previous image, with side-length $L = (1000,\,100,\,10)\,{\rm kpc}$ ({\em top}), $(1000,\,100,\,10)$\,pc ({\em middle}), $(1,\,0.1,\,0.01)$\,pc ({\em bottom}). The projection here is chosen to be face-on to the innermost central disk. Note the ``hole'' in the latter inside $r \lesssim 80\,$au is caused by our inner accretion boundary (dashed circle).
	\label{fig:images.faceon}}
\end{figure*} 

\begin{figure*}
	\centering\includegraphics[width=0.95\textwidth]{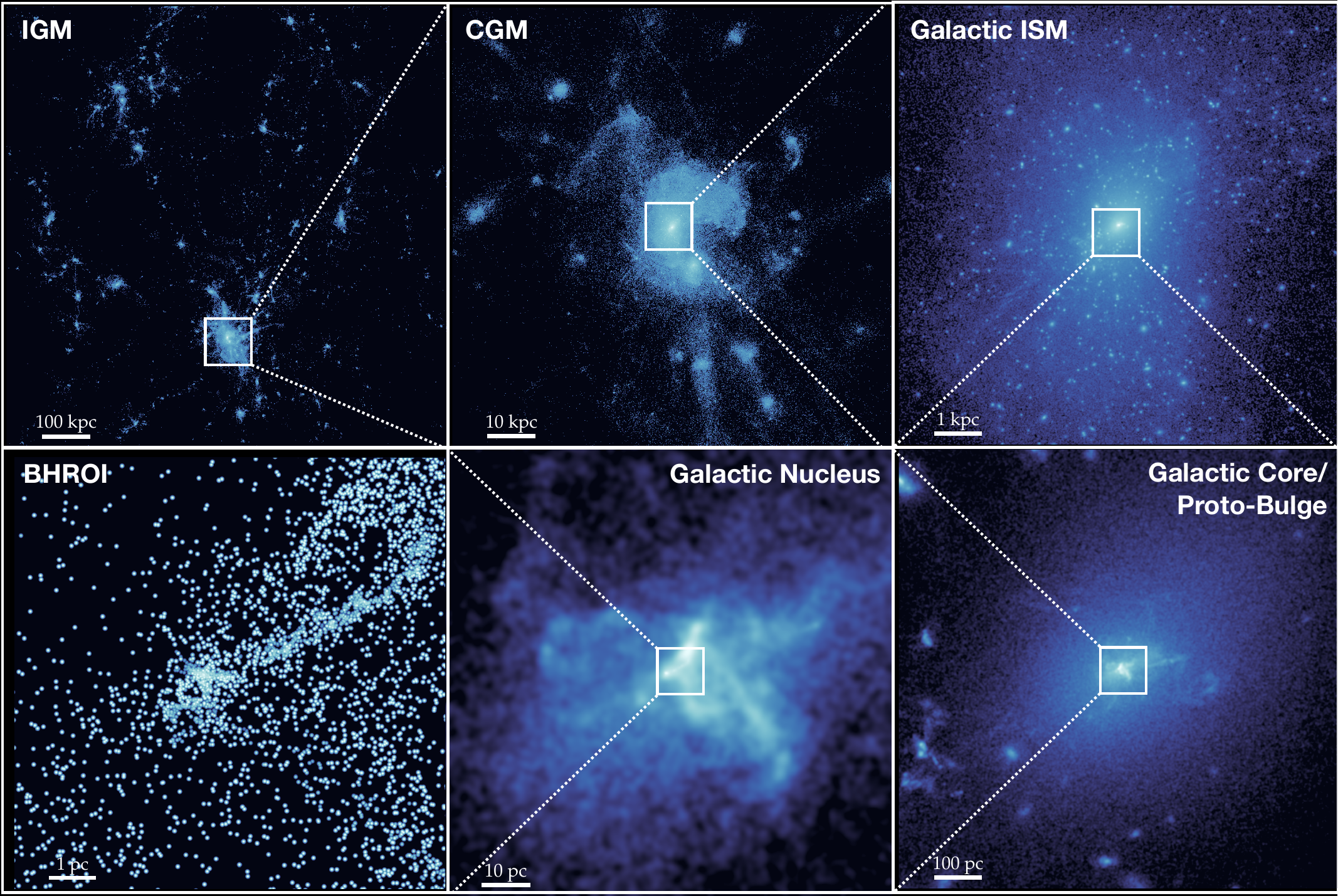} \\
	\caption{As Fig.~\ref{fig:images.faceon}, but in stars. The chaotic merger morphology at a few kpc, and clumpy, highly asymmetric stellar morphology driving gravitational torques on the gas is evident on all scales. In most panels we show a continuous projection of stellar density, but in the last panel this breaks down (the inter-stellar separation is no longer much smaller than a pixel) so we show {\em individual} O-stars. We do not show images at $\ll 1\,$pc because there is a negligible stellar mass compared to the gas on these scales.
	\label{fig:images.star.faceon}}
\end{figure*}

\section{Numerical Methods}
\label{sec:methods}

\subsection{Overview and Common Physics}
\label{sec:methods:overview}

Fundamentally, the simulation suite presented here combines two well-tested numerical physics implementations: the Feedback In Realistic Environments (FIRE) physics (specifically the FIRE-3 version from \citealt{hopkins:fire3.methods}), and STARFORGE physics \citep{grudic:starforge.methods}. Both of these physics modules have been extensively tested in the literature\footnote{For additional numerical validation tests of FIRE methods, we refer to  \citep{hopkins:2013.fire,ma:2015.fire.mass.metallicity,sparre.2015:bursty.star.formation.main.sequence.fire,garrisonkimmel:fire.subhalo.destruction,daa:BHs.on.FIRE,su:fire.feedback.alters.magnetic.amplification.morphology,escala:turbulent.metal.diffusion.fire,ma:fire2.reion.gal.lfs,orr:ks.law,hopkins:fire2.methods,chan:2018.cosmicray.fire.gammaray,garrison.kimmel:2019.sfh.local.group.fire.dwarfs,hopkins:cr.mhd.fire2,pandya:2021.loading.factors.of.fire,wetzel:fire2.public.release,wellons:2022.smbh.growth}, and for the same for STARFORGE, see \cite{grudic:sfe.cluster.form.surface.density,grudic:2022.sf.fullstarforge.imf,guszejnov:2018.isothermal.nocutoff,guszejnov:fire.gmc.props.vs.z,guszejnov:2020.mhd.turb.isothermal.imf.cannot.solve,guszejnov:2020.starforge.jets,guszejnov:2022.starforge.cluster.assembly,lane:2022.turbcloud}.} so we will only summarize what is included and refer to the relevant methods papers for each, in order to focus on what is novel here (how the two are integrated within our refinement scheme). An even more succinct high-level overview is provided in Table~\ref{tbl:physics}.

All of the relevant physics are implemented in the code {\small GIZMO}\footnote{A public version of {\small GIZMO} is available at \gizmourl} \citep{hopkins:gizmo}. The simulations evolve the radiation-magneto-hydrodynamics (RMHD) equations, using the meshless finite-mass MFM scheme (a mesh-free Lagrangian Godunov method). The ideal MHD equations are numerically integrated as described in \citet{hopkins:mhd.gizmo,hopkins:cg.mhd.gizmo} using the constrained-gradient method from \citet{hopkins:cg.mhd.gizmo} for greater accuracy\footnote{We note specifically that these methods have been shown to accurately capture the dynamics of the MRI in a wide variety of problems ranging from idealized test problems to global simulations of warped, asymmetric disks in the Hall MRI regime \citep{hopkins:mhd.gizmo,hopkins:cg.mhd.gizmo,hopkins:gizmo.diffusion,deng:2019.mri.turb.sims.gizmo.methods,deng:2020.global.magnetized.protoplanetary.disk.sims.gravito.turb.leads.to.large.B.saturation.vs.mri,deng:2020.parametric.instab.free.disks,deng:2021.magnetic.disk.frag.in.gizmo.small.planetesimals,deng:2022.warped.disk.dynamics.need.mfm.mfv,deng:2022.particle.transport.ppds}.}, with the addition of non-ideal terms including fully-anisotropic Spitzer-Braginskii conduction and viscosity \citep[implemented as in][including all appropriate terms needed to self-consistently apply them at arbitrary values of temperature or plasma $\beta$ and in both saturated and unsaturated limits]{hopkins:gizmo.diffusion,su:2016.weak.mhd.cond.visc.turbdiff.fx,hopkins:cr.mhd.fire2}, as well as ambipolar diffusion, the Hall effect, and Ohmic resistivity \citep{hopkins:gizmo.diffusion}. The RHD equations are integrated using the M1 moments method \citep[as tested and implemented in][]{lupi:2018.h2.sfr.rhd.gizmo.methods,lupi:2021.low.h2.ortho.para.ratio.gizmo.rhd.sims,lupi:2022.qso.feedback.gizmo.radiation.and.others.z6,hopkins:radiation.methods,hopkins:2019.grudic.photon.momentum.rad.pressure.coupling,williamson:2020.rad.hydro.agn.sims.with.gizmo,williamson:2022.gizmo.rhd.psph.sims.binary.smbh.torii.radiation.reduces.grav.torques,bonnerot:2021.gizmo.rhd.tde.sims} for each of five bands (H ionizing, FUV/photo-electric, NUV, optical-NIR, and an adaptive-wavelength blackbody FIR band).\footnote{This is the same RHD treatment as the default STARFORGE simulations, but is more sophisticated than the default LEBRON treatment used in most (but not all) previous FIRE simulations. However we stress that the FIRE physics are agnostic to the numerical RHD solver; they simply determine the ``look up tables'' used to inject radiation onto the grid, which can then be integrated via any of the RHD solvers implemented in {\small GIZMO} (see \citealt{hopkins:gizmo.public.release}). Moreover as shown in \citet{hopkins:radiation.methods} the two methods for integrating the radiation transport give similar results on large (galactic/CGM) scales.} As described in \citet{grudic:starforge.methods}, this includes the ability to evolve the effective wavelength or temperature of the IR radiation field so as to accurately handle wavelength/temperature-dependent opacities and emission from wavelengths of $\sim 0.1-1000\,{\rm \mu m}$. Also as described therein, we separately evolve the gas, dust, and radiation temperatures and different radiation bands, with the appropriate physical coupling/exchange terms between these, so that the code can self-consistently handle both limits where the various temperatures are arbitrarily de-coupled from one another and limits where they become closely-coupled \citep[e.g.][]{bonnerot:2021.gizmo.rhd.tde.sims,grudic:starforge.methods}. Note that compared to previous STARFORGE or FIRE RHD simulations, we greatly increase the reduced speed of light, with most runs here using $\tilde{c} = 0.1\,c$, though we have tested runs for a limited time with $\tilde{c}=0.01\,c$ and $=1\,c$ (i.e.\ no reduced speed of light at all) to validate that the radiation properties in the galaxy nucleus at $\ll 100\,$pc scales (the regime of greatest interest) are converged. Gravity is solved with an adaptive Lagrangian force softening matching hydrodynamic and force resolution for gas cells, and fixed softenings specified below for the collisionless particles, using a fifth-order Hermite integrator designed to accurately integrate ``hard'' gravitational encounters (e.g.\ close binaries) for the entire duration of our simulation \citep{grudic:starforge.methods}. We explicitly follow the enrichment, chemistry, and dynamics of 11 abundances \citep[H, He, Z, C, N, O, Ne, Mg, Si, S, Ca, Fe;][]{colbrook:passive.scalar.scalings}, allowing for micro-physical and turbulent/Reynolds diffusion \citep{escala:turbulent.metal.diffusion.fire}, as well as a set of tracer species. 

Thermo-chemistry is integrated with a detailed solver described in \citet{grudic:starforge.methods,hopkins:fire3.methods}: we follow all of the expected dominant processes at temperatures of $\sim 1-10^{10}$\,K including explicit non-equilibrium ionization/atomic/molecular chemistry as well as molecular, fine-structure, photo-electric, dust, ionization, cosmic ray, and other heating/cooling processes (including the effects of the meta-galactic background from \citealt{cafg:2020.uv.background}, with self-shielding). Crucially, the explicit radiation-hydrodynamics is coupled directly to the thermo-chemistry: radiative heating, ionization, and related processes scale directly from the local (evolved) radiation field, and cooling radiation is not simply ``lost'' (as assumed in many implementations of optically-thin cooling), but is emitted back into the evolved RHD bands appropriately \citep[see][for various tests demonstrating that this accurately captures the transition between optically thin and thick cooling regimes]{grudic:starforge.methods}. We assume a dust-to-gas ratio which scales as $f_{\rm dg} = 0.01\,(Z/Z_{\odot})\,\exp{(-T_{\rm dust}/1500\,{\rm K})}$, i.e.\ a standard dust-to-metals ratio at low dust temperatures with dust destruction above a critical dust temperature. This allows us to capture the most important dust transition at small radii in these simulations, namely dust destruction within the QSO sublimation radius. We stress that the thermo-chemistry modules are designed to self-consistently include essentially all processes which dominate radiative cooling and opacities from densities $n \ll 10^{-10}\,{\rm cm^{-3}}$ through $n \gg 10^{15}\,{\rm cm^{-3}}$ in proto-stellar disks (with or without dust). We separately account for the dust and gas opacities in each of the ionizing, photo-electric, NUV, optical-NIR, and gray-body IR bands, calculated as an appropriate function of the (distinct) dust and gas temperature and radiation temperature in each band including dust opacities from \citet{semenov:2003.dust.opacities}, bound-free/ionizing, free-free, Rayleigh and Thompson opacities for free $e^{-}$, HI, HII, H$^{-}$, H$_{2}$, CO, and partially-ionized heavy elements evolved, with the abundances of each of these species calculated in the chemical network (see e.g.\ \citealt{john:1988.hminus.opacity,glover:2007.low.metallicity.cooling.h2.tables} and other references in \citealt{hopkins:fire3.methods}). In addition to tests in the diffuse ISM and CGM/IGM limits \citep{hopkins:radiation.methods}, and the (dust-dominated) protostellar disk and molecular cloud limits \citep{grudic:starforge.methods}, we have also validated our opacities against those tabulated in \citet{lenzuni:1991.opacities} for metal-free gas with densities $n \sim 10^{12} - 10^{16}\,{\rm cm^{-3}}$.

We emphasize that all the physics and numerical methods above, including gravity, radiation transport, MHD, and thermochemistry, apply always and everywhere in the simulation: there is no distinction between FIRE and STARFORGE treatments.\footnote{Again note that some previous FIRE simulations using the ``default'' model in \citet{hopkins:fire2.methods} employ a simpler approximate LEBRON radiation-hydrodynamics solver, and FIRE-2 used a simpler thermochemistry module compared to the FIRE-3 version here. However these simplifications are not designed for handling extremely high densities or optically thin-to-thick transitions, so we adopt the more accurate M1 RHD and FIRE-3/STARFORGE thermochemistry detailed above. But we stress that these RMHD and thermochemistry modules have been used (and compared to those simpler modules) in multiple previous FIRE studies \citep[e.g.][and references therein]{hopkins:fire3.methods,hopkins:radiation.methods,hopkins:cr.spectra.accurate.integration,hopkins:cr.multibin.mw.comparison,shi:2022.hyper.eddington.no.bhfb,schauer:fire.dwarf.w.baryon.streaming.vels} as well as STARFORGE \citep{guszejnov:2020.starforge.jets,grudic:starforge.methods,grudic:2022.sf.fullstarforge.imf,guszejnov:2022.starforge.cluster.assembly,guszejnov:environment.feedback.starforge.imf,guszejnov:starforge.environment.multiplicity}, and many additional simulations using {\small GIZMO}, referenced above.} The {\em only} difference between the FIRE and STARFORGE limits in our simulations lies in how we treat ``stars.'' Specifically, when a gas cell is eligible for ``star formation'', we must decide whether to convert it into a ``single stellar population (SSP) particle'' which represents a {\em statistically-sampled ensemble} of multiple stars (the FIRE limit, relevant at lower resolution/large cell masses) or to convert it into a ``sink/single (proto)star particle'' which represents a {\em single} (proto)star (the STARFORGE limit, relevant at higher resolution/small cell masses). Those particles each then use their distinct appropriate (SSP or single-star) evolutionary tracks to calculate the mass/momentum/energy/cosmic ray/photon fluxes which are deposited back onto the grid, at which point that injected material is again evolved identically according to the algorithms above. Thus both operate simultaneously in the simulation.

\subsection{FIRE Treatment of Stars (Relevant for Large Cell Masses)}
\label{sec:methods:fire.mode}

FIRE was designed for galaxy-scale simulations, with resolution sufficient to resolve some phase structure in the ISM, but insufficient to resolve {\em individual} proto-star formation and stellar growth/evolution histories. As such, we apply the FIRE treatment of stars when the resolution is still sufficiently low (cell mass $>1\,M_{\odot}$), using the FIRE-3 implementation in \citep{hopkins:fire3.methods} and summarized above. In this limit, gas is eligible for star formation if it is locally self-gravitating at the resolution scale, Jeans unstable, and in a converging flow, as in \citet{hopkins:virial.sf,grudic:sfe.cluster.form.surface.density}. The intention here is not to resolve e.g.\ local ``peaks'' in the density field which will become {\em individual} stars, but rather to identify ``patches'' of the ISM where the fragmentation cascade becomes unresolved, so the gas should continue to fragment and ultimately form a population of stars. As such, for the cells which meet this criterion, we assume fragmentation on a dynamical time \citep[per][]{hopkins:fire2.methods} and convert them into ``single stellar population (SSP) particles'' -- i.e.\ collisionless particles which represent ensemble populations of stars which sample an assumed universal stellar IMF. 

Once formed, these SSP particles evolve as detailed in \citet{hopkins:fire3.methods} according to explicit stellar evolution models from the 2021 version of STARBURST99 \citep{2014ApJS..212...14L}, and return metals, mass, momentum, and energy to the ISM via resolved individual SNe (both Ia \&\ core-collapse) and O/B and AGB mass-loss as in \citet{hopkins:sne.methods}, with radiative heating and momentum fluxes determined from the stellar population spectra as in \citet{hopkins:radiation.methods}, appropriate for a \citet{kroupa:2001.imf.var} IMF. Cosmic rays are injected in fast stellar wind or SNe shocks as described in \citet{hopkins:fire3.methods}, using the approximate method from \citet{hopkins:2022.cr.subgrid.model}. Once injected onto the grid, gas/radiation/cosmic rays obey the physics in \S~\ref{sec:methods:overview}.

To deal with intermediate resolution cases, we employ the stochastic IMF sampling scheme from \citet{ma:2015.fire.escape.fractions,su:discrete.imf.fx.fire,wheeler:ultra.highres.dwarfs,grudic:2019.imf.sampling.fx.on.gmc.destruction}: when a SSP particle forms, we draw a quantized number of massive stars from an appropriate binomial distribution, from which the relevant feedback properties particular to (rare) massive stars (e.g.\ core-collapse SNe, ionizing radiation) scale. This allows us to at least approximately apply SSP particles down to gas mass resolution $\sim 1-10\,M_{\odot}$, where the resolution is still too poor to explicitly resolve individual (proto)star formation, but so high that the expected number of massive stars per star particle is $\ll 1$ and as such discreteness effects could be important \citep[see discussion in][]{ma:2015.fire.escape.fractions,ma.2016:binary.star.escape.fraction.effects,ma:2020.no.missing.photons.for.reion.supershells,grudic:2019.imf.sampling.fx.on.gmc.destruction}.

\subsection{STARFORGE Treatment of Stars (Relevant at Small Cell Masses)}
\label{sec:methods:starforge.mode}

STARFORGE, on the other hand, was designed for simulations which resolve individual (proto)star formation and evolution, e.g.\ simulations of individual molecular clouds, clumps, star clusters, or (proto)stellar disks. In this limit, each sink represents a single star, which obviously means it cannot meaningfully represent systems with resolution poorer than $\gtrsim 1\,M_{\odot}$. As such, we apply the STARFORGE treatment of stars when the resolution is sufficiently high (cell mass $<1\,M_{\odot}$). In this limit, gas is eligible for (proto)star formation if it meets a standard but more stringent set of seed criteria described in \citet{grudic:starforge.methods}, including a strict virial/self-gravity, Jeans, converging flow, fully-compressive tides, and local density/potential maximum criteria as well as restricting to gas cells without a pre-existing neighboring sink and requiring their collapse time be much shorter than the infall time onto the nearest sink (whatever its distance). If all of these criteria are met, a cell is immediately converted into a sink or {\em individual star} particle. 

Once formed, each sink accretes gas that is bound (accounting for thermal, kinetic, and magnetic energies) to it, and whose current and circularized radii fall within the sink radius (set comparable to the force softening), following a standard strict sink accretion model validated in a variety of idealized accretion problems (details in \citealt{grudic:starforge.methods}). The sinks evolve along combined proto and main-sequence stellar evolution tracks, explicitly following the stellar evolution physics versus time (e.g.\ contraction, heating, different burning stages) allowing for the dynamic accretion rate in every timestep \citep{Offner_2009_radiative_sim}. In the proto-stellar evolution stage, sinks radiate in all bands with the appropriate effective temperature, and launch collimated jets with a mass-loading proportional to the surface accretion rate onto the star and a launch velocity comparable to the escape velocity from the protostellar surface (details in \citealt{grudic:starforge.methods,guszejnov:2020.starforge.jets,grudic:2022.sf.fullstarforge.imf}). Main-sequence stars continue to emit jets and accretion luminosity if accretion continues, and radiate in all bands following their stellar evolution-track calculated effective temperatures and full spectra, while also emitting continuous stellar surface winds (assumed to be isotropic in the frame of the star, with a continuous main-sequence mass-loss rate given by \citealt{grudic:2022.sf.fullstarforge.imf} Eq. 1), as a function of the instantaneous stellar luminosity. At the end of their main-sequence lifetime stars can, if sufficiently massive, explode as SNe with $10^{51}\,{\rm erg}$ of ejecta kinetic energy. Again we refer to \citet{grudic:starforge.methods} for details. 

Importantly, we note that these physics have been validated by direct comparison to dense molecular gas properties, the observed stellar IMF and multiplicity statistics, and star cluster properties \citep{guszejnov:2020.starforge.jets,guszejnov:2022.starforge.cluster.assembly,grudic:2022.sf.fullstarforge.imf,lane:2022.turbcloud}, for typical Milky Way-like galaxy conditions. In \citet{hopkins:superzoom.imf} (henceforth \paperthree) we will study the predicted IMF from the simulations here around the SMBH; for now we note that although it becomes top-heavy very close to the SMBH ($\ll 1\,$pc), this analysis appears to justify the choice of a more universal \citet{kroupa:2001.imf.var}-like IMF at the larger radii $\gtrsim 10\,$pc assumed by our FIRE SSP particles.

\subsection{Initial Conditions and Refinement Choices}
\label{sec:methods:ics}

Our initial condition (see Fig.~\ref{fig:images.faceon}) is a fully-cosmological ``zoom-in'' simulation, evolving a large box from redshifts $z\gtrsim 100$, with resolution concentrated in a $\sim 10\,$Mpc co-moving volume centered on a ``target'' halo of interest (specifically, halo ``A1'' aka ``m12z4'' studied in \citealt{feldmann.2016:quiescent.massive.highz.galaxies.fire,feldmann:colors.highz.quiescent.massivefire.gals,oklopcic:clumpy.highz.gals.fire.case.study.clumps.not.long.lived,daa:BHs.on.FIRE,hopkins:cr.mhd.fire2,ma:2021.seed.sink.inefficient.fire,wellons:2022.smbh.growth}). While there are many smaller galaxies in that volume, for the sake of clarity we focus just on the properties of the ``primary'' (i.e.\ best-resolved) galaxy in the volume. The dynamic refinement scheme employed here is numerically identical to that used in many previous {\small GIZMO} studies,\footnote{Briefly, gas cells are assigned a target resolution $\Delta m^{\rm target}$ as a function of e.g.\ BH-centric distance: if they deviate by more than a tolerance factor $\sim 2$ shown below, they are refined in pairs (or de-refined via merging with their nearest neighbor if both are eligible for de-refinement) maintaining all numerically-integrated conserved quantities (mass, momentum, energy, etc.) to machine precision, with the refinement into pairs of equal-mass Lagrangian cells separated along the axis of lowest pre-existing neighbor cell number density (akin to an approximate local Voronoi-type remeshing). For details see \citet{hopkins:gizmo,su:2018.stellar.fb.fails.to.solve.cooling.flow,daa:20.hyperrefinement.bh.growth}. The full algorithm is provided in the public {\small GIZMO} code.} including examples which have refined to similar resolution around single or binary SMBHs, just without the use of the hybrid FIRE-STARFORGE physics described above \citep[see][]{orr:ks.law,hopkins:fire2.methods,su:2018.stellar.fb.fails.to.solve.cooling.flow,su:turb.crs.quench,su:2021.agn.jet.params.vs.quenching,benincasa:2020.gmc.lifetimes.fire,daa:20.hyperrefinement.bh.growth,franchini:2022.binary.bh.hyperzoom.dynamics}.

The simulation is run from $z\sim 100$ down to some redshift (here $z<4$), using a dynamic refinement scheme that adaptively varies the mass resolution, as a function of the minimum of either the thermal Jeans mass (ensuring it is resolved by $\sim 100$ cells) or a function of distance to the nearest SMBH particle (progressively moving from minimum refinement at $>100$\,kpc from the nearest SMBH to maximum refinement at $<10$\,kpc), between an imposed minimum refinement cell mass of $\approx 4000\,M_{\odot}$ and maximum refinement mass of $\approx 10^{6}\,M_{\odot}$. To ensure even low-density, non-self-gravitating multi-phase structure is reasonably resolved, once refined a cell cannot be de-refined unless it escapes far from the galaxy (though in practice, given the short simulation duration after reaching maximum resolution and relatively weak winds from the central sub-pc scales, very few cells of high refinement escape to interact with low-resolution cells). This allows us to resolve the galaxy at $\sim 4000\,M_{\odot}$ resolution through formation and initial growth of the SMBH, through a redshift of $z \approx 4$. 

We then select a specific time $t_{0}$ (at a redshift $z_{0} \approx 4.4$) from this original simulation just before a period where it (at its relatively modest resolution) identified rapid quasar-level SMBH growth, with one clearly-dominant SMBH particle within the galaxy nucleus. We will show the galaxy properties at this time in great detail below, but to summarize at $z\approx 4.4$ it has a dark matter halo mass of $\sim 3\times10^{12}\,{\rm M_{\odot}}$ inside $r<250\,$kpc, and a galaxy stellar mass of $2\times10^{10}\,{\rm M_{\odot}}$ (and very similar gas mass) inside $<10$\,kpc (with stellar half-mass radius of $\sim 1.5\,$kpc), and a nuclear SMBH mass of $\sim10^{7}\,{\rm M_{\odot}}$. We then re-start the simulation from this time, with an additional refinement layer: on top of the refinement scheme above, we apply a multiplier $f(r \equiv |{\bf x} - {\bf x}_{\rm SMBH}|,\,t-t_{0})$ to the ``target'' mass resolution, which is a continuous function of $r$ with $0<f\le 1$, where at small radii $f\propto r^{3}$ and at radii $\gtrsim 1\,{\rm kpc}$, $f=1$ exactly.\footnote{For comparison, \citet{daa:20.hyperrefinement.bh.growth} used a shallower $\propto r^{2}$ refinement with a maximum resolution of $\Delta m \sim 15\,M_{\odot}$ and $\sim 0.1\,$pc.} We reduce the minimum allowed/target cell mass to $\Delta m \lesssim 0.01\,M_{\odot}$. 

To avoid pathological behaviors, this refinement layer does not ``instantly'' activate but appears as a smooth function of time $t - t_{0}$, designed such that each concentric radius $r$ (beginning at $\sim 1\,$kpc interior to which the refinement begins) is evolved for a few local dynamical times before the next ``layer'' of refinement is applied interior to this. Specifically, the refinement is still continuous in both space and time. For each decade in $r$ (between $r_{\rm max}$ and $r_{\rm min} = r_{\rm max}/10$) until we reach our maximum target resolution, if the desired final refinement function $f(r,\,t\rightarrow \infty) \propto r^{\alpha_{f}}$, we take $f \propto r^{\alpha(t)}$ with $\alpha(t) = \alpha_{f}\,(1-\exp{\{ (t-t_{i}/\Delta t) \}})/(1-\exp{\{ (t_{f}-t_{i}/\Delta t) \}})$ after time $t_{i}$ equal to $t_{f}$ of the previous (decade-larger) interval, with $\Delta t \sim 1/\Omega$ (at the midpoint of the annulus) and $t_{f} - t_{i} \sim 5\,\Delta t$, with the normalization of $f$ set to the value from the previous annulus at $r_{\rm max}$ to ensure continuity, and $f=$\,constant at $r<r_{\rm min}$ for $t<t_{f}$. In \S~\ref{sec:resolution} below, we explicitly show and discuss the effective final ``ramp'' refinement functions in both space and time. We also enforce a requirement that no cell can refine or de-refine within less than $30$ numerical timesteps of its last refinement/de-refinement (where the numerical timesteps ensure this is some multiple of the relevant signal-crossing timescale). This gradual ``ramp'' of the refinement in both space and time is designed to reduce initial noise and spurious features imprinted by e.g.\ the low-resolution cell geometry (see discussion in \citealt{daa:20.hyperrefinement.bh.growth} and other references above). Note we have varied the ramp function ``speed'' in both space and time (and timestep interval) by factors of a few, and see no systematic effects from this. In practice, the biggest problem if one refines ``too quickly'' in space or time is not numerical interactions of different-mass cells (which the numerical methods here are specifically designed to be robust to, see \citealt{hopkins:gizmo}), but rather overloading the memory of the CPU nodes with the sudden creation of huge numbers of cells and radical restructuring of the gravity tree (occurring on multiple nodes simultaneously), leading to code crash or memory errors. The biggest problem if one refines too slowly in space or time is simply that one cannot practically reach the desired target resolution by the radii and times desired for zooming in on the formation of the quasar accretion disk. In summary, this means that the total duration of the simulation after the beginning of the ``hyper-refinement'' period is a couple of Myr, but after the highest resolution level (resolving orbits at $\lesssim 100\,$au around the SMBH with an orbital dynamical time of $\sim 10-20$\,days) is reached at a redshift closer to $z\sim 4$, we evolve for $\sim 10^{4}\,$years. 

At our highest resolution level, our mass resolution is $\Delta m \sim 0.001-0.01\,M_{\odot}$, with spatial resolution $\Delta x \sim 10^{-5}-10^{-4}\,$pc, so we can resolve densities up to $\sim 10^{13}-10^{15}\,{\rm cm^{-3}}$ and timescales down to $\sim 1\,$day, with an ``effective'' grid resolution across our box equivalent to $N_{\rm eff} \sim (10^{13})^{3}$. More details are shown in \S~\ref{sec:resolution} below.

\begin{figure}
	\centering\includegraphics[width=0.98\columnwidth]{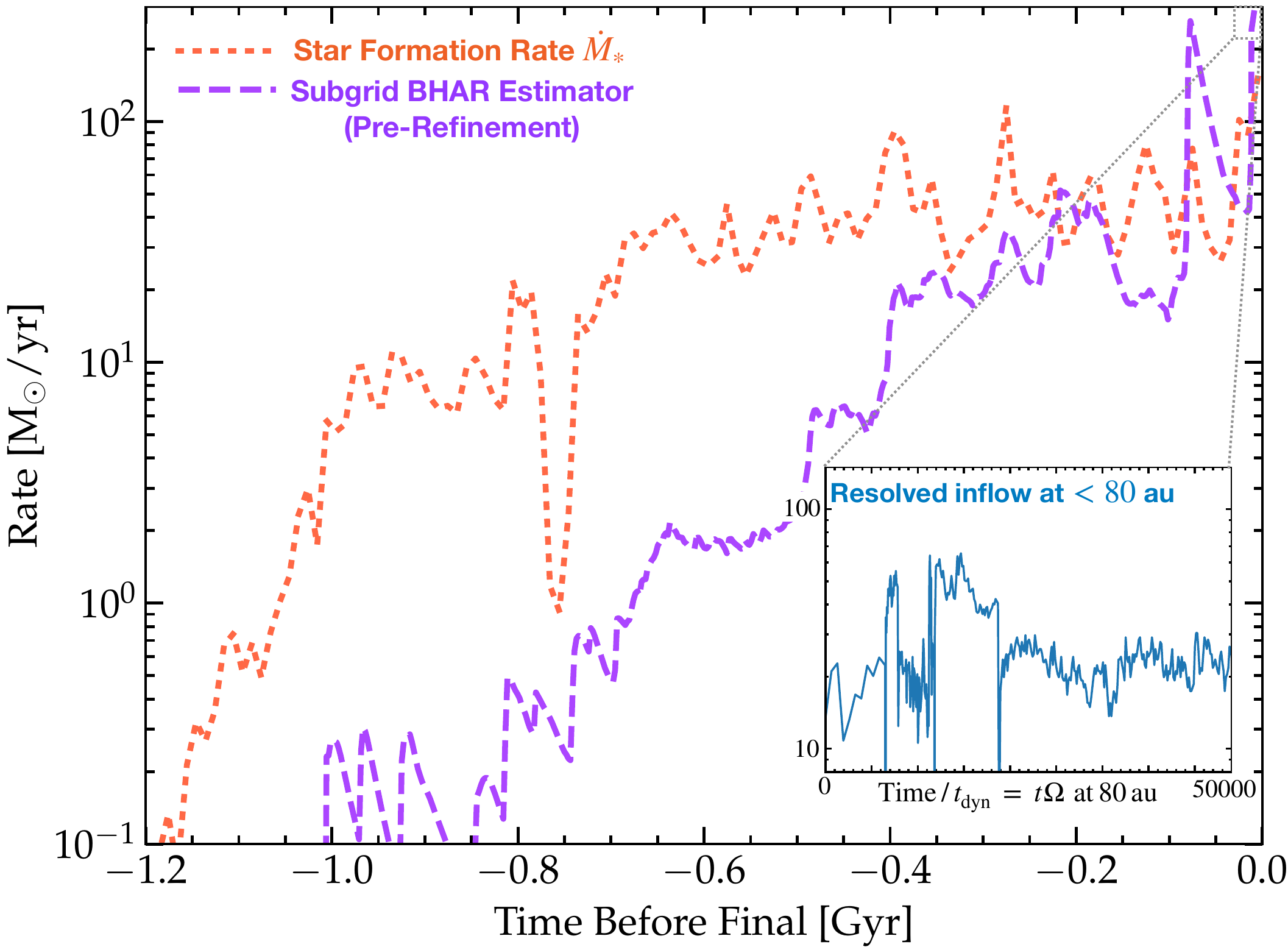}
	\caption{Illustration of the time evolution of the main galaxy in our simulation before our hyper-refinement. We plot the galaxy-integrated SFR $\dot{M}_{\ast}$, and the {\em sub-grid} estimated BH accretion rate (BHAR) from the model -- which scales approximately as $\dot{M}_{\rm subgrid} \sim \eta\,M_{\rm gas}\,\Omega$ at the low-resolution limit of $\sim 10-100\,$pc before hyper-refinement is turned on -- as a function of time prior to the time of refinement (at $z\sim 4.4$). The inset shows the {\em resolved} gas inflow rate into the central sink resolution around the SMBH of $<80\,$au, as a function of time in units of the dynamical time $t_{\rm dyn}\equiv 1/\Omega$ at this resolution scale ($\sim 80\,$au), over the final $\sim 1500\,$yr of our simulation duration (well after it reaches the maximum refinement level everywhere). Though this is a very short relative timescale (compared to the order Hubble time evolution on large scales), we see that the inflow rate into the inner accretion disk is quite stable over tens of thousands of dynamical times in the center, for a given set of conditions at larger radii. We also see extremely high inflow rates, as expected based on the high nuclear gas masses and densities in the ``parent'' simulation which motivated the choice of this particular moment in time to ``zoom in.''
	\label{fig:previous.sim}}
\end{figure}

\subsection{Types of Cells/Particles}
\label{sec:methods:particles}

In summary, our simulations include five types of cells or particles:
\begin{enumerate}[labelindent=0pt,labelwidth=10pt,labelsep*=0pt,leftmargin=!,align=parleft]
\item{{\bf Gas/Radiation Cells:} These define the effective mesh on which the equations of radiation-magneto-hydrodynamics are solved, including thermochemistry and all the physics described above. The mesh resolution (for gravity and all other forces) is adaptive with spatial resolution $\Delta x = (\rho/\Delta m)^{1/3}$, and $\Delta m$ ranges smoothly from $\lesssim 0.01\,M_{\odot}$ in the high-resolution region around the SMBH (after the ``hyper-refinement'' phase begins) to a median of $\sim 5000\,M_{\odot}$ in the $\sim$\,kpc-scale ISM of galaxies to $\sim 10^{6}\,M_{\odot}$ in the diffuse IGM. The physics/equations integrated for the gas are independent of resolution -- the choice of FIRE or STARFORGE physics only appears as a choice of whether cells should form SSP or sink/single-star particles.}

\item{{\bf Dark Matter Particles:} Dark matter is represented in standard fashion by collisionless particles which interact only via gravity. In the low-resolution cosmological box (well outside of the high-resolution region) these particles have lower resolution in factor-of-two increments (with the poorest resolution $\sim 4\times10^{10}\,M_{\odot}$ in the $\sim 100\,$Mpc region), but following standard practice we have confirmed that within $\sim 500\,$kpc of the ``target'' galaxy of interest, there are zero low-resolution dark matter particles. The high-resolution dark matter particles have $\Delta m \sim 10^{6}\,M_{\odot}$ and a force softening equivalent to $\Delta x \approx 200\,$pc. While crucial for cosmological evolution and galactic-scale dynamics, the dark matter contributes negligibly to the nuclear dynamics (contributing just $\sim 6\%$ of the total mass inside $< 200\,$pc and a vanishingly small fraction of the mass inside $<20\,$pc),\footnote{These factors are taken from the resolved simulation, but we obtain similar results if we extrapolate a steeper-than-NFW profile $\rho_{\rm DM} \propto r^{-3/2}$ from $1\,$kpc to the same radii. Comparison to the stellar and gas mass profiles implies the dark matter density would have to rise much more steeply at small $r \ll 1\,$kpc in order to contribute significantly to the nuclear dynamics.} and the force softening is sufficiently large that the worst-scale $N$-body deflection from an encounter between a high-resolution baryonic cell and DM particle would be no larger than the acceleration/deflection for a gas cell with density $n \lesssim 1\,{\rm cm^{-3}}$ (vastly lower density/acceleration scales than those of interest in the galaxy nucleus).\footnote{We validated this directly in post-processing, calculating the acceleration and torques on nuclear gas from all particle types.}}

\item{{\bf SMBH Particles:} SMBHs are represented by collisionless sink particles. In the ``pre-simulation'' phase (running from redshift $z\sim 100$ to the time $t_{0}$ ($z\sim4.4$) when we begin our hyper-refinement), the BHs are formed and evolve according to the default sub-grid FIRE-3 seeding, dynamics, accretion, and feedback models described in \citet{hopkins:fire3.methods}. But this is only relevant in that it gives us a plausible initial condition for our hyper-resolution run. Once the hyper-refinement phase begins, we disable all ``sub-grid'' models for BH growth and accretion: the BHs are represented by sinks that follow normal gravitational dynamics. Any cells/particles which fall inside of the SMBH sink radius set to $\approx 80\,{\rm au} \sim  300\,R_{s}$ (where $R_{s}=2\,G\,M_{\rm BH}/c^{2}$) are immediately captured (removed from the domain and added to the sink mass).\footnote{At this capture radius ($\approx 80\,$au), the escape velocity is $\sim 2 \times10^{4}\,{\rm km\,s^{-1}}$, and all the accreted material is tightly bound to the SMBH.} We do not include SMBH ``feedback'' from within the sink radius during this phase. For simplicity, we choose a time where the primary galaxy of interest contains just one SMBH: a sink with mass $\approx 1.3\times10^{7}\,M_{\odot}$.}

\item{{\bf Single Stellar Population (SSP) Particles:} Unresolved fragmentation to stars formed in the ``FIRE limit'' (\S~\ref{sec:methods:fire.mode}) produce SSP particles which represent stellar populations. The mass of an SSP particle is fixed to the mass of the gas cell from which it forms, so varies smoothly with radius accordingly. The force softening is fixed for each SSP particle after it forms but depends on its mass with the scaling $\Delta x \approx 0.6\,{\rm pc}\,(\Delta m/100\,{\rm M_{\odot}})^{1/3}$, equivalent to a pre-formation gas density $\approx 2\times10^{4}\,{\rm cm^{-3}}$ typical of where cells of this resolution form stars.\footnote{This ensures the ``worst-case'' N-body deflection is never stronger than typical encounters between gas and star-forming gas clumps/clouds, and is much smaller than any interactions with gas cells at the median gas density inside $\lesssim 100\,$pc. Given the chaotic, turbulent nature of the galaxy we follow, the discrete $N$-body heating rate estimated as in \citet{hopkins:fire2.methods} or \citet{ma:2022.discrete.df.estimator} is several orders of magnitude smaller than the typical turbulent dissipation rates in the simulation.} These particles act on the surrounding medium via continuous stellar mass-loss, radiation, and SNe, sampling from a universal stellar IMF, as described in \S~\ref{sec:methods:fire.mode}.}

\item{{\bf Sink/Single (Proto)Star Particles:} Resolved collapse to {\em individual} stars in the ``STARFORGE limit'' (\S~\ref{sec:methods:starforge.mode}) produces sink or individual (proto)star particles, each of which represents a single stellar object or system. The mass of a sink reflects the current mass of the star: at formation, the initial mass equals that of its progenitor gas cell but it grows via resolved accretion. The force softening/sink radius for these is set to a fixed value with a Plummer-equivalent $\epsilon \approx 25$\,au ($\Delta x \approx 40\,$au). These stars follow individual evolution tracks and act on the surrounding medium via radiation, jets, mass-loss, and SNe as described in  \S~\ref{sec:methods:fire.mode}.}

\end{enumerate}

\begin{figure*}
	\centering\includegraphics[width=0.88\textwidth]{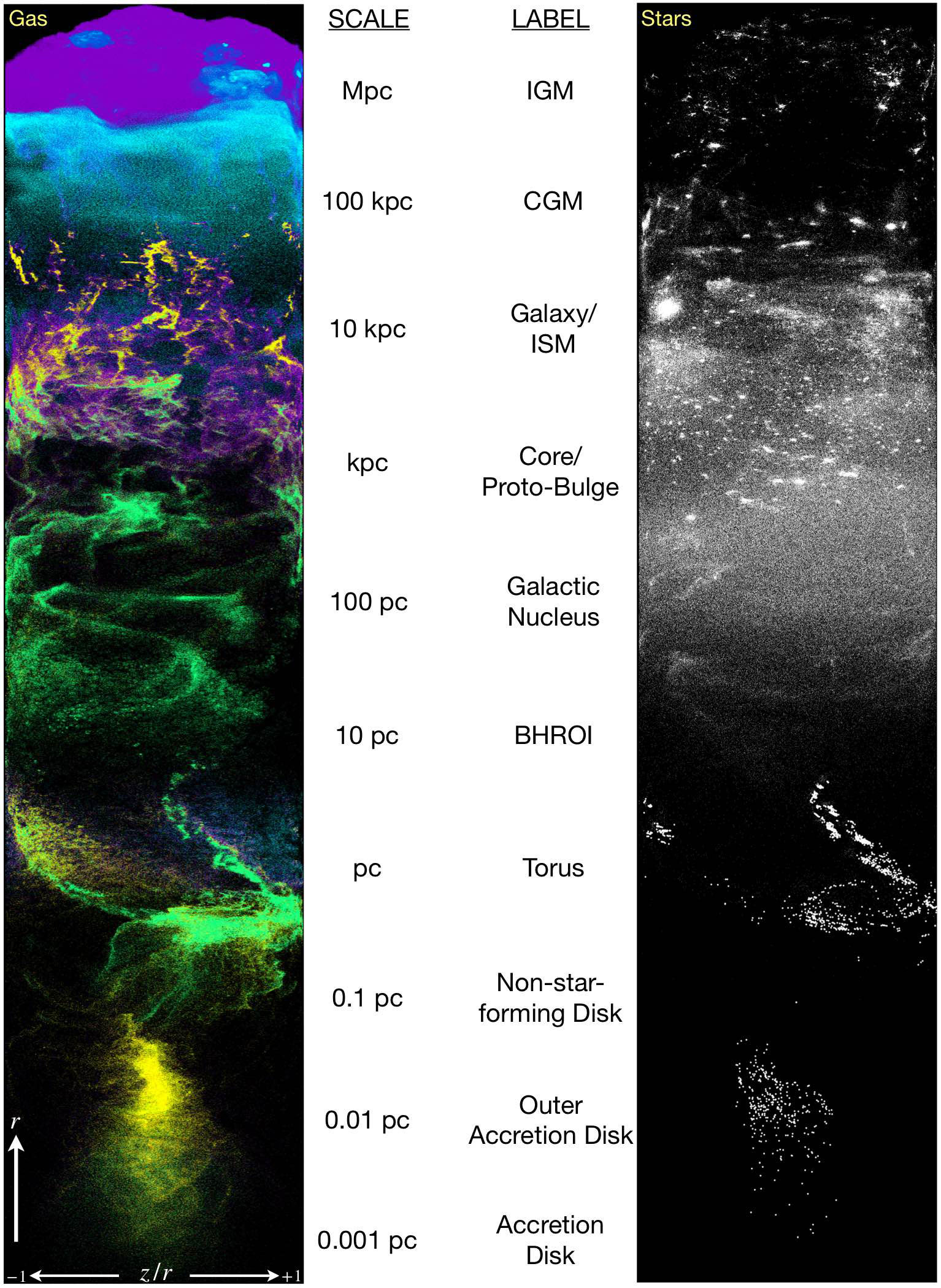}
	\caption{Images of the gas ({\em left}) and stars ({\em right}) with different scales and their approximate naming label conventions from \S~\ref{sec:scales} shown. The images show BH-centric radius $r$ increasing from bottom-to-top (the vertical axis) on a logarithmic scale as labeled. The horizontal axis shows $\cos{\theta} \equiv z/r$ from $-1$ to $+1$ (defined so $z=0$ corresponds to the midplane of the inner disk), in a wedge of azimuthal opening angle $\sin{\phi} < 0.3$. For gas ({\em left}) colors denote different phases $T<10^{3}\,$K ({\em green}), $10^{3}<T<10^{4}$\,K ({\em yellow}), $10^{4}<T<10^{5}\,$K ({\em magenta}), $10^{5}<T<10^{6}\,$K ({\em purple}), $T>10^{6}\,$K ({\em cyan}). We see other galaxies on IGM scales, the virialized CGM with accretion in warm clumps/filaments, the highly clumpy/inhomogeneous/asymmetric and multi-phase structure in the galaxy and (thermally colder, primarily atomic+molecular) galaxy nucleus, settling into the more ordered (but still visibly thick and turbulent) non-star forming disk and BH accretion disk on sub-pc scales.
	\label{fig:scale.labels}}
\end{figure*}

\section{Results}
\label{sec:results}

We summarize some of the results for our fiducial simulation in Figs.~\ref{fig:images.faceon}-\ref{fig:images.star.faceon}, showing images of the simulation on scales from $>$\,Mpc to $< 100$\,au. Specifically, Figs.~\ref{fig:images.faceon} \&\ \ref{fig:images.star.faceon} show images of the projected simulation gas and stellar mass densities, viewed from the same viewing angle (chosen to be face-on to the accretion disk in the center), on a range of scales. Fig.~\ref{fig:previous.sim} illustrates the pre-history of the large-scale ``parent'' simulation, for reference.

\subsection{Different Characteristic Scales/Regimes}
\label{sec:scales}

We can clearly see in Figs.~\ref{fig:images.faceon}-\ref{fig:images.star.faceon} that the dynamic range spanned by the simulation is enormous -- a factor of $> 10^{9}$ in black-hole centric radii, and more like $\sim 10^{13}$ if we compare our smallest spatial resolution at radii $\sim 10^{-3}-10^{-2}$\,pc from the SMBH, to the size of our entire cosmological box. Fig.~\ref{fig:scale.labels} shows an alternative illustration, plotting the gas and stars on a logarithmic scale and identifying the different scales with the labels below, in an attempt to visualize the qualitative structures and phases of gas on each scale. It is difficult to actually describe so many orders of magnitude in scale at once, so we break the scales from $10^{-3} - 10^{6}\,$pc in SMBH-centric radius down into each order-of-magnitude and both assign a characteristic label for these scales and describe some of the key physics and processes occurring. From largest to smallest scales, we follow gas inflows as follows:

\begin{itemize}[labelindent=0pt,labelwidth=10pt,labelsep*=0pt,leftmargin=!,align=parleft]

\item{IGM $\rightarrow$ CGM:} On scales $\gg 100\,$kpc, the IGM is ``cool'' (temperatures $\sim 10^{4}$\,K), diffuse ($\rho \ll 10^{-2}\,m_{p}\,{\rm cm^{-3}}$), quasi-spherical ($H/R \sim 1$), dark-matter dominated, weakly-magnetized ($\beta_{\rm plasma}\equiv P_{\rm thermal}/P_{\rm magnetic} = n\,k_{B}\,T/(|{\bf B}|^{2}/8\pi) \gg 100$, with $|B|\sim 1-10\,{\rm nG}$), with weak outflows and strong primarily-radial (so $\Pi_{rr} \equiv \langle \rho\,v_{r}\,v_{r} \rangle $ dominates the kinetic stress tensor) super-sonic inflows of $\sim 300\,{\rm M_{\odot}\,yr^{-1}}$ onto the halo. Essentially gas is in free-fall collapsing with dark matter via the cosmic web. Since this has been well-studied and resolved in many previous simulations, it is not our goal to study this regime in detail here, but what we see is consistent with the most previous studies with the FIRE simulations \citep{hafen:2018.cgm.fire.origins,hafen:2019.fire.cgm.fates,butsky:2020.cr.fx.thermal.instab.cgm,hopkins:2020.cr.outflows.to.mpc.scales,ponnada:fire.magnetic.fields.vs.obs,ji:fire.cr.cgm,ji:20.virial.shocks.suppressed.cr.dominated.halos,li:2021.low.z.fire.cgm.probes,stern21_ICV,esmerian:2021.fire.precipitation.threshold.in.halos,kim:2022.hot.gas.cgm,butsky:2022.cr.kappa.lower.limits.cgm} as well as results from other codes and semi-analytic models \citep{hummels:resolution.vs.cgm.abs,pandya:2022.cgm.sam.model.reservoirs.taken.from.fire} and standard observational inferences \citep[see][reviews]{tumlinson:2017.cgm.review,chen:2020.kbss.vs.fire.mocks,lan:2020.cgm.b.fields.rm}.

\item{CGM $\rightarrow$ Galactic ISM:} On scales $\sim 10-100\,$kpc, the volume-filling gas in the CGM is shock-heated to virial temperatures $\sim 10^{6}\,$K, with $\beta_{\rm plasma} \sim 100$ and trans-sonic or sub-sonic turbulence, mostly ionized, with thermal pressure comparable to the total pressure and gravity. But the gas is multi-phase, with accretion and outflows of comparable magnitude, with outflows prominent in the diffuse/volume filling phases and inflows dominated by accretion of ``cool'' ($\lesssim 10^{5}\,$K) gas along filaments with densities $\sim 100$ times larger than the median background hot gas, lower $\beta_{\rm plasma} \sim 10$, and velocities of order the free-fall speed in a dark-matter dominated potential. This is essentially the classic ``cold flows in hot halos'' picture, again consistent with many previous theoretical studies \citep{keres:hot.halos,dekelbirnboim:mquench,brooks:2009.coldflow.disk.assembly,keres:cooling.revised,keres:fb.constraints.from.cosmo.sims,fgkm2011,sijacki:2011.gadget.arepo.hydro.tests,keres:2011.arepo.gadget.disk.angmom,vogelsberger:2011.arepo.vs.gadget.cosmo,stern2020} and more recent observations \citep{ribaudo:2011.cold.flow.evidence,kacprzak:2012.accretion.flow.evidence,vayner:2022.cold.flows.onto.quasars}.

\item{Galactic ISM $\rightarrow$ Galactic Core/Proto-Bulge:} On scales $1-10$\,kpc in the galaxy, the gas is highly multi-phase with self-shielding of the UV radiation field ($\Sigma_{\rm gas}\gtrsim 10\,{\rm M_{\odot}\,pc^{-2}}$) allowing formation of ``cold'' neutral medium (CNM) and molecular medium with $T \ll 10^{4}$\,K, alongside hot gas with $T\gtrsim 10^{7}$\,K from SNe, while gas densities range from $\lesssim 10^{-2}\,m_{p}\,{\rm cm^{-3}}$ to $\gg 10\,m_{p}\,{\rm cm^{-3}}$ in cold cloud complexes (and $\beta_{\rm plasma}$ similarly ranges from $\sim 0.1$ in cold phases to $\sim 1-10$ in warm phases and $\sim 100$ in the most diffuse volume-filling phases, and $|B|\gtrsim $\,a few $\mu {\rm G}$). These cold complexes maintain most of the SF, with a SFR inside $<10\,$kpc of $\sim 50-100\,{\rm M_{\odot}\,yr^{-1}}$ (over the last $\sim 100\,$Myr).  The potential becomes dominated by stars inside a few kpc (the galaxy effective radius). The turbulence is mildly super-sonic (sonic $\mathcal{M}_{s} \sim 1-$\,a few) in a volume-averaged sense (with the volume-average dominated by warm ionized media [WIM] and warm neutral media [WNM] at $\sim 10^{4}$\,K), but highly super-sonic ($\mathcal{M}_{s} \sim 10-100$) in the ``cold'' phases. Most of the gas is atomic or molecular. While turbulence maintains an effective volume-averaged $Q\sim$\,a few as at all larger radii, the {\em thermal} Toomre $Q$ parameter drops to $\ll 1$ in the cold phases, in particular, meaning that fragmentation via self-gravity is rapidly promoted, with the characteristic ``most unstable'' fragment masses expected to contain most of the power in the fragment mass spectrum (e.g.\ the largest self-gravitating complexes) ranging from $\sim 10^{7}$ to a few $10^{9}\,M_{\odot}$ (larger than in low-redshift galaxies, owing to the massive gas content of this dense, high-redshift galaxy, similar to complexes observed at high redshift). Again this is broadly consistent with previous theoretical \citep{noguchi:1999.clumpy.disk.bulge.formation,bournaud:chain.gal.model,agertz:disk.fragmentation.model,dekel:2009.clumpy.disk.evolution.toymodel,ceverino:2010.clump.disks.cosmosims,hopkins:clumpy.disk.evol,oklopcic:clumpy.highz.gals.fire.case.study.clumps.not.long.lived} and observational \citep{elmegreen:2004.chain.gal,martinez-sansigre:high.qsos.in.submm,kriek:z2.hubble.sequence,daddi:highz.gal.high.fgas,forster-schreiber:2011.hiz.gal.morph,newman:z2.clump.winds} studies of massive star-forming and quasar-host galaxies at redshifts $z\gtrsim 2$. The system is extremely inhomogeneous, with non-axisymmetric mode amplitudes $|a_{1}| \sim 0.1-1$ and large clump and cloud complexes and star clusters visible. At the time of this particular simulation, the torques from $\sim 1-10$\,kpc clearly involve large non-axisymmetries which are visually dominated by a large minor merger (with the companion at $\sim 10\,$kpc, having just passed pericenter). 

\item{Galactic Core/Proto-Bulge $\rightarrow$ Galactic Nucleus:} On scales $\sim 0.1-1$\,kpc, most of the ``Galactic ISM'' intuition applies, and a significant fraction of the SFR ($\sim 20\,{\rm M_{\odot}\,yr^{-1}}$ averaged over the last $\sim 100\,$Myr, but closer to $\sim 100\,{\rm M_{\odot}\,yr^{-1}}$ in the last $\sim 1\,$Myr) comes from these radii. The gas is primarily molecular (by mass) and the turbulence is super-sonic and super-\Alf{ic} ($\mathcal{M}_{s} \sim \mathcal{M}_{A} \sim 3-10$ ). The potential is deeper, and the density and surface density scales are higher, with the surface density approaching the scales $\sim 10^{3}\,{\rm M_{\odot}\,pc^{-2}}$ at which stellar feedback becomes less efficient \citep{fall:2010.sf.eff.vs.surfacedensity,grudic:sfe.cluster.form.surface.density,grudic:2019.imf.sampling.fx.on.gmc.destruction,grudic:mond.accel.scale.from.stellar.fb,2018ApJ...859...68K,hopkins:2021.bhs.bulges.from.sigma.sfr} so the outflows weaken again relative to inflow, and the most massive cloud complexes are more like $\sim 10^{6}-10^{7}\,{\rm M_{\odot}}$. For the first time the radial $\Pi_{rr}$ component does not strongly dominate the stress, as various clumps and accreting gas are not on primarily radial orbits as they are when accreting at larger radii but on quasi-isotropic (often tangential) orbits, having quasi-circularized though with different angular momenta (as there is no coherent disk). Accretion is strongly dominated by gravitational torques with $|a_{1}| \sim 1$ -- i.e.\ order-unity asymmetries in the potential dominated by the {\em stellar} structure (since this dominates the mass) leading to the gas structures shocking and losing angular momentum on a timescale comparable to the dynamical/orbital time \citep[see][]{levine2008:nuclear.zoom,hopkins:zoom.sims,hopkins:inflow.analytics,hopkins:qso.stellar.fb.together,angles-alcazar:2013.bh.growth.vs.accretion.prescription,angles.alcazar:grav.torque.accretion.cosmo.sim.implications,daa:20.hyperrefinement.bh.growth,prieto:2016.zoomin.sims.to.fewpc.hydro.cosmo.highz,prieto:2017.zoomin.sims.agn.fueling.sne.fb}. The system begins to be optically thick to cooling radiation in NIR/optical/NUV/UV bands, so the IR radiation energy density begins to rise. 

\item{Galactic Nucleus $\rightarrow$ Black Hole Radius of Influence (BHROI):} On scales $\sim 10-100$\,pc, the increasing density and surface density scale ($\rho \sim 10^{2}-10^{6}\,m_{p}\,{\rm cm^{-3}}$, $\Sigma_{\rm gas} \sim 10^{3}-10^{4}\,{\rm M_{\odot}\,pc^{-2}}$) means that the gas cools rapidly and the ``hot'' and ``warm ionized'' phases vanish rapidly so by $10$\,pc the gas has an average temperature $\lesssim 1000$\,K (primarily in relatively ``warm'' molecular gas). This also means $\beta_{\rm plasma}$ drops from $\sim 1$ at the outer range of these radii to $\lesssim 0.01$ at the inner radii. The system has moved above the critical $\Sigma_{\rm eff}$ at which stellar feedback becomes highly inefficient and we see outflows diminish. But strong instability, fragmentation to more ``GMC-like'' mass scales, inhomogeneity with $|a_{1}|\sim1$, and highly super-sonic (and still super-\Alf{ic}) turbulence ($\mathcal{M}_{\rm s} \sim 30$ even in the volume-filling phases) persists. The region is still actively star-forming but given the smaller area and mass contributes only $\sim$\,a few ${\rm M_{\odot}\,yr^{-1}}$ in ``steady-state'', but this is spiking to $\sim 10-100\,{\rm M_{\odot}\,yr^{-1}}$ in the $\sim 1-10$\,Myr immediately preceding the snapshots analyzed owing to the rapid inflows. At the time of our hyper-refinement, the inflows through these scales are dominated by one large gaseous complex (itself torqued by the ongoing mergers above) which has a close passage with the BHROI\footnote{Defined formally as the radius interior to which the BH dominates the gravitational acceleration.} allowing the BH to tidally capture the gas -- gravitational torques still clearly dominate in this regime. The system is beginning to become more optically thick at some wavelengths but still has cooling times much shorter than dynamical times and is not in a black-body like state (the dust, radiation, and gas temperatures are all significantly different). 

\item{BHROI $\rightarrow$ ``Torus'':} On scales $\sim 1-10$\,pc, the BH begins to dominate the potential, though stars still strongly dominate over gas in the local fluctuations in the potential (since the density of stars is much higher than gas). Because the system is now ``fully'' optically thick to its own cooling radiation, we begin to see a clear inversion of the density-temperature relation, with denser gas being warmer (in both its kinetic, gas, and radiation temperatures, even though these are not yet all in equilibrium with one another) in a quasi-adiabatic manner (as opposed to the usual case at larger radii where denser gas is colder), with $\beta_{\rm plasma} \sim 0.01$.  The densities in the midplane and dense gas phases begin to exceed $\gtrsim 10^{6}\,m_{p}\,{\rm cm^{-3}}$, at which point the dust temperature starts to couple appreciably to the gas kinetic temperature so the two begin to approach one another, but the large inhomogeneity of the medium and much shorter dynamical times (compared to e.g.\ the conventional case in molecular clouds) mean this coupling is still relatively weak/gradual and incomplete. Despite inflow rates still as large as $\sim 100\,{\rm M_{\odot}\,yr^{-1}}$, the SFRs averaged over the last $(100,\,10,\,1)$ Myr are $\sim(0.5,\,5,\,25)\,{\rm M_{\odot}\,yr^{-1}}$, indicating its highly non-global-equilbrium nature. Turbulence remains highly super-sonic (in primarily warm molecular gas) but becomes only mildly super-\Alf{ic} with $\mathcal{M}_{A} \sim 1-3$. Again, gravitational torques clearly dominate the visual structure (with large coherent gas asymmetries and $|a_{1}|\sim1$): the major change from larger radii is that instead of being incoherent/clumpy structures, the increasingly Keplerian nature of the potential means that coherent, non-linear $m=1$-like perturbations related to torques between gas and stars dominate \citep{tremaine:slow.keplerian.modes,zakamska:eccentricity.wave.propagation,hopkins:zoom.sims,hopkins:inflow.analytics,hopkins:cusp.slopes,hopkins:slow.modes}. 

\item{``Torus'' $\rightarrow$ Non-Star-Forming Disk:} As shown in \S~\ref{sec:plasma}, on scales $\sim 0.1-1$\,pc, the temperatures rise to a few $10^{3}$\,K, and molecules begin to dissociate into atomic gas at these warmer temperatures (though a non-negligible molecular fraction remains). The gas is still dust-laden (we are outside the sublimation radius). We see an important morphological transition in Figs.~\ref{fig:images.faceon}-\ref{fig:images.star.faceon}: the stream of gas tidally torn from the external gas complex and fueling the accretion disk begins to circularize and self-intersect, forming a more coherent disk. Compared to larger radii, we see (\S~\ref{sec:sf}) an accompanying rapid rise in $Q_{\rm turb}$ and $Q_{\rm mag}$. At $r \gtrsim $\,pc, we still have $Q_{\rm thermal} \lesssim 1$ (and a cooling time short compared to the dynamical time), and as shown in previous studies (see e.g.\ \citealt{hopkins:2013.turb.planet.direct.collapse} and discussion below) a disk with these conditions is still unstable to gravitational fragmentation within ``patches'' even if it is {\em statistically} marginally stable with turbulent+magnetic support (the system still has $\beta \ll 1$, with trans-\Alf{ic} turbulence), so some star formation continues. Indeed, the SFR averaged on $(100,\,10,\,1,\,0.1)$\,Myr timescales inside $\sim 1\,$pc is $\sim(0.01,\,0.1,\,1,\,10)\,{\rm M_{\odot}\,yr^{-1}}$ (to the extent that it can be defined in any meaningful way on these small timescales), as the individual protostars and main-sequence stars formed since the gas arrived at these radii are still accreting. However, as we approach $\lesssim 0.1$\,pc, the system dramatically changes and star formation effectively ceases.

\item{Non-Star-Forming Disk $\rightarrow$ ``Accretion Disk'':} ($\sim 0.01-0.1\,$pc) Just outside $\sim 0.1$\,pc, a crucial transition occurs as $Q_{\rm thermal}$ increases to $\gtrsim 1$ with $Q_{\rm mag} \gg 1$ dominated by increasingly-organized toroidal fields. Meanwhile, the characteristic maximal fragment mass $\sim \pi\,\Sigma_{\rm gas}\,H^{2}$ starts to drop into the stellar mass range. As a result (discussed in detail below), star formation shuts down. The SFR inside $<0.1\,$pc averaged over the entire duration of our simulation is $\lesssim 10^{-3}\,{\rm M_{\odot}\,yr^{-1}}$, compared to inflow rates of  $\sim 10-100\,{\rm M_{\odot}\,yr^{-1}}$. With this transition, the disk mass inside $<0.1\,$pc is now locally gas-dominated instead of stellar-dominated, so gravitational torques rapidly become inefficient \citep{hopkins:inflow.analytics}, but we still see $m=1$ modes propagate into these radii  and gravito-turbulent behavior, but now a combination of Maxwell and Reynolds stresses take over as the dominant provider of the torques maintaining a similar global bulk inflow rate. As we go to smaller scales still, the deepening potential means the scale height becomes somewhat smaller and the disk becomes increasingly well-ordered. The disk is strongly-magnetized, with $\beta_{\rm plasma} \lesssim 10^{-3}$, and the turbulence becomes modestly sub-\Alf{ic} at smaller radii.

\item{``Accretion Disk'' $\rightarrow$ ISCO:} On scales $\ll 0.01\,$pc ($\ll 10^{4}$ gravitational radii), the disk is essentially in the regime of a ``traditional'' $\alpha$-like accretion disk in many ways. It is optically thick, geometrically thin or ``slim,'' radiating in increasingly black-body-like fashion, nearly-Keplerian and close to circular, gravitationally stable ($Q_{\rm thermal} \gtrsim 1$ with $Q_{\rm mag} \gg 1$), with maximum fragmentation mass scale $\lesssim 1\,M_{\odot}$ so it is not able to fragment efficiently at all. But there are many important differences between the disk here and what is usually assumed in accretion disk studies (to be studied in detail in \citealt{hopkins:superzoom.disk}, henceforth \papertwo). Even though the optical depth is large, the effective black-body cooling time is much shorter than the dynamical time (by a factor $\sim 10^{-3}$), and the turbulence is supersonic, so it maintains a quasi-isothermal, relatively cool global structure. The disk is strongly magnetized with $\beta_{\rm plasma} \sim 10^{-4}$ ($|B|\gtrsim 100\,$G primarily toroidal fields), sustained by flux-freezing from the flux it is fed from the ISM (hence a ``flux-frozen'' and/or ``flux-fed'' accretion disk) and modestly sub-\Alf{ic} (hence highly-supersonic) turbulence. 

\end{itemize}

\begin{figure}
	\centering\includegraphics[width=0.95\columnwidth]{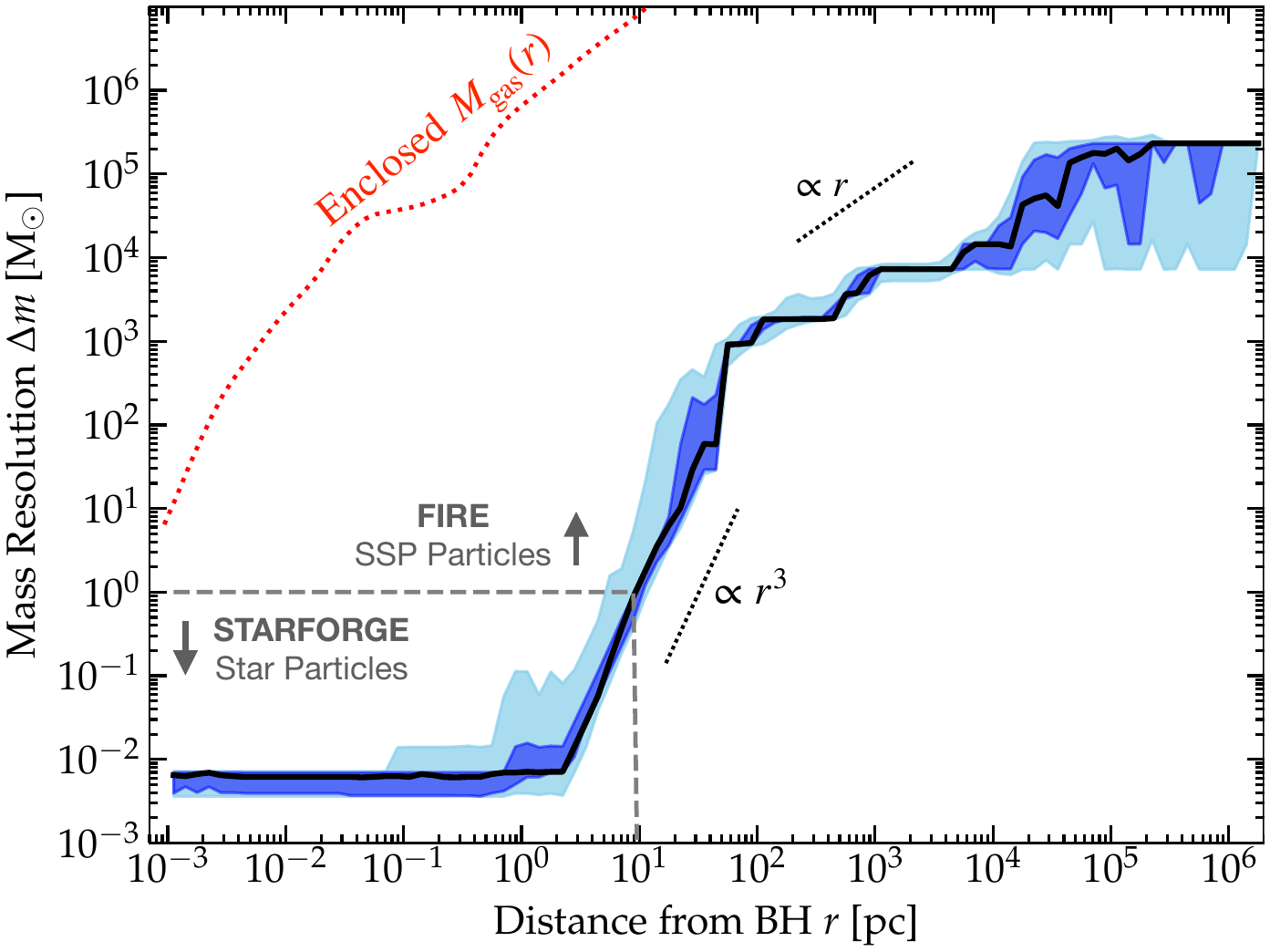}
	\centering\includegraphics[width=0.95\columnwidth]{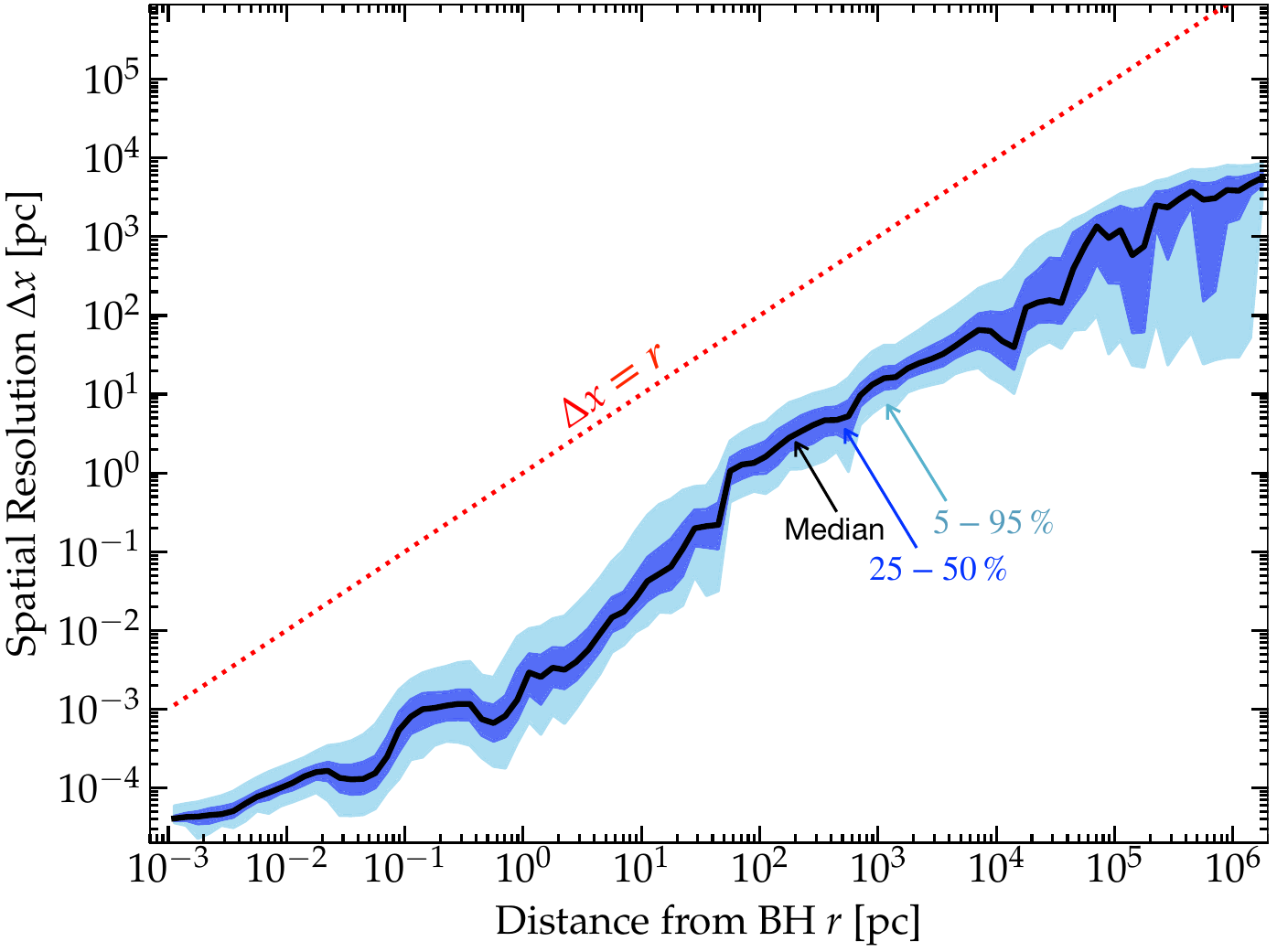}
	\centering\includegraphics[width=0.95\columnwidth]{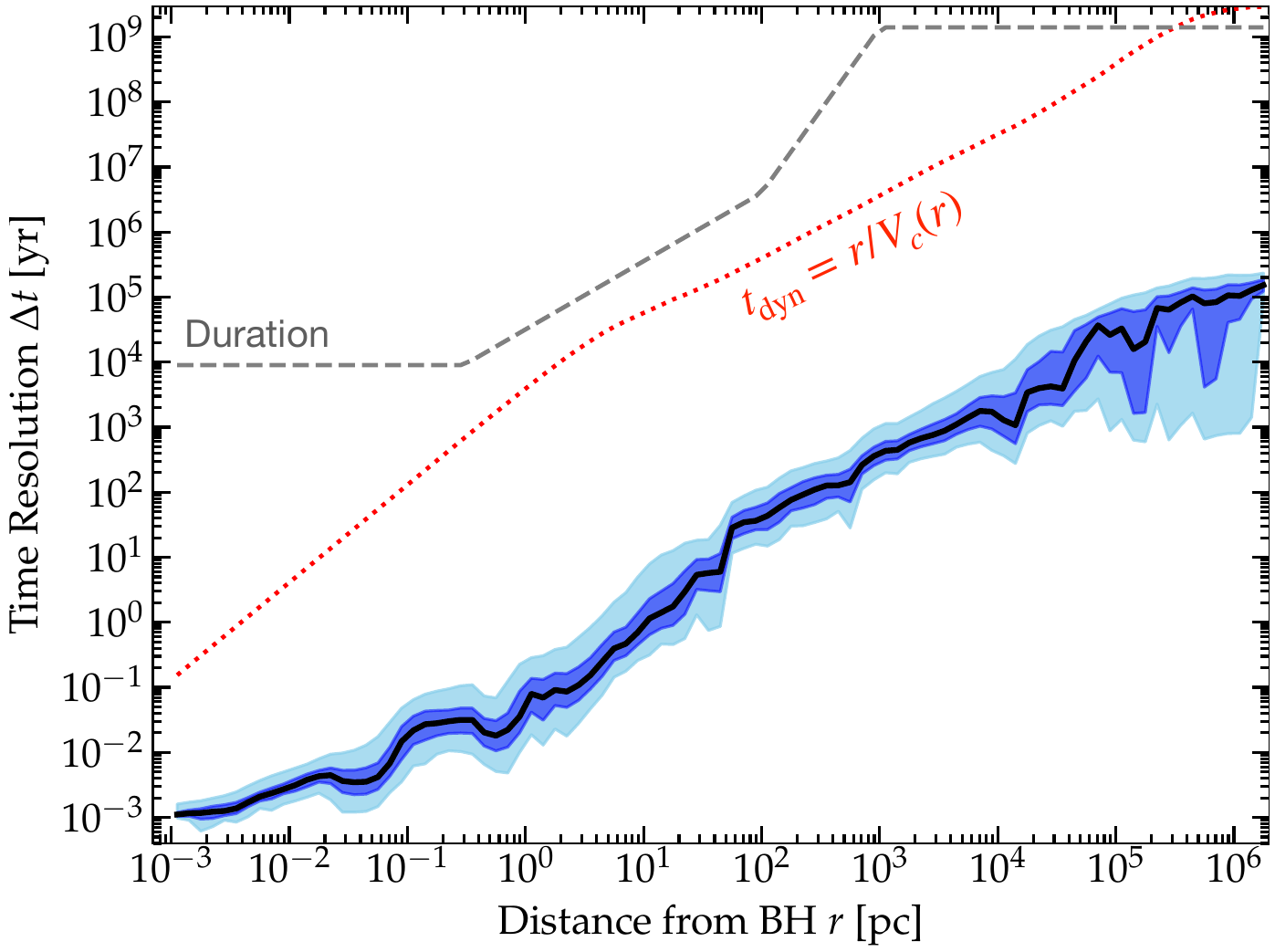}
	\caption{Effective resolution of the simulation as a function of radial distance from the BH ($r$). We show the median, $50\%$ and $90\%$ inclusion intervals ({\em shaded}) of the mass resolution (gas cell mass $\Delta m$; {\em left}), spatial resolution (equivalent cell size $\Delta x \equiv (\Delta m/\rho)^{1/3}$), and time resolution (numerical timestep $\Delta t$), at each $r$. The bulk of the galaxy is resolved with $\Delta m<10^{4}\,M_{\odot}$; from $1-100$\,pc the resolution is rapidly refined as $r$ decreases, with the target resolution of $\Delta m<0.01\,M_{\odot}$ and $\Delta x < 10^{-3}$\,pc reached for the particles at $r \lesssim $\,pc scales. 
	For $\Delta m$, $\Delta x$, and $\Delta t$ we compare the global enclosed gas mass $M_{\rm gas}(r)$, radius $r$, and dynamical time $t_{\rm dyn}$ at each $r$. In $\Delta m$, we also denote the region where the simulation will form STARFORGE single-star/sink particles, as opposed to FIRE SSP particles. In $\Delta t$, we also denote as ``duration'' approximately how long the simulation was run after reaching the maximum refinement level at each radius.
	\label{fig:res}}
\end{figure}

\subsection{Effective Resolution of the Simulation}
\label{sec:resolution}

Fig.~\ref{fig:res} shows the effective mass, spatial, and time resolution of the fiducial simulations as a function of BH-centric radius $R$ at the times where we study it. This reflects the target resolution discussed in \S~\ref{sec:methods} above. In brief: at galactic radii ($\lesssim 10\,$kpc) the resolution is uniformly better than $\sim 10^{4}\,M_{\odot}$ as given by our target refinement criterion before the ``hyper-refinement'' is activated; this then achieves the desired radial refinement with $\Delta m$ smoothly decreasing from $\sim $\,kpc to $\sim\,$pc scales before we saturate at our target resolution inside $\lesssim 1\,$pc of $\Delta m <0.01\,M_{\odot}$. This also lets us clearly identify where the simulation lies in different ``limits'' with regard to star formation per \S~\ref{sec:methods:fire.mode}-\ref{sec:methods:starforge.mode}: at $>10\,$pc scales, the resolution is always in the ``FIRE'' limit (forming SSP particles), and at $< 1\,$pc scales, the resolution is always in the ``STARFORGE'' limit (forming single-star particles). As shown in \citet{grudic:2019.imf.sampling.fx.on.gmc.destruction}, the ``transition'' between these limits, with mass resolution $\sim 1-10\,{\rm M_{\odot}}$, is awkward in the sense that individual stars and the IMF are clearly not well-resolved, but neither can one effectively sample massive stars from the IMF with a single particle (and resulting detailed stellar feedback efficiencies can depend on the statistical method used for IMF sampling at the factor $\sim 2-3$ level),\footnote{We technically address this by simply adopting properties for the IMF sampling for whatever mass star ``should'' exist according to the draw from the IMF as per \citet{su:discrete.imf.fx.fire,wheeler:ultra.highres.dwarfs}, but limiting the returned mass to the total star particle mass to retain mass conservation. More sophisticated schemes exist, which will be explored in future work (Steinwandel et al., in prep.), but in practice this only occurs for a negligibly small fraction of star particles around $\sim 10\,$pc in the simulation here, and as a result we see no effects re-running briefly with a simpler model where we adopt IMF-averaged properties without sampling for these particles.} so we intentionally design our refinement scheme to interpolate through this intermediate resolution over a narrow range of radii (\S~\ref{sec:methods:ics}). We can also compare to the total enclosed gas mass in the simulation, to demonstrate that there are always $N\gg 1$ gas cells in each radial annulus.

Likewise, we see the spatial resolution at small radii is uniformly $\ll H$ (the scale height of the gas at that radius, defined below), reaching $\Delta x \sim 10^{-5}\,$pc, and the time resolution is always much shorter than the local dynamical time. Our timesteps reach extremely small values $\Delta t <1$\,day in the central regions at the maximum refinement level: even with hierarchical timestepping (obviously necessary for such large dynamic range) this limits how long the simulations can be run. Here we evolve for $\sim 10^{4}\,$yr after the finest refinement level was activated: roughly $3\times10^{5}$ dynamical times ($t_{\rm dyn} = 1/\Omega$) at our innermost boundary condition (the ``excision radius'' around the SMBH of $80\,$au or $4\times10^{-4}\,$pc).

\begin{figure*}
	\centering\includegraphics[width=0.48\textwidth]{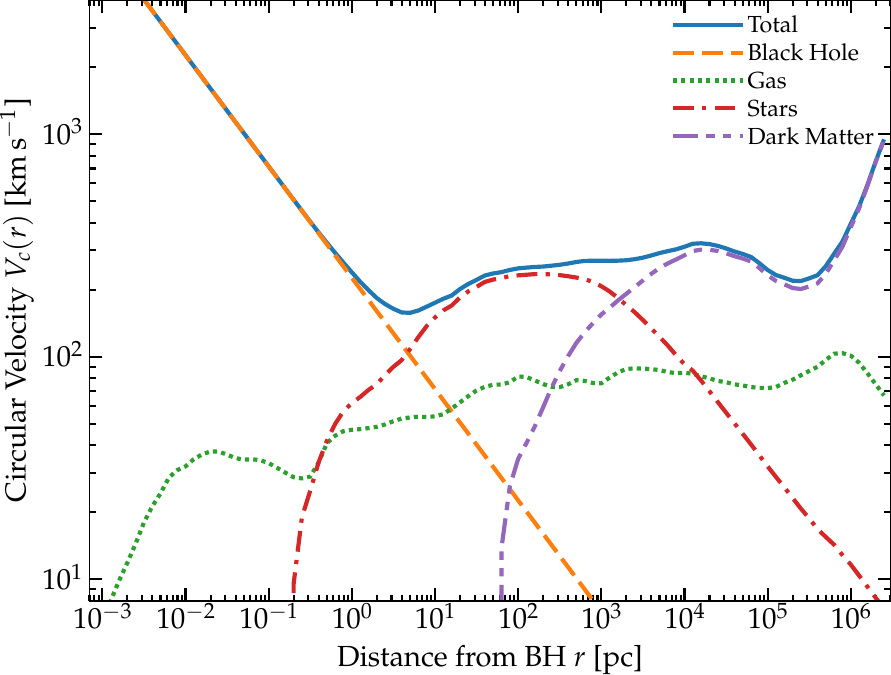}
	\centering\includegraphics[width=0.48\textwidth]{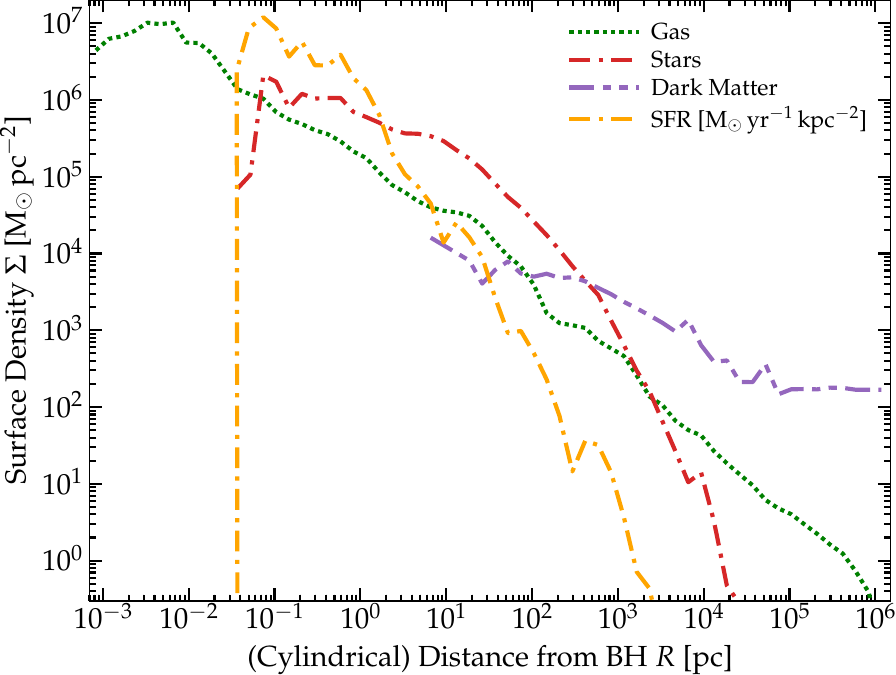} \\
	\centering\includegraphics[width=0.48\textwidth]{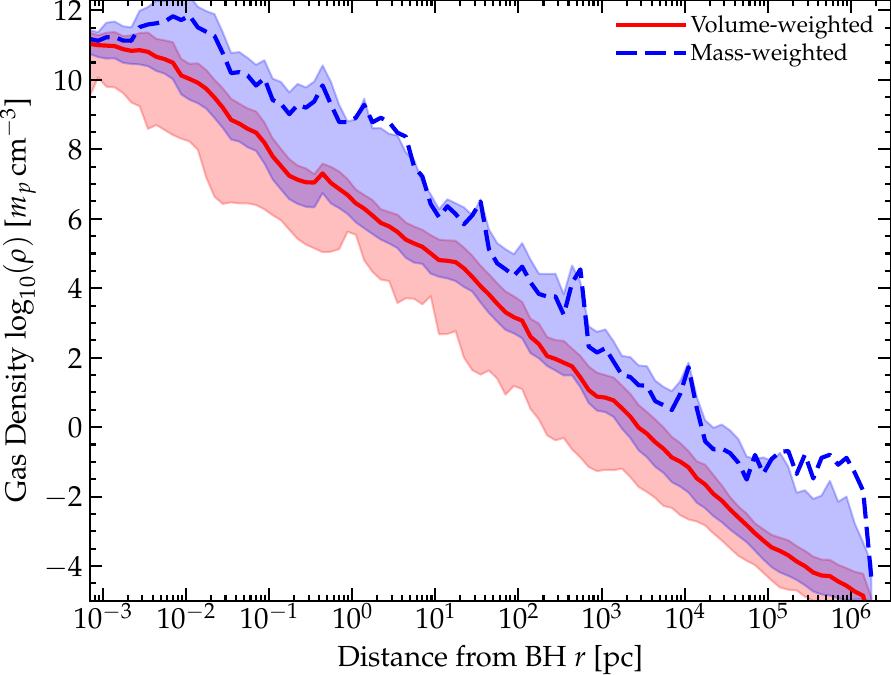}
	\centering\includegraphics[width=0.48\textwidth]{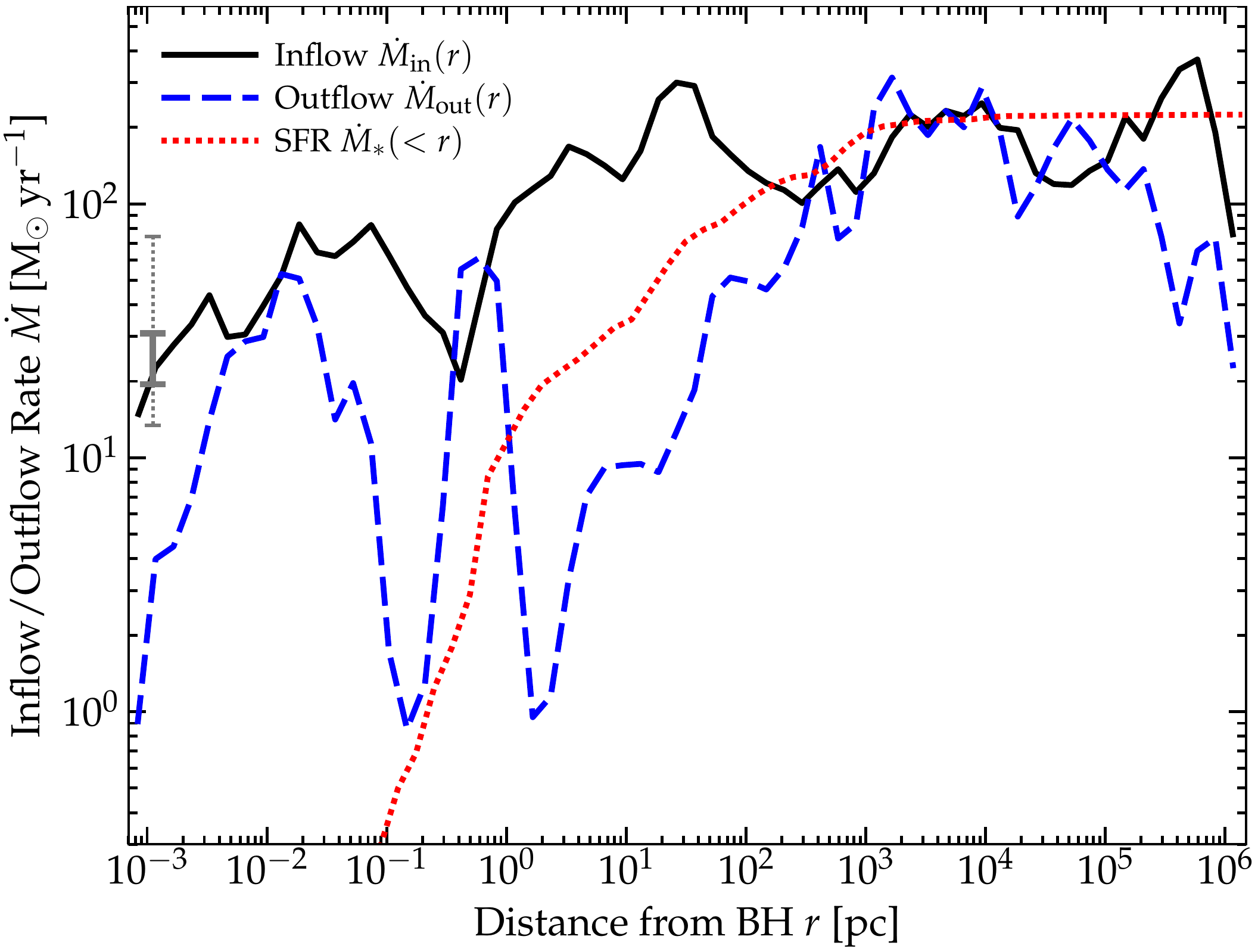} \\
	\caption{{\em Top Left:} Circular velocity profile $V_{\rm c}\equiv \sqrt{G\,M_{\rm enc}(<r)/r}$ versus spherical distance from the SMBH $r$, with contributions from different mass components. The BH excision radius truncates the gas mass distribution at $\lesssim 0.001\,$pc. We clearly see a transition from dark matter dominating outside the galaxy, stars dominating within the galaxy, until the BHROI at $\sim$\,a few pc, and the cessation of star formation giving a gas-dominated disk at $\lesssim 0.1\,$pc.
	{\em Top Right:} Projected mean gas, stellar, dark matter, and ``star formation rate'' (defined by mass of stars formed in the last $t_{\rm dyn} \equiv 1/\Omega$ at each radius) surface density $\Sigma$, in cylindrical shells of different projected radii $R$ (plotted down to radii where we have at least $>1000$ particles of each type). We see the same transitions. While there are large deviations at any radius, $\Sigma_{\rm gas} \propto r^{-1}$ approximates the scaling reasonably well from $\sim 10^{-3}-10^{6}$\,pc. In this and subsequent plots we independently determine the ``midplane'' in each annulus from the angular momentum vector of the gas in the annulus. The suppression of $\Sigma_{\rm SFR}$ at small radii is discussed below (\S~\ref{sec:sf:stopping}).
	{\em Bottom Left:} Three-dimensional gas density $\rho$ versus spherical radius $r$. We show both the volume-weighted density mean ({\em line}) in shells and mass-weighted mean (shaded range shows the volume-weighted $90\%$ inclusion interval; the mean can exceed this upper limit if the distribution has large tails). The gas density profile is crudely isothermal (in a very order-of-magnitude sense), $\rho \propto r^{-2}$, from $\sim 10^{-3}-10^{6}$\,pc. Mass-weighted densities are systematically enhanced relative to volume-weighted but also vary more owing to clumping and phase structure. 
	{\em Bottom Right:} Mass flow rate $\dot{M}$, showing the total inflow rate $\dot{M}_{\rm in}$ and total outflow rate $\dot{M}_{\rm out}$ through each annulus, and the cumulative SFR summed within each radius $\dot{M}_{\ast}(<r)$, versus spherical $r$. Grey bar shows the $50\%$ ({\em solid}; $\sim18-30\,{\rm M_{\odot}\,yr^{-1}}$) and $90\%$ ({\em dotted}; $\sim15-73\,{\rm M_{\odot}\,yr^{-1}}$) range of inflow rates at $<100\,$au over the duration of our highest-resolution simulation. Lack of spherical symmetry and non-equilibrium dynamics mean that inflow and outflow co-exist and can change relative sign (e.g.\ the ``outflow'' at $\lesssim$\,pc scales is mostly just coherent eccentric motion), but despite this a remarkably stable order-of-magnitude inflow rate to the SMBH of $\dot{M}_{\rm in} \sim 10-100\,{\rm M_{\odot}\,yr^{-1}}$ persists.
	\label{fig:profile.mass}}
\end{figure*}

\begin{figure*}
	\centering\includegraphics[width=0.483\textwidth]{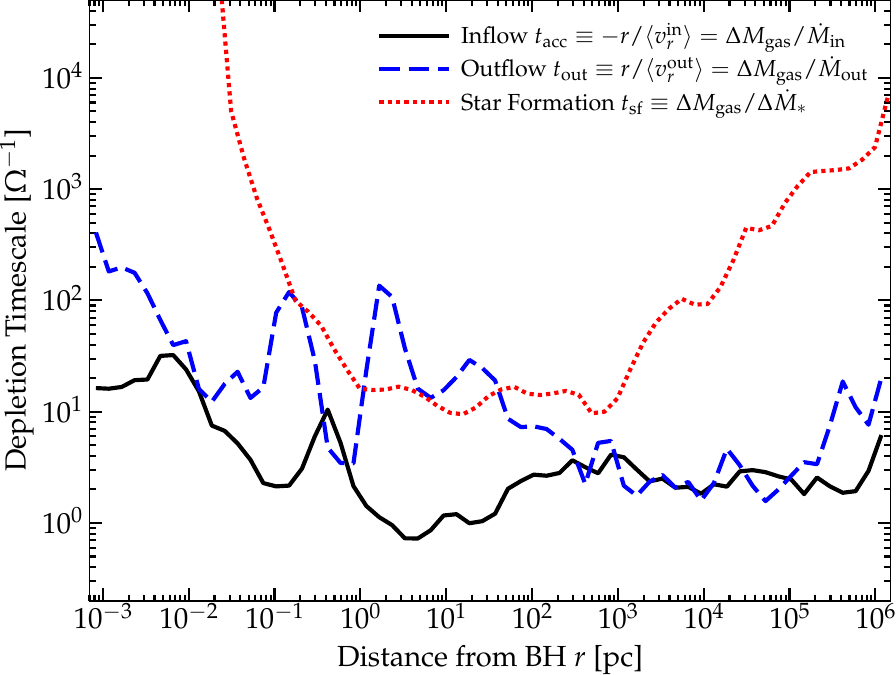}
	\centering\includegraphics[width=0.481\textwidth]{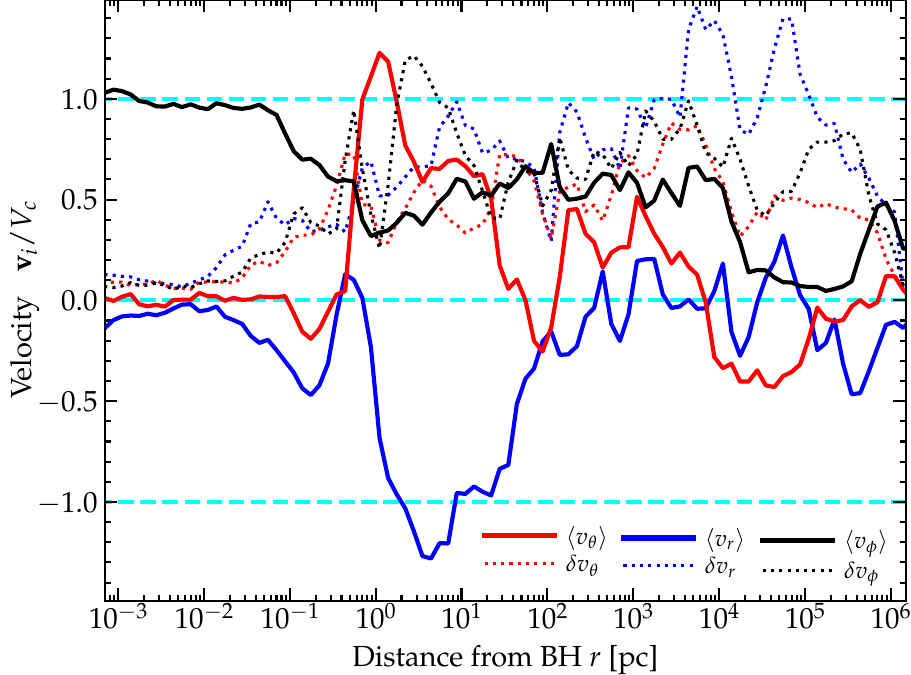} \\
	\caption{Radial profiles of related timescales and bulk flows as Fig.~\ref{fig:profile.mass}. 
	{\em Left:} ``Depletion'' (or accretion) timescales for inflow/accretion, outflow, and star formation, in different radial annuli, defined as shown (with $\Delta M_{i} \equiv \partial M_{i} / \partial \ln{r}$), in units of the galactic dynamical time ($t_{\rm dyn} \equiv \Omega^{-1}$) at each radius. Star formation is efficient at Galactic radii from $\sim$\,pc to $\sim$\,kpc, as expected. Inflows are dynamical at essentially all radii, and produce net inflow at sub-kpc scales during this strong-inflow phase.
	{\em Right:} Gas mass-weighted mean radial ($v_{r}$), azimuthal/rotational ($v_{\phi}$, with the $\hat{\phi}$ axis defined in each annulus by the net angular momentum vector as Fig.~\ref{fig:profile.mass}), and polar ($v_{\theta}$) velocities, and velocity dispersions $\delta v_{i}$, in units of $V_{c}$. We see a mix of inflow/outflow motions (large $\delta v_{r}$) dominate at large radii ($\gg$\,kpc), partial rotation in a kinematically hot/thick galaxy configuration at Galactic radii ($\sim 0.1-10\,$kpc), and near free-fall at the radii interior to the BHROI where the GMC-like gas complex fueling the SMBH is tidally disrupted and captured ($\sim 1-10\,$pc), which then circularizes at sub-pc scales to form a kinematically cold, rotation-dominated disk at $\lesssim 0.1\,$pc scales. 
	\label{fig:profile.mass.timescales.kinematics}}
\end{figure*}

\subsection{Mass and Accretion Rate Profiles}
\label{sec:mdot}

Figs.~\ref{fig:profile.mass} \&\ \ref{fig:profile.mass.timescales.kinematics} more quantitatively examines the radial profiles of various quantities related to the mass and mass flows: the circular velocity (defined as $V_{\rm c} \equiv \sqrt{G\,M_{\rm enc}(<r)/r}$) and its contribution from the SMBH, gas, stars, and dark matter; the radial profile of surface density $\Sigma_{\rm gas}$ and mid-plane three-dimensional density $\rho$, and the inflow and outflow rates $\dot{M}$ through each annulus.\footnote{We define $\dot{M}_{\rm in} \equiv \langle v_{\rm in} \rho \rangle_{r}\,4\pi\,r^{2} \equiv \Delta r^{-1}\,\int_{r-\Delta r/2}^{r+\Delta r/2} \Theta(-v_{r})\,\rho({\bf r})\,r^{2}\,dr\,d\Omega$, where $\Theta(x)=0$ or $1$ for $x<0$, $x>1$ identifies all the inflowing gas, and likewise for $\dot{M}_{\rm out}$ (with $\Theta(v_{r})$), in some sufficiently narrow annulus $\Delta r$ (here $\sim 0.1$\,dex). This allows for there to be both inflow and outflow through a given radius, given the lack of spherical symmetry. We also define the SFR as the total mass locked into stars (via new (proto)star formation and accretion onto stars) in the last dynamical time in each annulus, as this gives a better sense of the ``instantaneous'' rate at small radii where the dynamical time is much shorter than usual fixed timescales like $\sim 10-100\,$Myr used observationally to define ``recent'' star formation.}

Because $V_{c}^{2}$ at some $r$ is just proportional to the enclosed mass,\footnote{Note at very large radii outside the halo virial radius, $V_{c}$ defined here as $\sqrt{G\,M_{\rm enc}/r}$ begins to rise as it should while the (mean) dark matter and gas densities approach constant values near cosmic mean, so $M_{\rm enc} \propto r^{3}$.} we can clearly read off from Fig.~\ref{fig:profile.mass} where different components dominate the potential and the local matter distribution. The BH dominates inside the BHROI at a few pc, and we see the local density is gas-dominated only interior to the radii where star formation shuts down ($\ll 0.2\,$pc here), while stars dominate the local density from $\sim 0.2\,$pc to $\sim 2\,$kpc, and dark matter dominates the density at much larger scales $\gg$\,kpc (while the steep profile of stars and gas mean dark matter is completely negligible at smaller radii, for any reasonable dark matter profile). While there are clearly very large local fluctuations in gas density, it (rather remarkably) appears to follow an approximately isothermal-sphere-like $\rho_{\rm gas} \propto r^{-2}$ profile {\em on average} over nine decades in radius.\footnote{We quantify both the volume-weighted $\langle \rho_{\rm gas}(r)\rangle \equiv d{\rm M_{\rm gas}} / 4\pi\,r^{2}\,dr$ in concentric shells, and the ``midplane'' gas density defined as the mass-weighted mean gas density within $<10\%$ of the midplane defined by the net gas angular momentum vector within each concentric shell. The latter much more obviously shows large variance owing to phase structure, satellite galaxies (at large radii), and other forms of inhomogeneity, and is (as expected) systematically larger than the volume-weighted mean by $\sim 1-3$\,dex, but the broad trends are similar.} This leads to a gas surface density profile scaling approximately as $\Sigma_{\rm gas} \propto R^{-1}$ (an actual power-law fit gives a very slightly shallower slope). 

The density profiles and $V_{c}$ profiles of stars, gas, and dark matter also clearly demonstrate that, despite the chaotic dynamics of the galaxy at this time, there is a reasonably well defined galaxy ``center'' (with the SMBH within it) on $\gtrsim 10$\,pc scales (i.e.\ the stellar and gas and dark matter densities rise monotonically down to these radii), interior to which the SMBH itself dominates the potential and defines the ``center.'' As discussed in \citet{daa:20.hyperrefinement.bh.growth,hopkins:2021.bhs.bulges.from.sigma.sfr} and \citet{byrne:2023.fb.lim.bh.growth.center.formation.critical}, this is important for establishing the pre-conditions for a rapid accretion event, enabling gas flows to efficiently reach the SMBH radius of influence from much larger scales, so it should not be surprising to see here (given our selection of a massive accretion event to ``zoom in'' onto).

Recall, the duration of our simulation at its highest resolution is still short compared to the global dynamical/evolution timescales on $\gtrsim 1-10\,$pc scales, so (as expected) these profiles are robust in time over the duration of the simulations. Even at the smallest radii, where we run for many dynamical times, after the initial refinement period they remain consistent within the scatter shown, as they are determined by the boundary conditions from larger radii.

In terms of accretion rates, we also see a surprisingly close-to-constant $\dot{M}_{\rm in}(r)$ from radii of $\sim$\,Mpc down to $\lesssim 10^{-3}\,$pc. This is especially surprising given (a) the wildly different characteristic dynamical times on these scales, and (b) as noted from the morphologies above and some kinematic discussion below, that many radii are strongly out-of-equilibrium. The latter does produce some of the large ``wiggles'' in $\dot{M}_{\rm in}$, but seems to produce much more dramatic variation in the outflow rates $\dot{M}_{\rm out}$ at different radii. That is consistent with e.g.\ the behavior seen in \citet{daa:20.hyperrefinement.bh.growth}, especially when we consider the time variability shown in Fig.~\ref{fig:profile.mass} over the dynamical time at each radius, but we caution that we focus on much smaller spatial scales for a much shorter overall period of time, compared to their study. Interestingly however, comparing the different simulations considered in \citet{daa:20.hyperrefinement.bh.growth}, the (weak) variation in $\dot{M}_{\rm in}(r)$ we see is most similar to their ``full-QSO'' simulation (the simulation with the largest sustained inflow, most similar to the case here). That suggests the radial and time variability may be much larger at lower accretion rates (which is plausible, as e.g.\ star formation and outflows and other potential ``bottlenecks'' may play a much larger role limiting gas supply at low $\dot{M}$). We do, on average, see some systematic decline in $\dot{M}_{\rm in}$ from the largest to smallest radii, as expected (material can ``stall'' and simply cease inflow without efficient angular momentum transport mechanisms, or be ejected in outflows, or go into star formation, at each radius), but this is weak, especially at the smallest radii $\ll$\,pc where star formation has ceased (again notably weaker than simulations modeling systems with orders-of-magnitude lower mass inflow rates like M87, see e.g.\ \citealt{guo:2022.superzoom.riaf.in.m87.nosf.nomhd.etc}). And we see outflow rates are order-of-magnitude comparable to inflow rates at most radii; but even where $\dot{M}_{\rm out} > \dot{M}_{\rm in}$ locally (which again clearly indicates out-of-equilibrium behavior, and is much more transient in these simulations) inflows are sustained over the entire duration of the simulation. It is also the case that even at large radii both the inflow/outflow rates are generally much larger than star formation rates within a given annulus (except right around $\sim 1$\,kpc), as shown in Fig.~\ref{fig:profile.mass.timescales.kinematics}, owing to feedback self-consistently regulating star formation to be relatively slow (with an average efficiency $\sim 1\%$ per free-fall time; see \citealt{hopkins:rad.pressure.sf.fb,orr:ks.law}), as expected from previous studies of gas-rich, star-forming galaxies. Finally we can also see where star formation ceases at small radii in both Figs.~\ref{fig:profile.mass} \&\ \ref{fig:profile.mass.timescales.kinematics}.

Looking at different times in our simulations in Fig.~\ref{fig:previous.sim}, we see that while there are some significant (factor of a few to order of magnitude) variations in the accretion rates into the central $<80\,$au over the duration of the simulation, the accretion rates at these radii are quite slowly-evolving in a dynamical sense (these variations occur over tens of thousands of local dynamical times $t_{\rm dyn} \sim 1/\Omega$ at the smallest radii). Thus it is reasonable to consider the inner regions to be in some kind of statistical quasi-steady-state in terms of accretion and dynamics, at given large-scale time in the galaxy. Of course, there are other peaks in the accretion rate into the central $\sim 10-100\,$pc in the simulation before hyper-refinement; we select the strongest inflow for refinement, but in future work hope to explore these other episodes.

\begin{sidewaysfigure}
	\centering\includegraphics[width=1.0\linewidth]{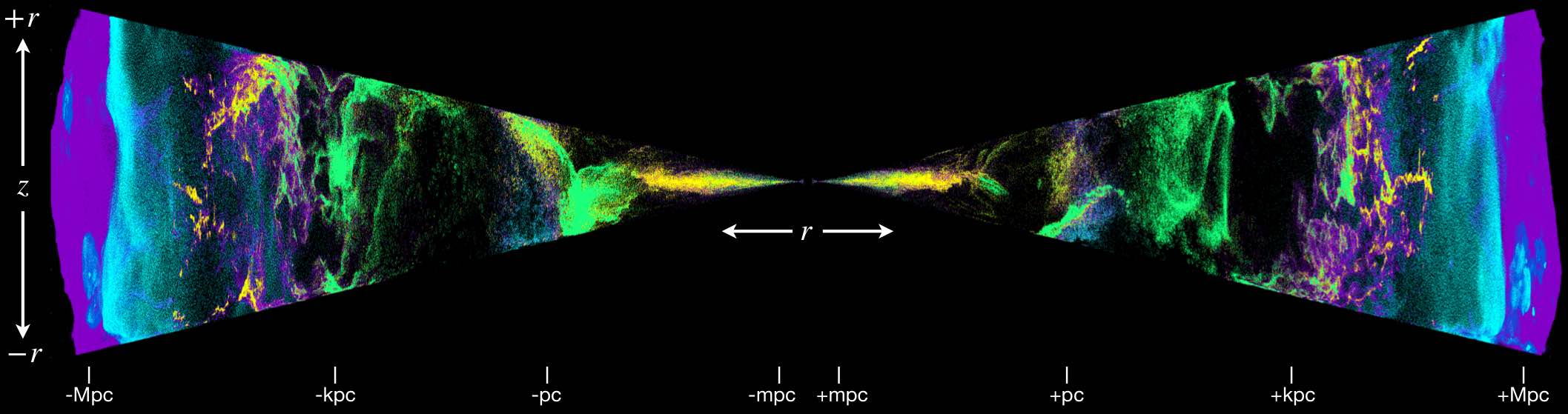}
	\caption{Images of the gas phase structure as Fig.~\ref{fig:scale.labels}, but showing distance from the BH along the accretion-disk midplane $\hat{r}$ direction (horizontal axis, in a wedge with azimuthal angle $|\sin{\phi}|<0.3$) versus $\hat{z}$ (vertical axis), with both $r$ and $|z|$ stretched logarithmically (so both expand at the same rate with distance from the origin). Here we can see the highly-globally-asymmetric (and multi-phase) structures driving inflow on a wide range of scales.
	\label{fig:edgeon.triangle.phases}}
\end{sidewaysfigure}

\begin{sidewaystable}
\begin{center}
\footnotesize
\caption{Key Physics on Different Scales.\label{tbl:physics.vs.scale}}
\begin{tabular}{| l | l | l | l | l | l | }
\hline
Scale Range & Gravity & Magnetic Fields & Thermo-Chemistry & Radiation & Star Formation/Feedback \\
\hline\hline
$10^{-5}\,{\rm pc}\lesssim r\lesssim 0.01\,$pc & Stable orbit integration & Critical for disk structure & Quasi-adiabatic, well-ionized & Radiation pressure critical & Mostly suppressed \\
ISCO$\rightarrow$Accretion Disk & Global mode self-gravity & \&\ angular momentum & (except metal-line opacities)  & Black-body like, simple opacities & (no global dynamical effects) \\
\hline\hline
$0.01\,{\rm pc}\lesssim r\lesssim 1\,$pc & Self-gravity essential & Important for fragmentation,  & Coupled to RHD \&\ CRs & Essential for thin-to-thick transition & ``Single star'' formation \&\ proto+main sequence \\
Accretion Disk$\rightarrow$BHROI & (fragmentation, SF) & IMF, \&\ small-scale structure & Key for IMF, not global dynamics  & Opacities with \&\ without dust & evolution (jets, radiation, winds, SNe) \\
\hline\hline
$1\,{\rm pc} \lesssim r \lesssim 10^{6}\,$pc & Self-gravity essential & Relatively minor effects & Optically-thin cooling sufficient & Needed for FB \&\ UVB & ``Stellar Population'' SF rules \&\ feedback needed \\
BHROI$\rightarrow$IGM & (fragmentation, SF, ISM) & on dynamics & (tabulated $\Lambda(T,\,Z,\,n,\,...)$) & but simple models sufficient & (radiation, winds, SNe) \\
\hline
\end{tabular}
\end{center}
\end{sidewaystable}

\subsection{Plasma and Thermo-Chemical Properties}
\label{sec:plasma}

\begin{figure*}
	\centering\includegraphics[width=0.48\textwidth]{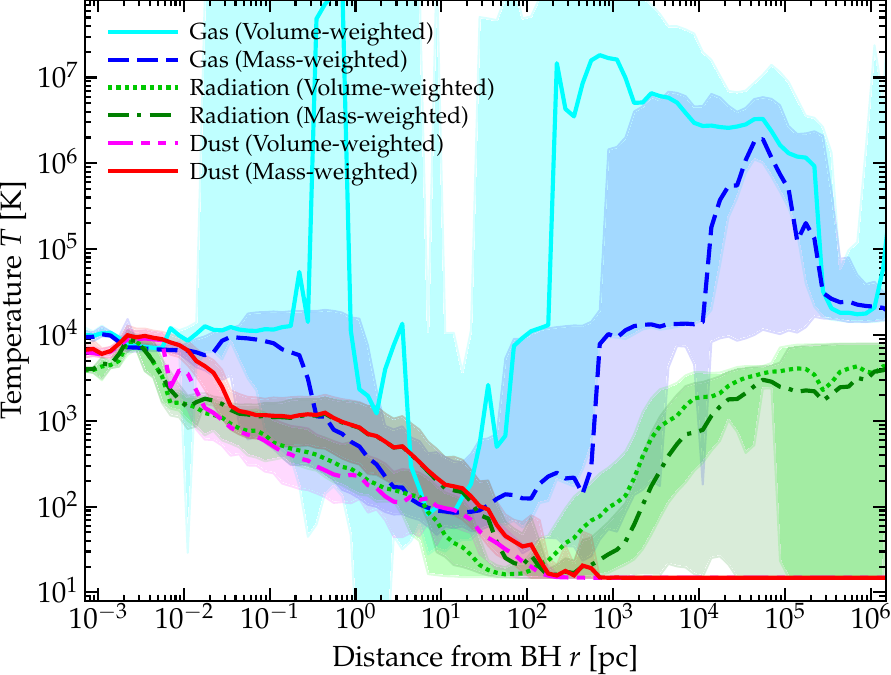}
	\centering\includegraphics[width=0.48\textwidth]{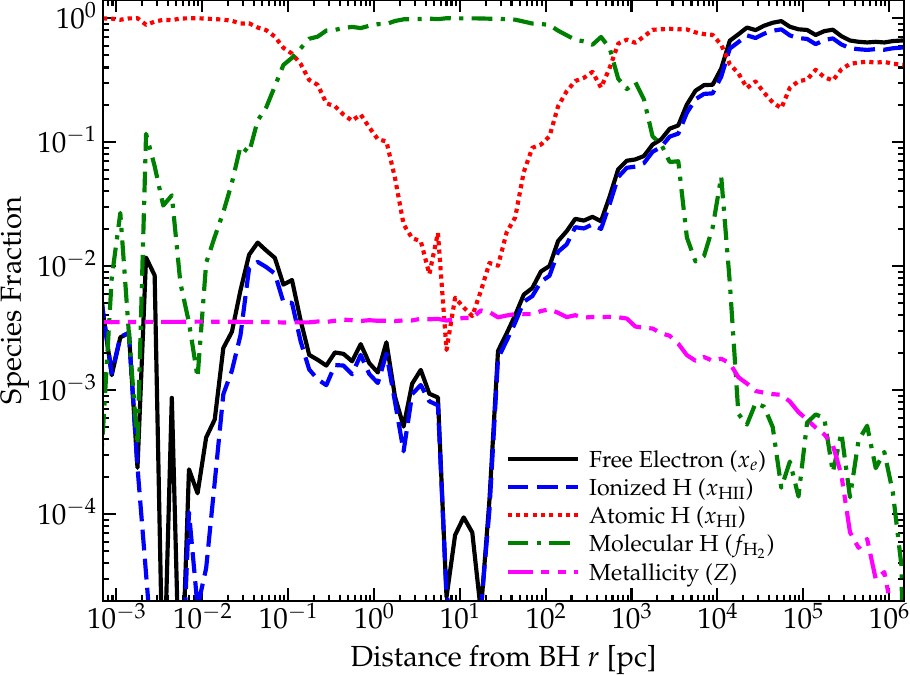} \\
	\centering\includegraphics[width=0.48\textwidth]{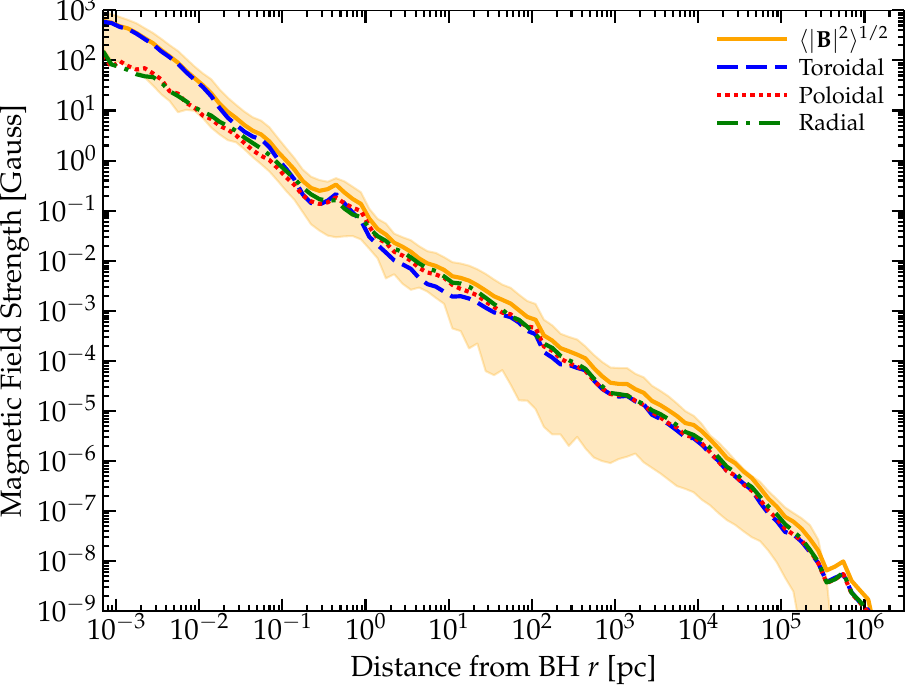}
	\centering\includegraphics[width=0.48\textwidth]{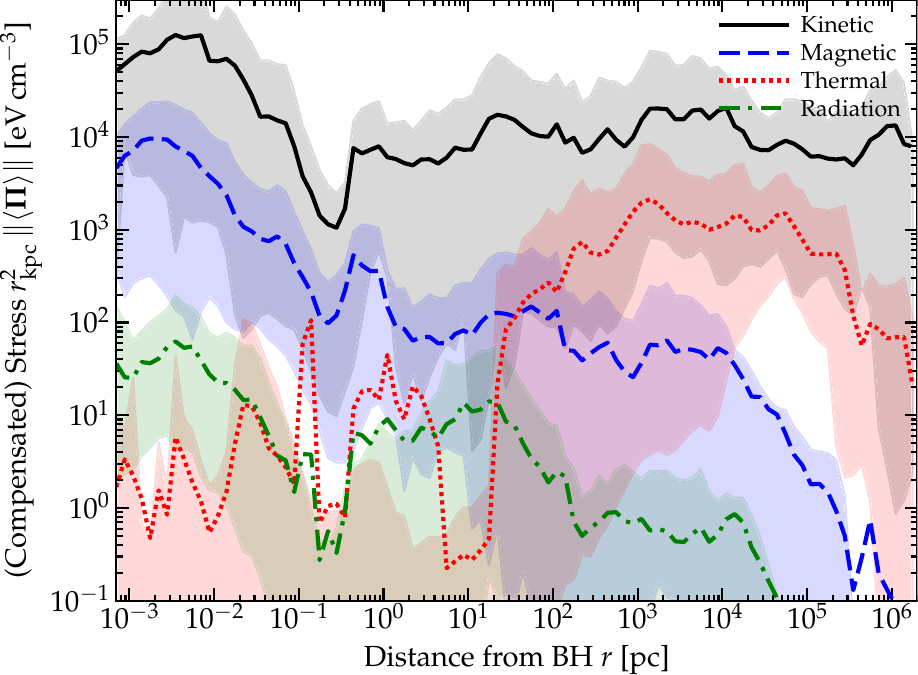} \\
	\caption{{\em Top Left:} Thermal properties: we plot the volume-weighted and mass-weighted mean gas temperatures (and their range, {\em shaded}) in each annulus, together with the effective radiation temperature (integrating over all bands evolved, but excluding the CMB which dominates at $r\gtrsim 200\,$pc) and dust temperature (also dynamically evolved), all versus distance from the BH in spherical annuli. 
	{\em Top Right:} Chemical properties: average free electron ($x_{e}$), ionized hydrogen ($x_{\rm HII}$), molecular hydrogen mass ($f_{\rm H_{2}}$), atomic hydrogen ($x_{\rm HI}$), and metal ($Z$, so solar $\sim0.01$) mass fractions. We see the highly multi-phase, optically-thin medium collapse into a cool atomic+molecular medium with warm dust in the galactic nucleus, then eventually see the system converge towards black-body like behavior with $T_{\rm rad} \sim T_{\rm dust} \sim T_{\rm gas}$ in the center, with the dust sublimating. The scatter at a given $r$ in e.g.\ $x_{e}$ is large, and shown below.
	{\em Bottom Left:} Magnetic field strengths: we plot the energy-weighted rms field strength, and radial/toroidal/poloidal components. To crude approximation $\langle |{\bf B}|\rangle^{1/2} \propto r^{-1}$ from $\sim 10^{-3}-10^{6}\,$pc, with crudely isotropic (some mildly radial-dominant reflecting inflows) fields at most radii until the inner disk forms, and the field becomes primarily toroidal (see \papertwo). 
	{\em Bottom Right:} Radial profile (compensated by $r^{2}$ to make comparison easier) of different components of the pressure/stress tensor (for anisotropic components, we plot $\| \langle \boldsymbol{\Pi} \rangle \|$ for the stress $\boldsymbol{\Pi}$). At all radii, the kinetic energy density/ram pressure is important, and it is largely isotropic with mild radial bias at most radii until the disk forms where it becomes tangential. We see the transition from plasma $\beta \gg 1$ at large radii to $\ll 1$ at small radii. Radiation (again excluding the CMB term here) is generally sub-dominant at all radii, and other stresses (viscous, cosmic ray) are even smaller.
	\label{fig:profile.thermochem}}
\end{figure*}

\begin{figure*}
	\centering\includegraphics[width=0.98\textwidth]{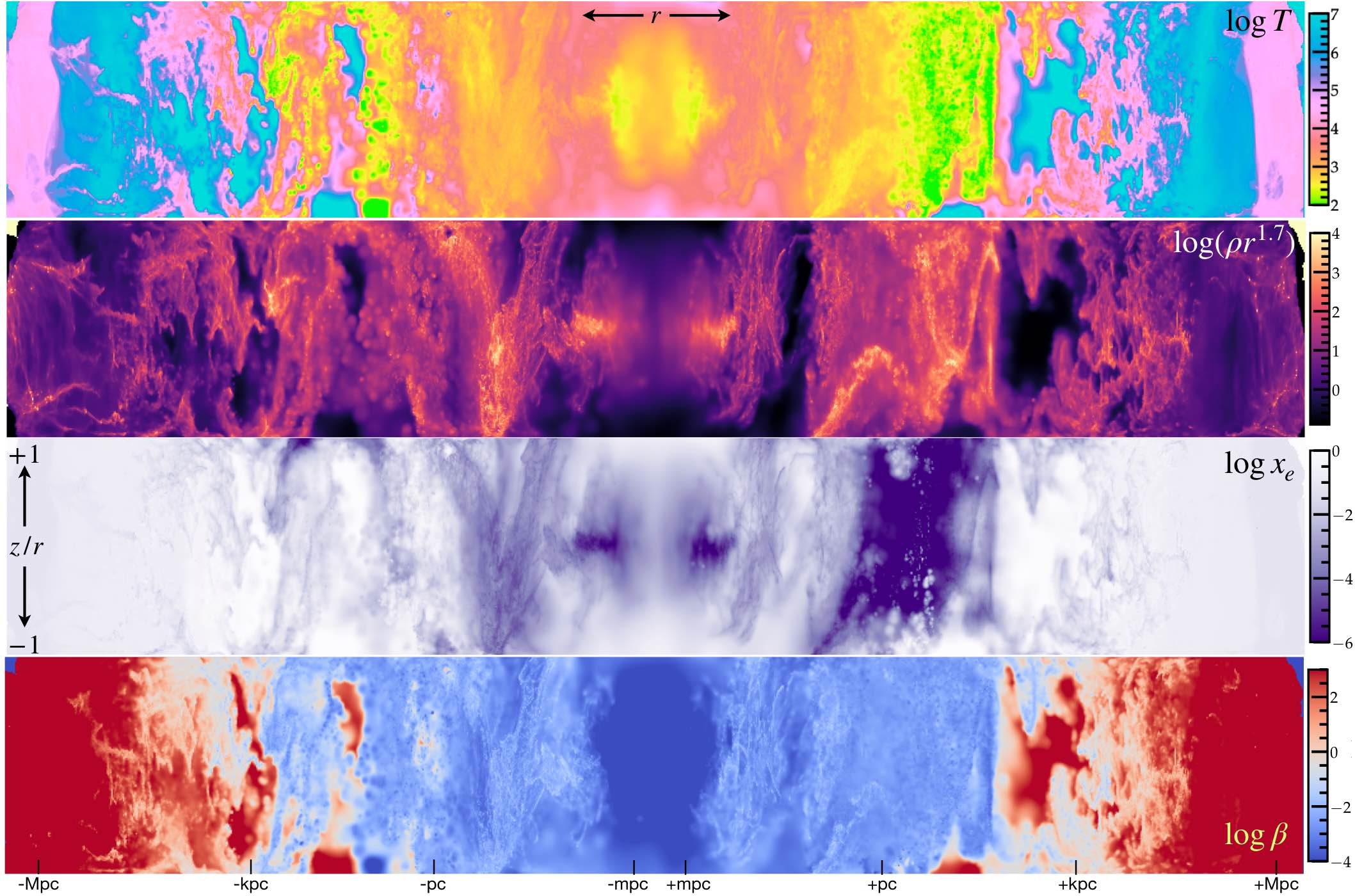}
	\caption{Two-dimensional projections of the mass-weighted mean thermochemical properties (temperature $T$ in ${\rm K}$; density $\rho$ compensated by $r^{1.7}$ for the sake of visualization, in units of ${\rm cm^{-3}\,{\rm kpc}^{1.7}}$; free electron fraction $x_{e}$; plasma $\beta$) from Fig.~\ref{fig:profile.thermochem}. Horizontal axis shows distance $r$ from the SMBH, log-scaled, along a wedge with opening angle $|\sin{\phi}|<0.3$, as Fig.~\ref{fig:edgeon.triangle.phases}, while the vertical axis shows the height $z/r$ as Fig.~\ref{fig:scale.labels}. The wedge is oriented along the axis of the inner accretion disk. The radial trends from Fig.~\ref{fig:profile.thermochem} are evident as is highly multi-phase structure at most radii.
	\label{fig:images.thermochem}}
\end{figure*}

In Fig.~\ref{fig:edgeon.triangle.phases} we illustrate some of the multi-phase structure of the simulation more explicitly, alongside the (highly inhomogenous) stellar distribution. For comparison, Fig.~\ref{fig:profile.thermochem} illustrates the average radial profiles (smoothing out these local variations) in various plasma and thermo-chemical properties of the medium. 

We see that the mean temperature jumps from typical $\sim 10^{4}\,$K IGM values to much warmer $\gtrsim 10^{6}\,$K (comparable to the virial temperature) inside the virial radius of the dark matter halo as the gas shocks, although as shown in Fig.~\ref{fig:edgeon.triangle.phases} much of the accretion onto the galaxy can still be in the form of warm/cold filaments and clumps. At these large radii the radiation and dust temperatures are largely determined by the CMB (if we specifically ignore the CMB, the radiation temperature of the residual radiation instead just reflects the meta-galactic UV background), because the medium is optically thin without significant sources on these scales. As expected, the gas is largely a mix of ionized and atomic phases. Inside the galaxy, we see an even more dramatic multi-phase structure (evident in e.g.\ the separation between mass and volume-weighted mean temperature), with large amounts of gas at $\sim 10^{4}-10^{5}$\,K (both ionized and warm atomic), but some cold neutral and molecular star-forming gas and a hot phase at $\gtrsim 10^{6}\,$K. In the galaxy nucleus at scales, $\lesssim 100\,$pc, the mean temperature drops as the densities are so high that there is very little hot phase, and the medium becomes primarily molecular. As we go to smaller radii and the star formation rate density becomes higher and infrared optical depths become appreciable, we see the dust temperature rise from CMB values up to $\sim 100\,$K, very similar to the typical values observed in low-redshift starburst nuclei and circum-nuclear disks around AGN where the molecular gas and SFR densities are comparable to those predicted here \citep[see e.g.][and references therein]{narayanan:2005.co32.lirgs,evans:agn.host.sfr,iono:ngc6240.nuclear.gas.huge.turbulence,hopkins:cusps.mergers,hopkins:cusps.fp,casey:highz.ulirg.pops,wang:highz.qso.ir,izumi:nuclear.disk.plus.outflow.equals.mass.acc,lelli:2022.qso.co.similar.to.superzoom.warm.co.high.inflow.smbh.mass.similar}. We also see a significant ``warm'' molecular component at $\sim 1000\,$K begin to appear at $\sim$\,pc. At $r \lesssim\,$pc, most of the medium is at warm phase temperatures $\sim 10^{3}-10^{4}\,$K, and molecules begin to be dissociated again, and at $\lesssim 0.1\,$pc the optical depths to cooling radiation become large while the densities are sufficiently large ($\gg 10^{6}\,{\rm cm^{-3}}$) that the dust and gas and radiation temperatures all begin to couple to one another (rapidly converging to broadly similar values by $\sim 0.01\,$pc). 

Meanwhile, similar to what we saw with the density field, the mean magnetic fields follow a profile $\langle |{\bf B}| \rangle \propto r^{-1}$ (becoming slightly steeper in the CGM/IGM, see \citealt{ponnada:fire.magnetic.fields.vs.obs}).\footnote{We follow standard practice and initialize a uniform, trace seed field with comoving strength $\sim 10^{-15}\,{\rm cG}$ at redshift $z\sim 100$ in the ``pre-refinement'' simulation initial conditions, in order to source the simulation magnetic fields. This is rapidly amplified self-consistently and at all but the most extreme diffuse IGM at radii $\gg$\,Mpc, the predicted (saturated) magnetic field strength here is independent of the trace field \citep[see e.g.][]{su:2016.weak.mhd.cond.visc.turbdiff.fx,rieder.teyssier:2017.turb.dynamo.saturation.few.percent.fraction.b,2018MNRAS.479.3343M}.} Note that, because of the extreme dynamic range here, the fact that $|{\bf B}|$ scales slightly steeper than $r^{-1}$, while $\rho$ scales slightly shallower than $r^{-2}$ means that the mean ideal-MHD \Alf\ speed ($v_{A} = ({|\bf B}|^{2}/4\pi\rho)^{1/2}$) is not exactly constant but increases gradually from tens of ${\rm km\,s^{-1}}$ on ``galactic'' scales $\sim 1-10\,$kpc to hundreds of ${\rm km\,s^{-1}}$ at scales $\ll 0.01\,$pc, but this is consistent with a very weak trend $\langle v_{A} \rangle \propto r^{-0.15}$ or so. At radii $\gg\,$pc we see large variance in $|{\bf B}|$ reflecting the multi-phase structure of the gas. We also see that the energy-weighted typical plasma $\beta \equiv c_{s}^{2}/v_{A}^{2} \gg 1$, as expected and observed in the ISM and CGM of typical galaxies \citep[e.g.][]{mao:2012.lmc.bfields,2017ARA&A..55..111H,mao:2018.ska.Galactic.bfield.constraints,seta:2021.turb.sims.vs.rm.dm.obs,vandevoort:2021.illustris.bfields.veryhigh.numerical.amplification.likely,prochaska:2019.weak.magnetization.low.Bfield.rm.massive.gal.frb,lan:2020.cgm.b.fields.rm,ponnada:fire.magnetic.fields.vs.obs}. However it is important to note that this is phase-dependent: as usual, in the coldest phases in a multi-phase ISM (e.g.\ the molecular or cold neutral medium), $\beta \ll 1$ almost by definition 
(see e.g.\ \citealt{crutcher:cloud.b.fields,alina:2019.magnetic.field.alignment.analysis.evidence.weak} for observations or \citealt{ostriker:2001.gmc.column.dist,padoan:2011.new.turb.collapse.sims,su:2016.weak.mhd.cond.visc.turbdiff.fx,hopkins:cr.mhd.fire2,guszejnov:2020.mhd.turb.isothermal.imf.cannot.solve} for theoretical discussion). Where we see the mean temperature/thermal pressure of the medium drop sharply as the mass collapses into cold phases at $\lesssim 100\,$pc, we therefore see a sharp transition from a total-energy or volume-weighted $\beta \gg 1$ to $\beta \ll 1$ at smaller radii. The magnetic field evolution through this point is relatively smooth; it is the thermal phase structure which changes much more rapidly.  We can also see that outside of the nuclear disk at $\gtrsim 0.1\,$pc, the magnetic fields are order-of-magnitude isotropic (no single component strongly dominates $|{\bf B}|$) and exhibit a large dispersion, reflecting the disordered and super-\Alf\ turbulent morphology of the gas and multi-phase structure (there is at some radii a small bias towards radial ${\bf B}$ fields, reflecting inflows and outflows). Inside $\lesssim 0.1\,$pc where the ordered, thin, nuclear disk forms, we see this produces a much more ordered, predominantly toroidal field. This will be studied in more detail in \papertwo, where we will examine the time-dependence of the field and its amplification mechanisms in detail.

At all radii, the kinetic energy density of gas is non-negligible (whether primarily ordered or disordered), as expected. The radiation energy density is always sub-dominant to kinetic, magnetic+thermal, and gravitational energy densities ($\sim \rho\,V_{c}^{2}$) -- this is expected at large radii where the galaxy is optically thin, but is surprising at the smallest radii, where again it will be discussed in greater detail in \papertwo. However the radiation energy density we see in the simulation at small radii is expected (it is approximately that of a simple black-body if we set the cooling luminosity equal to the accretion luminosity $\sim \dot{M}_{\rm in}\,V_{c}^{2}(r)$ at each $r$) -- it is simply that the magnetic and kinetic energy densities are much larger. The relative composition of the radiation energy density is unsurprising: at large radii the broad NUV band dominates as expected for an optically thin young stellar population, whereas at small radii our adaptive ``IR'' (but really just any re-radiated light) band dominates when the medium becomes optically thick to NUV and optical emission. The cosmic ray energy density is also small (except perhaps at the very largest radii) compared to others in such a dense environment, as expected and discussed in more detail below.

\begin{figure*}
	\centering\includegraphics[width=0.48\textwidth]{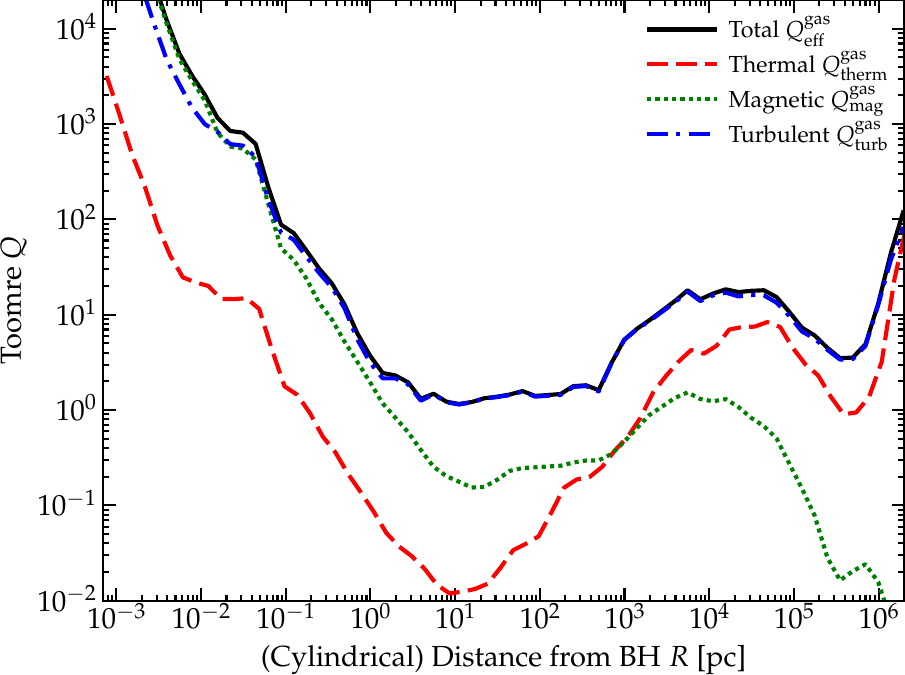}
	\centering\includegraphics[width=0.48\textwidth]{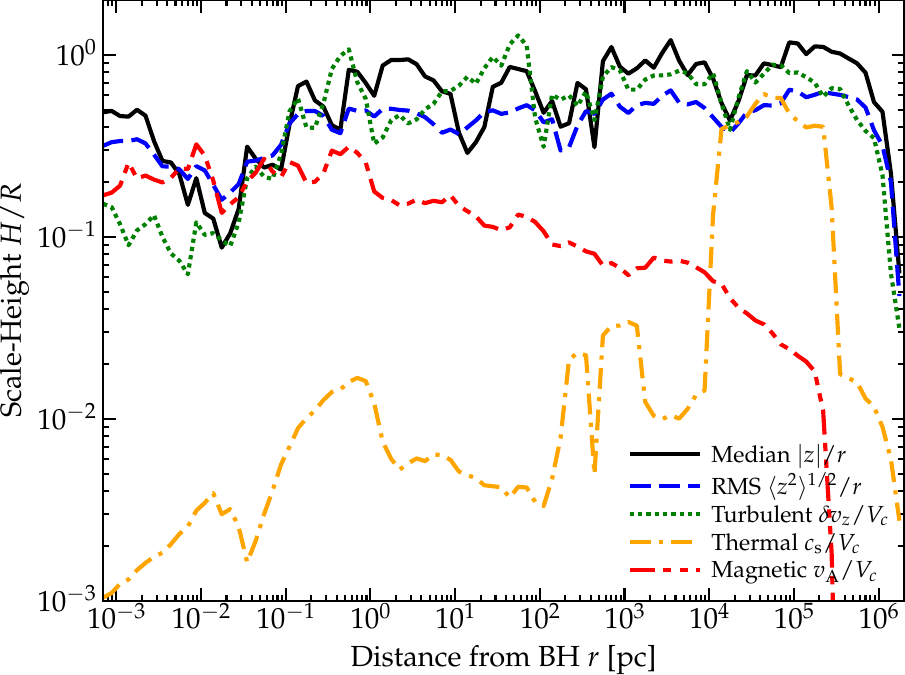} \\
	\centering\includegraphics[width=0.48\textwidth]{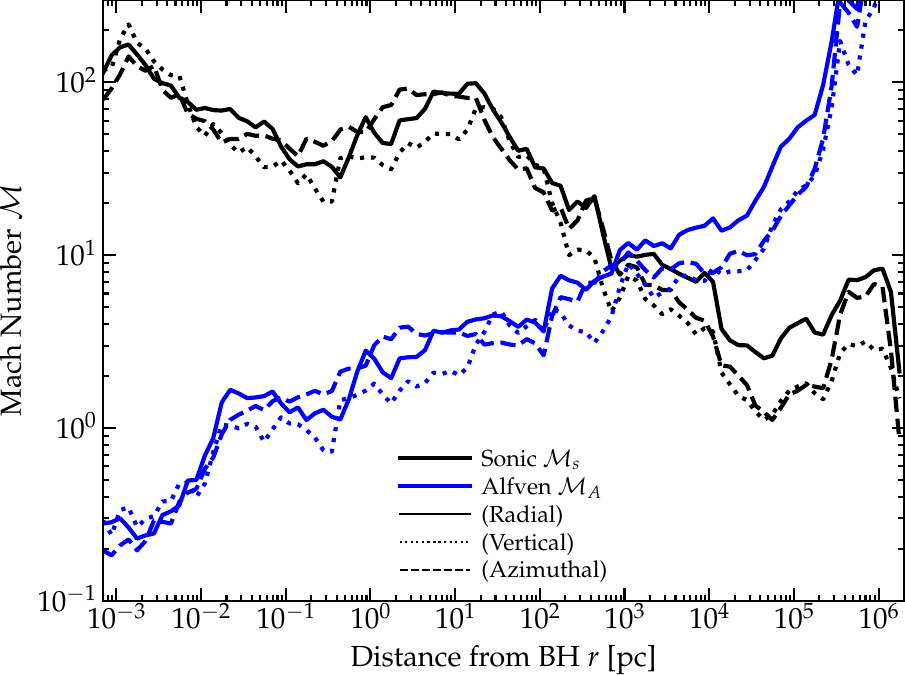}
	\centering\includegraphics[width=0.48\textwidth]{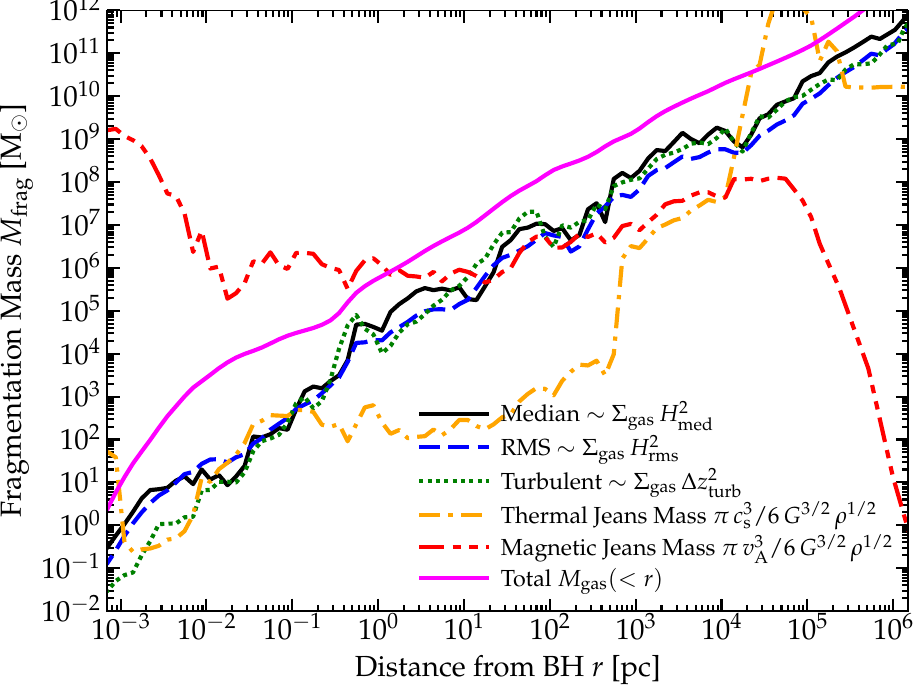} \\
	\caption{{\em Top Left:} Toomre $Q$ parameter of the gas (accounting for a multi-component potential) in annuli $R$, accounting for the thermal ($Q_{\rm therm}$), magnetic ($Q_{\rm mag}$), turbulent ($Q_{\rm turb}$), or combined support of the gas. The ISM/CGM are semi-stable as expected, the galactic ISM is thermally unstable and with marginal turbulent stability, until $\lesssim 0.1\,$pc when stability leads to cessation of star formation.
	{\em Top Right:} Scale height $H/R$ of the gas (mass-weighted), directly measured in each annulus as the median or rms $|z|$ after rotating to the angular momentum axis in that annulus, compared to different velocities (thermal sound speed $c_{s}$, \Alf\ $v_{A}$, or vertical turbulent $\delta v_{z}$) relative to $V_{c}$. At most scales, kinetic/turbulent support dominates, with trans-sonic thermal support in the CGM and magnetic support taking over at $\lesssim 0.01\,$pc.
	{\em Bottom Left:} Sonic ($\mathcal{M}_{s} \equiv \delta v_{\rm turb}/c_{s}$) and \Alf{ic} ($\mathcal{M}_{A} \equiv \delta v_{\rm turb}/v_{A}$) Mach numbers in each annulus (mass-weighted), for each component of the random motions. The velocity dispersions are broadly isotropic at most radii but inflow/outflow leads to a mild radial bias at $\gtrsim \,$kpc scales. The CGM/IGM are trans-sonic (sub-sonic in the diffuse gas, but the average here is dominated by dense substructure), galactic and smaller scales highly super-sonic; we see a clear trend of $\beta$ decreasing at small $r$ so the accretion disk is modestly sub/trans-\Alf{ic}.
	{\em Bottom Right:} Enclosed gas mass inside $r$, versus characteristic maximum gravitational fragmentation (Hill/Toomre) mass $\sim \Sigma_{\rm gas}\,H^{2}$, maximal turbulent fragmentation mass from \citet{hopkins:frag.theory}, and mass-weighted (so biased to much lower values versus volume or thermal-energy weighted) thermal or magnetic Jeans masses. At galactic radii, the gas-rich galaxy produces massive clump complexes with masses $\gtrsim 10^{9}\,M_{\odot}$. Between $\sim 0.1-1\,$pc, the thermal Jeans mass approaches $\Sigma_{\rm gas}\,H^{2}$ (equivalent to $Q_{\rm thermal} \gtrsim 1$), and the magnetic Jeans mass exceeds $M_{\rm gas}(<r)$  (equivalent to all possible scales being magnetically sub-critical). 
	\label{fig:profile.dynamics}}
\end{figure*}

Briefly, it is worth noting that for a reasonable estimate of the UV luminosity from un-resolved ($\ll 100\,$au) scales around the SMBH given the accretion rate here, we might expect the broad line region (BLR) to reside at radii $\sim 20-150$ light-days \citep{kaspi:blr.size}, or $\sim 0.01-0.1$\,pc. It is notable that the gravitational velocities at these radii are $\gtrsim 1000\,{\rm km\,s^{-1}}$, and (per Fig.~\ref{fig:profile.dynamics}) the disk covering fraction or $H/R$ is relatively large $\sim 0.03-0.1$ and increasing at these radii, where it is broadly comparable to the fraction of the UV/optical quasar continuum emitted in the broad lines \citep{vandenberk01:composite.qso.seds,richards:seds}. This is highly suggestive, but more quantitative comparisons (and conclusions related to the physical nature of the BLR ``clouds'' in these simulations) will require detailed post-processing radiative line transfer, which we hope to explore in future work. 

As noted previously in \S~\ref{sec:mdot}, the profiles here remain stable in time (well within their fairly large scatter) over the duration of the highest-resolution simulation, though they can, of course, evolve on larger radii (pre-refinement) on much longer timescales (of order many galaxy dynamical times). The time evolution of the innermost magnetic field structure, and its relation to amplification mechanisms, will be studied in \papertwo.

\subsection{Star Formation \&\ Fragmentation Dynamics On Different Scales}
\label{sec:sf}

We next turn to examining more dynamical properties of the simulation, in order to better understand what drives fragmentation and star formation (or the lack thereof) and inflows on various scales.

\begin{figure*}
	\centering\includegraphics[width=0.95\textwidth]{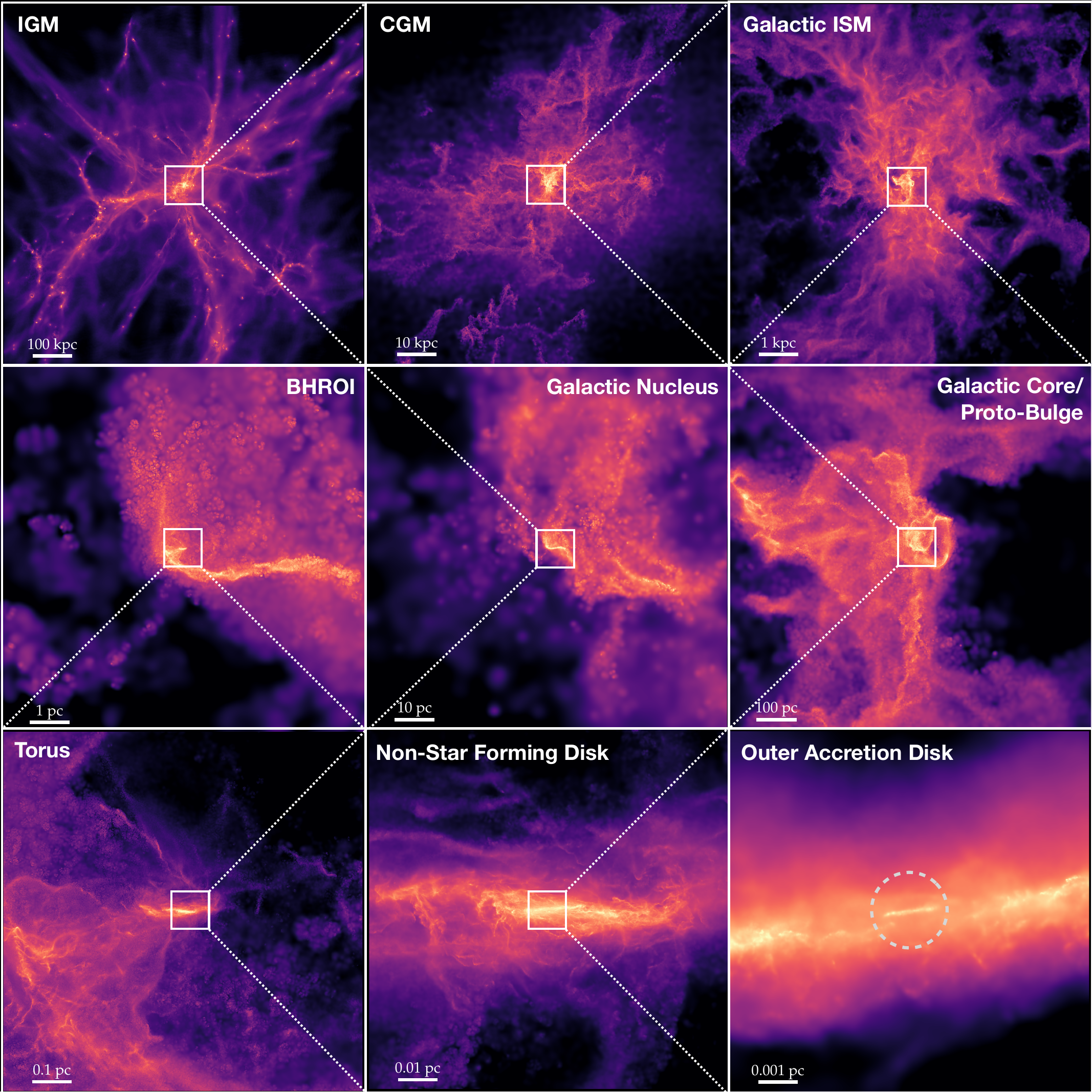} \\
	\caption{As Fig.~\ref{fig:images.faceon}, but showing the edge-on projection to the inner disk at the same time. We can more clearly see the formation of the thin disk at sub-pc scales from capture via filamentary, misaligned inflow from a highly chaotic ISM/CGM at larger radii. 
	\label{fig:images.edgeon}}
\end{figure*} 

\begin{figure}
	\centering\includegraphics[width=0.98\columnwidth]{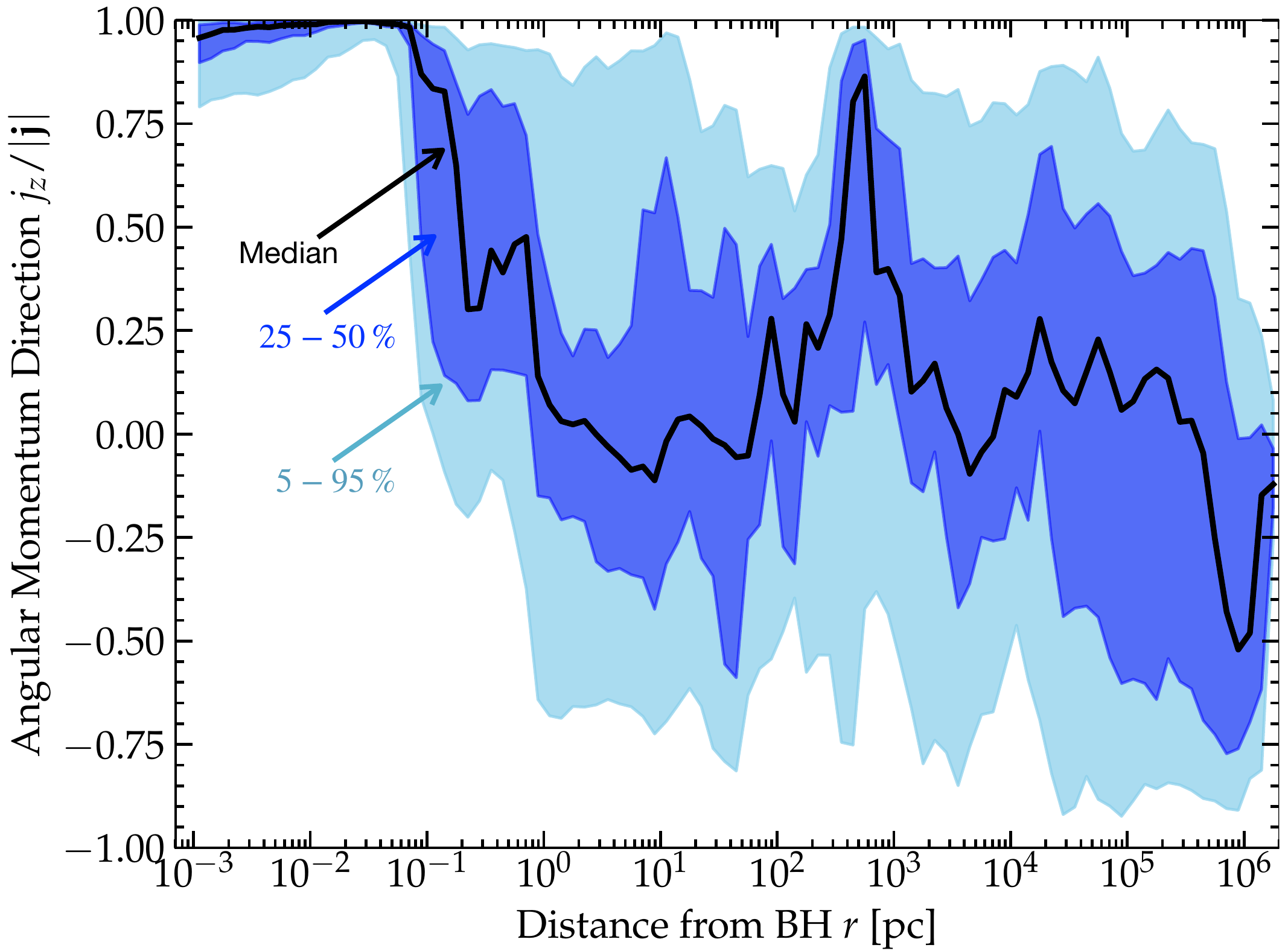}
	\caption{Angular momentum direction of the gas ($\cos{\theta} \equiv \hat{\bf j}_{z} \equiv \hat{\bf j} \cdot \hat{\bf j}_{\rm inner}$), relative to the mean angular momentum direction of the inner disk ${\bf j}_{\rm inner}$ (averaged within $<0.01\,$pc), as a function of BH-centric radius $r$. We clearly see the aligned, dynamically cold disk ($\cos{\theta} \approx 1$, with relatively little variation) in the central $\lesssim 0.1\,$pc, with an un-aligned, and much more spherical/kinematically hot (broad distribution of $\cos{\theta}$) at larger radii.
	\label{fig:profile.angmom}}
\end{figure}

\subsubsection{Definitions of ``Disk'' Dynamical Properties}
\label{sec:sf:disk}

Fig.~\ref{fig:profile.dynamics} shows radial profiles (as Figs.~\ref{fig:profile.mass}-\ref{fig:profile.thermochem}) for different dynamical properties of the simulation. We show the Toomre $Q$ parameter of the gas, defined as $Q_{x} \approx \sigma_{x}\,\kappa / \pi\,G\,\Sigma_{\rm disk}$ where $\sigma_{x} = c_{s}$ for the thermal $Q_{\rm thermal}$, $=v_{\rm A}$ for the magnetic $Q_{\rm mag}$, $=\delta v_{\rm turb}$\footnote{More formally we follow \citet{orr:non.eqm.sf.model,orr:2021.fire.cmz.analog} and define $Q$ using the expressions for a multi-component disk from e.g.\ \citealt{romeo:1992.two.component.dispersion}, using the appropriate {\em mass-weighted} integrals over the distribution function (similar to defining $\langle \sigma_{x} \rangle^{-1} \sim (\Delta A)^{-1} \,\int (\rho/\sigma_{x})\,d A\,dz$) to define the dispersion/sound speed in the gas (since the system is multi-phase). Note this gives significantly lower $Q$ (so is more conservative for our purposes here -- essentially measuring $Q$ in the ``most unstable'' gas phases) than using e.g.\ a thermally-weighted average or a simple rms value of $\delta v$, which can be strongly biased by outflow motions or small amounts of gas in hot phases populating the ``tails.''} for the turbulent $Q_{\rm turb}$, and $\sigma_{\rm eff}^{2} = c_{s}^{2} + v_{\rm A}^{2} + \delta v_{\rm turb}^{2}$ for the total effective $Q_{\rm eff}$. We also show how the different thermal/magnetic/turbulent components contribute to the vertical support of the gas and the gas scale height, the sonic $\mathcal{M}_{s} \equiv \delta v_{\rm turb}/c_{s}$ and \Alf{ic} $\mathcal{M}_{\rm A} \equiv \delta v_{\rm turb} / v_{\rm A}$ Mach numbers, and the characteristic fragmentation scales of the disk determined by the characteristic maximum/dominant fragment mass $\sim \pi\,\Sigma_{\rm gas}\,H_{i}^{2}$ \citep[see][]{hopkins:frag.theory} and minimal Jeans mass $\sim (\pi/6)\,\sigma_{i}^{3}\,G^{-3/2}\,\rho^{-1/2}$. Unless otherwise specified, these are mass-weighted averages in each radial annulus.

\subsubsection{Fragmentation and Star Formation}
\label{sec:sf:frag}

In the CGM/IGM, we see the gas is thermally stable against self-gravity ($Q_{\rm therm} \gtrsim 1$), the turbulence is trans-sonic (or sub-sonic in the hottest phases), and the gas is quasi-spherical ($H\sim R$), all as expected. On galaxy scales from $\sim 1$\,pc to $\sim 10\,$kpc, the gas is not thermally stable, but has a $Q_{\rm therm} \ll 1$, so it fragments and forms stars, which in turn maintain super-sonic turbulence (with turbulent dispersion dominating the effective scale-height, i.e.\ $H \sim \delta v_{\rm turb,\,z}/\Omega$) with an approximately constant (self-regulating) turbulent $Q_{\rm turb} \sim 1$, as observed in both nearby galaxies \citep{leroy:2008.sfe.vs.gal.prop} and high-redshift starburst and quasar host systems \citep{forsterschreiber:z2.sf.gal.spectroscopy,fisher:toomre.self.reg.starburst.galaxies,reichardt:2022.star.form.reg.outflows.starburst.q.disk}, and seen in previous simulations with similar physics \citep{hopkins:rad.pressure.sf.fb,hopkins:disk.settling,orr:non.eqm.sf.model,orr:2021.fire.cmz.analog}. The galaxy size ($\sim\,$kpc) and compactness are also reasonably similar to observed massive galaxies at these redshifts \citep[compare][]{bezanson:massive.gal.cores.evol,damjanov:red.nuggets,van-dokkum:compact.e.evol.w.profile.changes,hopkins:density.galcores,hopkins:maximum.surface.densities}. The characteristic fragment mass at $\sim$\,kpc scales is relatively large, $\sim 10^{8}\,M_{\odot}$, and this corresponds to the mass of the large star-forming cloud complexes or ``clumps'' seen in the gas morphology in Fig.~\ref{fig:images.faceon} -- these are more massive than Milky Way GMCs as expected because, for constant $Q_{\rm turb}$, the clump mass scales as the gas fraction $\propto f_{\rm gas}^{3}$ and this galaxy (with a gas fraction of $\sim 30\%$ at $\sim1\,$kpc) is a factor of $\sim 6$ more gas-rich than the Milky Way (so we expect a maximal clump mass $\sim 200$ times larger than the largest GMC complexes in the Milky Way), as studied in more detail for similar systems in \citet{oklopcic:clumpy.highz.gals.fire.case.study.clumps.not.long.lived}. Given the large gas fractions and $Q_{\rm turb}$, the system is still thick, with $H/R \sim 0.3-0.5$ at these radii. All of these behaviors are consistent with many previous studies of the star-forming ISM in both idealized and cosmological galaxy simulations \citep{noguchi:1999.clumpy.disk.bulge.formation,bournaud:disk.clumps.to.bulge,ceverino:2010.clump.disks.cosmosims,hopkins:clumpy.disk.evol,hopkins:2013.accretion.doesnt.drive.turbulence}, and again generically expected. As the ISM becomes thermally cold at $\ll 100\,$pc, we see the turbulence go from super-\Alf{ic} to trans or even mildly sub-\Alf{ic}, and highly super-sonic, and $c_{s}$ contributes negligibly (even in a volume-weighted sense) to the vertical disk support. 

It is important to stress that even though star-forming systems with {\em thermal} $Q_{\rm therm} \ll 1$ can and do self-regulate to turbulent (and magnetic) $Q_{\rm turb} \sim 1$, these are not locally stable against fragmentation and star formation (see references above). In fact, \citet{hopkins:2013.turb.planet.direct.collapse} show that such systems will {\em always} produce more fragmentation on small scales as $Q_{\rm turb}$ increases if $Q_{\rm therm}$ remains constant, owing to shocks and compressions generating locally overdense regions with $Q_{\rm eff} \ll 1$ (including the local turbulent, magnetic, and thermal energy densities). This is of course implicitly necessary for star formation to explain their turbulent self-regulation.

\subsubsection{The Cessation of Star Formation at Small Radii}
\label{sec:sf:stopping}

At smaller radii inside the BHROI, we see (1) $Q$ begins to rise for all components, owing to the steep rise in $\Omega$, and the in particular the thermal $Q_{\rm therm} \gtrsim 1$; (2) the disk begins to become thinner (as for relatively slowly-varying $v_{\rm A}$, $V_{\rm c}$ begins to rise); (3) correspondingly the turbulence becomes somewhat weaker (more sub-\Alf{ic}); (4) as the optical depth increases and gas becomes more thermally homogeneous at warm temperatures (Fig.~\ref{fig:profile.thermochem}) the minimum Jeans mass stabilizes\footnote{Note that the thermal Jeans mass plotted in Fig.~\ref{fig:profile.dynamics} is defined as a gas-mass-weighted median, so it is effectively tracing the cold, dense phases (a simple mean or volume-weighted median is much larger), so we can see the behavior of the coldest, most fragmentation-prone phases at each radius. Note also that in a single-component, homogeneous slab disk with only thermal support, the ratio of thermal Jeans mass to $\Sigma_{\rm gas}\,H^{2}$ scales as $\sim Q^{3/2}$, but we see that this precise correspondence is broken for the more complex systems here.} while the characteristic upper fragmentation mass\footnote{This is defined as $\sim \Sigma_{\rm gas}\,H^{2}$, dimensionally the same as the most-unstable mass or Toomre mass or maximum Hill mass in a disk, but with the $\mathcal{O}(1)$ pre-factor taken from \citet{hopkins:2013.turb.planet.direct.collapse} as the value above which the mass function of fragments becomes exponentially suppressed.} decreases into the stellar-mass range and becomes comparable to the minimum thermal Jeans mass (implying all scales are thermally stable); (5) the magnetic field becomes more ordered and dominated by a coherent toroidal component as the disk becomes more organized; (6) the ``magnetic Jeans mass'' becomes larger than the enclosed gas mass, so all scales are magnetically sub-critical. The combination of these effects leads to the cessation of star formation. 

When $Q_{\rm therm} \gtrsim 1$, the system is nominally locally ``stable'' in a formal sense. Still, since the cooling time is short compared to the free-fall time at these radii, we see $Q_{\rm therm}$ remains modest (not $\gg 10$, for example) at all but the smallest radii $\lesssim 0.01\,$pc. As a result, one might expect an intermediate gravitoturbulent regime \citep[e.g.][]{gammie:2001.cooling.in.keplerian.disks}, and indeed the gas morphology in Fig.~\ref{fig:images.faceon} appears consistent with this. While fragmentation in the gravitoturbulent regime is not ``catastrophic'' in the same sense as in the galactic regime (where $Q_{\rm thermal} \ll 1$ and gas locally fragments on its free-fall time), it could in principle still produce efficient fragmentation if we neglected some of the other effects above \citep{meru:2011.frag.crit.planet.disk,meru:2011.nonconvergence.disk.frag.time,paardekooper:2011.edge.fx.disk.frag,meru:2012.nonconvergence.disk.frag,hopkins:2013.turb.planet.direct.collapse,deng:gravito.turb.frag.convergence.gizmo.methods,fletcher:2019.gizmo.vs.many.codes.comparison.migration.orbital.dynamics.detailed.differences.in.gap.accretion.dynamics.from.sink.and.numerical.viscosities,zier:2022.gravitoturb.sims.validating.stochastic.frag}. Most importantly, we are in a regime with thermal $Q_{\rm therm} \gtrsim 1$ but magnetic $Q_{\rm mag} \gg 1$, i.e.\ $\beta \ll 1$. Idealized experiments have shown that this strongly stabilizes gravitoturbulence against fragmentation (even a modest $\beta \lesssim 1$ is usually sufficient for this, let alone the extremely small values of $\beta$ we see here; as argued analytically in \citealt{lizano:2010.magnetized.ppd.linear.considerations,lin:2014.linear.stability.magnetized.selfgrav.ppds,jafari:2019.mhd.ppd.review} and in simulations in e.g.\ \citealt{riols:2016.mhd.ppd.gravitoturb,riols:2018.mhd.sims.ppd.gravitoturb.fx.on.mri,forgan:2017.mhd.gravitoturb.sims}). Moreover, the field geometry being toroidal is essentially the ``most stable'' against local self-gravitational fragmentation.\footnote{In a pure radial field, pure radial modes assumed in $Q$ would not feel magnetic forces, and in a pure poloidal field initial vertical collapse could occur. But in a toroidal field, only pure azimuthal collapse can occur unresisted by magnetic fields; however any such mode with a finite radial width/wavenumber -- i.e.\ containing finite mass -- will be strongly resisted by differential rotation/shear and rapidly sheared out.} And the combined action of gravitoturbulence with these fields can create a dynamo or locally mix/re-order the field lines in a manner which further suppresses local collapse \citep{deng:2020.global.magnetized.protoplanetary.disk.sims.gravito.turb.leads.to.large.B.saturation.vs.mri,riols:2021.gravitoturb.mhd.dynamo.ppds}.

A closely related and more formal statement of magnetic stability comes from our comparison of the magnetic Jeans mass\footnote{We plot magnetic Jeans mass $M_{\rm J}^{\rm B}$ rather than magnetic critical mass $M_{\Phi} \equiv c_{\Phi}\,\Phi_{B}/G^{1/2} = c_{\Phi}\,|{\bf B}|\,A/G^{1/2}$ because the latter can only be defined by reference to a specific area ${\bf A}$ or size, whereas $M_{\rm J}^{\rm B}$ can be defined locally. But note that the dimensionless ratio $M_{\rm J}^{\rm B}/M$ for some volume enclosing mass $M$ is simply proportional to $(M_{\Phi}/M)^{3}$, so for our purposes the two can be treated the same as we only focus on their relative scaling.} $M_{\rm J}^{\rm B} \propto v_{A}^{3}/G^{3/2}\,\rho^{1/2}$ to the total enclosed mass $M_{\rm gas}(<r)$. While the magnetic Jeans mass in and of itself is not a strong determinant of star formation in the same way as the thermal Jeans mass, and the ``magnetic $Q$ parameter'' $Q_{\rm mag} \gg 1$ likewise does not alone formally ensure local stability \citep{lynden-bell:1966.magnetic.spiral.structure}, when $M_{\rm J}^{\rm B}$ exceeds $M_{\rm gas}(<r)$ (which occurs here at $\ll1\,$pc), it is equivalent to the statement (for a homogeneous disk or spheroid) that any perturbation of any wavelength/size $\le r$ is magnetically sub-critical (has a mass-to-flux ratio sufficiently low that it cannot collapse), i.e.\ that fragmentation is strongly suppressed \citep{armitage:protoplanetary.disk.review}.

A second, less important but still non-negligible barrier to fragmentation at these radii is the strong torques which we see are producing angular momentum loss and inspiral on a timescale of order the orbital time (discussed in detail below, but this can be read directly off from Fig.~\ref{fig:profile.mass.timescales.kinematics}, or inferred from Fig.~\ref{fig:profile.mass} by simply noting $\dot{M}_{\rm in} \sim 0.1\,\Sigma_{\rm gas}\,R^{2}$). Akin to the problem of giant planet formation which was the focus of many historical gravito-turbulence experiments above \citep[see e.g.][]{armitage:protoplanetary.disk.review,kratter:grav.instab.review}, even if we neglect the magnetic fields, vigorous gravitoturbulence would lead to an initial collapse of a perturbation by a factor of a few in density, at which point (since the cooling time, while shorter than dynamical, is not completely negligible, and the clumps of order the most unstable wavelength in size are optically thick to their cooling radiation) its cooling would proceed more slowly and it would contract quasi-adiabatically on a cooling time, but this must occur before inspiral. For planet formation one might have an inspiral time of millions of orbits in the gas-rich disk; here, one has only a few orbits. As a result, we see that with magnetic fields present so fragmentation is already suppressed, most of the mildly-overdense clumps that do form (e.g.\ one evident in Fig.~\ref{fig:images.faceon}) spiral inwards and are tidally sheared out upon reaching smaller radii (or simply accrete into our sink particle SMBH) before they can reach even order-of-magnitude overdensities (let alone become anywhere near dense enough to approach star formation).

It is worth noting that none of the above effects {\em completely} eliminate all fragmentation and star formation. We only see that occur on even smaller scales, $r \ll 0.01\,$pc, where the thermal Toomre $Q_{\rm thermal}$ rises extremely rapidly to values $\gtrsim 1000$ by $r\lesssim 0.001\,$pc (much more strongly suppressing gravitoturbulence). However, it is sufficient to ensure the star formation rate and total gas accretion rates onto stars are negligible compared to the gas inflow rates, and therefore that star formation (as well as stellar feedback, at least for the duration of this simulation at highest resolution) plays an essentially negligible role in the global dynamics of the system on $\ll$\,pc scales. More detailed properties of the star formation at these innermost radii (including the IMF), exploring how the rare stars that do form are influenced by their environment (and how their feedback does or does not influence that environment locally) will be studied in \paperthree.

We stress that this suppression of star formation is only possible because of the unique circum-SMBH environment, and is not a generic property of magnetically-dominated ($\beta \ll 1$) media. The suppression of fragmentation by strong shear ($\Omega$ and $Q$ rising rapidly), and the increasing order of the toroidal-dominated mean-field (which efficiently suppresses collapse in the radial and vertical directions) are directly related to the increasingly-strong differential rotation of the disk in an external (BH-dominated) potential, and as discussed in \paperthree\ do not arise generically in environments like the circum-quasar gas outside the BH radius of influence at radii $\gg$\,pc, let alone more typical GMCs.

\subsection{Torques and Inflow Driving at Different Radii}
\label{sec:torques}

\begin{figure*}
	\centering\includegraphics[width=0.48\textwidth]{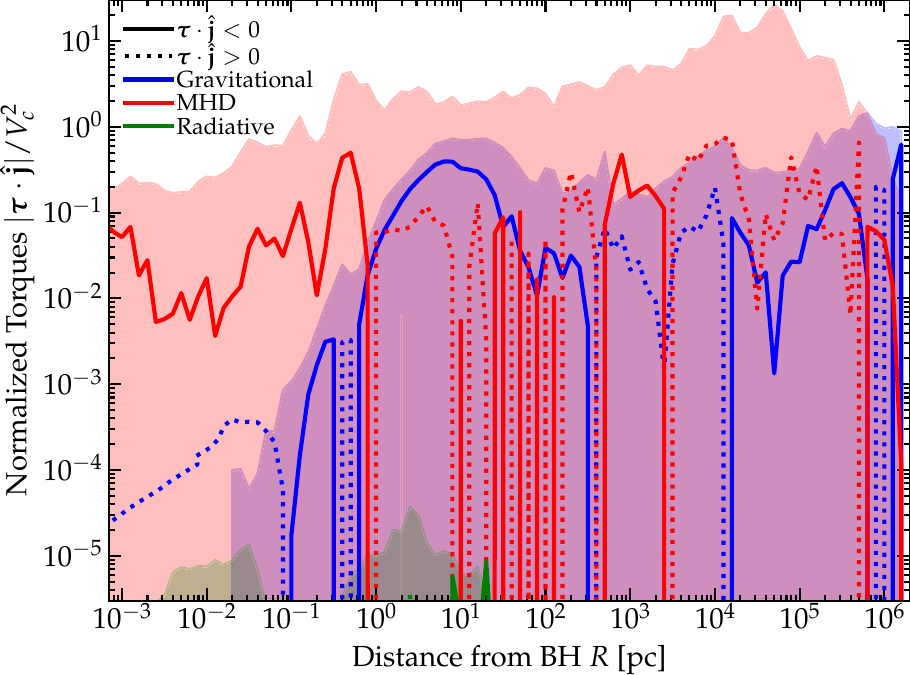}
	\centering\includegraphics[width=0.48\textwidth]{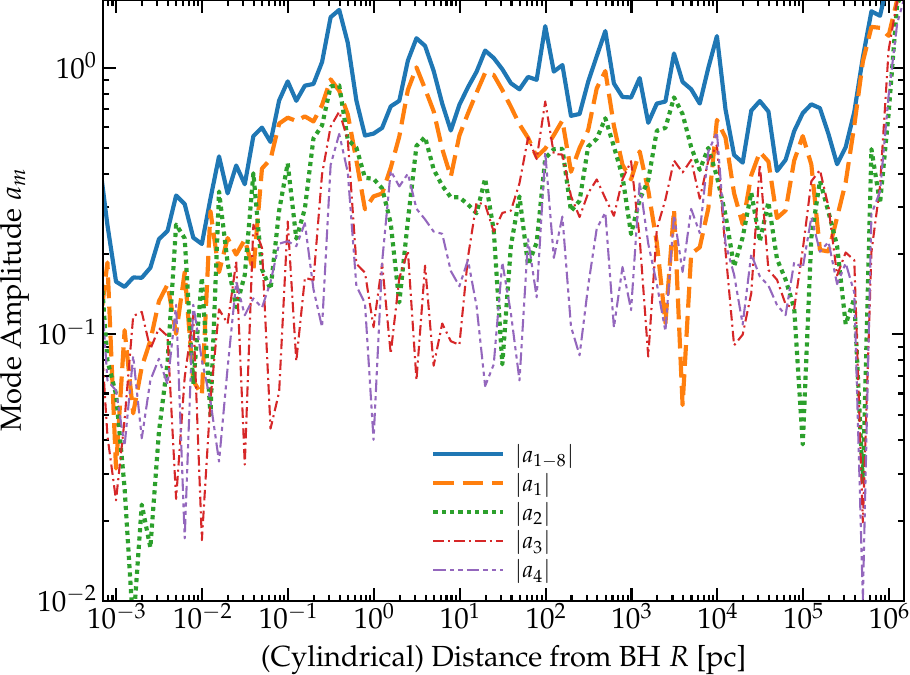} \\
	\centering\includegraphics[width=0.48\textwidth]{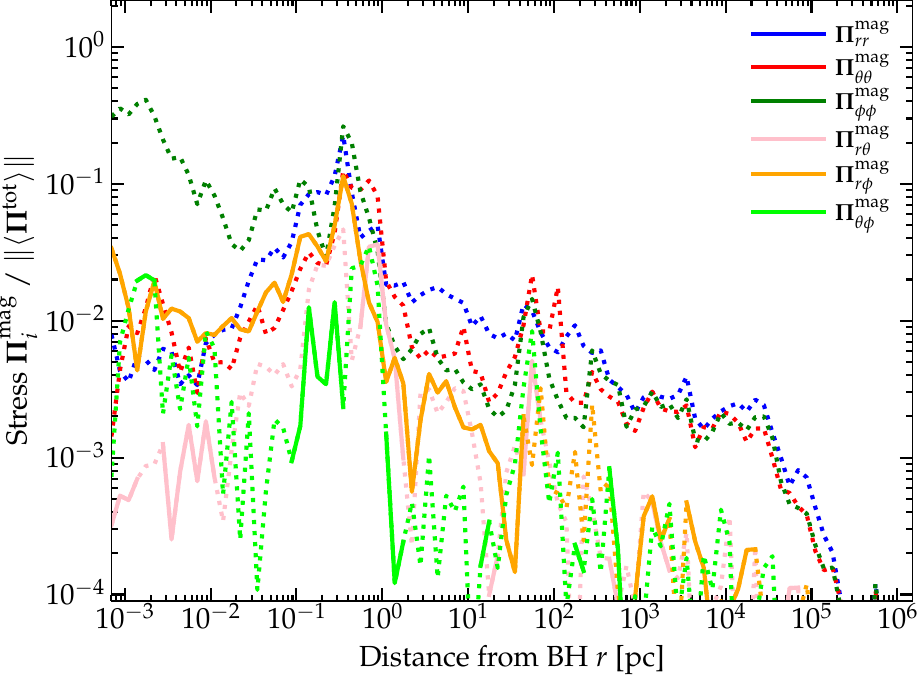}
	\centering\includegraphics[width=0.48\textwidth]{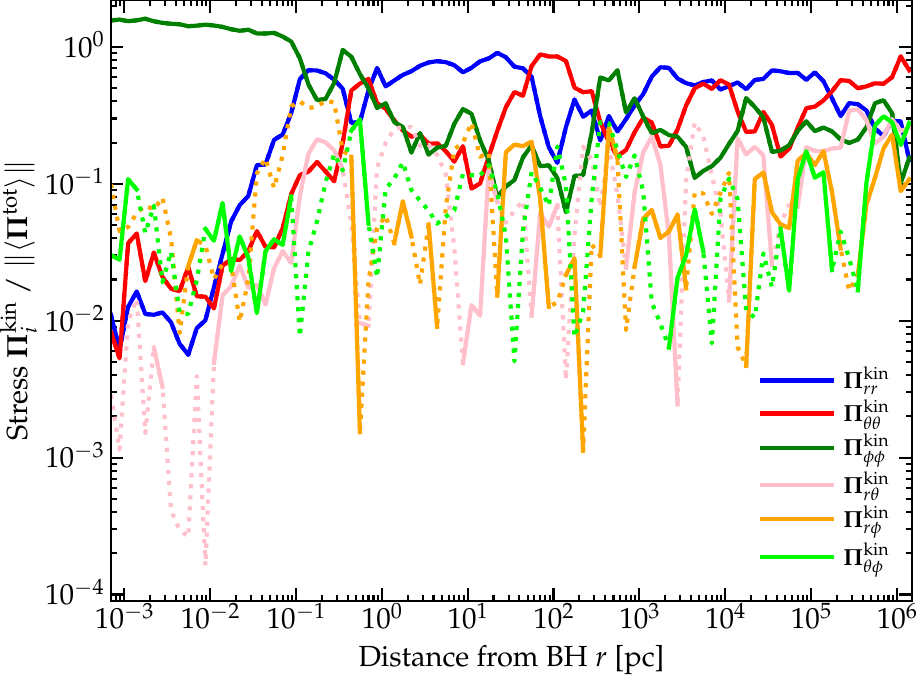} \\
	\caption{{\em Top Left:} Instantaneous (gas mass-weighted) mean torques $\tau$ in the direction of the mean angular momentum vector $\hat{\bf j}$ within each radial annulus (normalized to $r\,V_{c}\,\Omega = V_{c}^{2}$) in shells as a function of distance $r$. We restrict to cool gas with $T < 10^{4.5}\,$K but the trends are qualitatively similar regardless. Shaded region shows the $\sim 90\%$ inclusion interval. We plot the torques directly from the simulation from gravitational forces, radiation pressure forces, and MHD (magnetic, thermal, turbulent) forces. At all radii torques are efficient, implying angular momentum loss in a few orbital times. At large radii gravitational torques from stars on gas, plus MHD torques driven by stellar feedback, dominate. Where there are few stars, magnetic and turbulent/Reynolds torques take over.
	{\em Top Right:} Fractional (Fourier) mode amplitudes of asymmetric modes in the face-on projected gas surface density ($\Sigma(r,\,\phi) \equiv \Sigma_{0}\,(1 + \sum_{m} a_{m}\,\cos{(m\,\phi + \phi_{0,\,m})})$) in cylindrical annuli. At radii $\gg 0.1\,$pc there are order-unity asymmetries, often dominated by the lowest-$m$ modes (global asymmetries rather than just small-scale clumping). At small, increasingly Keplerian radii $|a_{m}|$ decreases but only modestly, to $\sim 0.1$ at $\lesssim 10^{-3}\,$pc.
	{\em Bottom Left:} Profile of different (volume-weighted) mean components of the magnetic stress tensor $\boldsymbol{\Pi}^{\rm mag} \equiv (1/4\pi)\,( |{\bf B}|^{2}\,\mathbb{I} - {\bf B}{\bf B}/2)$, in spherical coordinates, versus radius. We normalize to the mean value of the total stress tensor $\boldsymbol{\Pi}$ at each radius. Solid ({\em dotted}) line correspond to $\boldsymbol{\Pi}>0$ ($\boldsymbol{\Pi} < 0$). 
	{\em Bottom Right:} Same, for the (total) kinetic stress $\boldsymbol{\Pi}^{\rm kin} \equiv \rho\,{\bf v}\,{\bf v}$ (note this is distinct from e.g.\ a Reynolds stress). 
	\label{fig:profile.torque}}
\end{figure*}

\subsubsection{Different Contributions to the Torques}
\label{sec:torques:breakdown}

In Fig.~\ref{fig:profile.torque}, we now turn to understanding the torques driving gas inflows in more detail. We first simply plot the actual torques in the simulation. For every gas cell, we calculate the specific torque vector $\boldsymbol{\tau} \equiv {\bf r} \times {\bf a}$, where ${\bf a}$ is the acceleration from various sources (recording the value directly computed in-code), and we consider the component along the existing specific angular momentum direction $\mathbf{j} \equiv {\bf r} \times {\bf v}$ (where ${\bf r}$ is defined as the vector distance to the SMBH), and plot this in units of $V_{c}^{2} = |{\bf r}|\,V_{c}(r)\,\Omega(r)$, so that a value of $\boldsymbol{\tau} \cdot \hat{\mathbf{j}} = \epsilon\,V_{c}^{2}$ corresponds to the torque removing all of the angular momentum from an initially circular orbit in a time $\Delta t \approx (\epsilon\,\Omega)^{-1}$. We separately quantify this for the acceleration from MHD forces (the Riemann problem in the code), cosmic ray forces \citep[using the full expressions from][which allow for both the tight-coupling and free-streaming limits]{hopkins:m1.cr.closure,hopkins:cr.multibin.mw.comparison}, radiation forces \citep[likewise allowing for both limits and anisotropic radiation tensors following][]{hopkins:radiation.methods}, and gravitational forces. 

Closely related to this, we quantify different components of the stress tensor in the code. The momentum equation solved in the simulation can be written: ${\partial (\rho {\bf v} )}/{\partial t} + \nabla \cdot \boldsymbol{\Pi}^{\ast} = {\bf S}$, where in the source term ${\bf S}$ includes e.g.\ non-hyperbolic terms from radiation and cosmic rays (see references above) and other terms relevant in the weak-coupling limit, and $\boldsymbol{\Pi}^{\ast}$ is the stress tensor which can be decomposed into: 
\begin{align}
\boldsymbol{\Pi}^{\ast} &\equiv \boldsymbol{\Pi}_{\rm internal} + \boldsymbol{\Pi}_{\rm grav} = \\
\nonumber & \boldsymbol{\Pi}_{\rm kin} + \boldsymbol{\Pi}_{\rm mag} + \boldsymbol{\Pi}_{\rm therm} + \boldsymbol{\Pi}_{\rm visc} + \boldsymbol{\Pi}_{\rm cr} + \boldsymbol{\Pi}_{\rm rad} + \boldsymbol{\Pi}_{\rm grav} 
\end{align}
representing the sum of kinetic (including turbulent) $\boldsymbol{\Pi}_{\rm kin}$, magnetic $\boldsymbol{\Pi}_{\rm mag}$, thermal $\boldsymbol{\Pi}_{\rm therm}$, viscous $\boldsymbol{\Pi}_{\rm visc}$, cosmic ray $\boldsymbol{\Pi}_{\rm cr}$, radiation $\boldsymbol{\Pi}_{\rm rad}$ stress tensors constituting the usual ``total stress tensor'' $\boldsymbol{\Pi}=\boldsymbol{\Pi}_{\rm internal}$, plus gravitational $\boldsymbol{\Pi}_{\rm grav}$ forces. These terms are defined as:
\begin{align}
\boldsymbol{\Pi}_{\rm kin} &\equiv \rho {\bf v} {\bf v} \\
\boldsymbol{\Pi}_{\rm rad} &\equiv \int \frac{e_{{\rm rad},\,\nu}}{3}\,\mathbb{D}_{\nu}\,{\rm d}\nu \\
\boldsymbol{\Pi}_{\rm therm} &\equiv P_{\rm therm} {\bf I} \equiv n k_{\rm B} T {\bf I} \\
\boldsymbol{\Pi}_{\rm mag} &\equiv \boldsymbol{\Pi}_{\rm B,pressure} + \boldsymbol{\Pi}_{\rm B,\,tension} \equiv \frac{{\bf B}\cdot{\bf B}}{8\pi} {\bf I} - \frac{{\bf B}{\bf B}}{4\pi}  \\ 
\boldsymbol{\Pi}_{\rm visc} &\equiv \frac{\nu_{\rm visc}}{3}\,\left( 3 \hat{\bf B} \hat{\bf B} - {\bf I} \right) \left(3 \hat{\bf B} \hat{\bf B} - \bf{I} \right) : \left(\nabla {\bf v} \right) \\
\boldsymbol{\Pi}_{\rm cr} &\equiv \int {\bf p}_{\rm cr}\,{\bf v}_{\rm cr}({\bf p}_{\rm cr}) f_{\rm cr}({\bf p}_{\rm cr}) \, d^{3}{\bf p}_{\rm cr} \\
\boldsymbol{\Pi}_{\rm grav} &\equiv \frac{1}{4\pi\,G}\,\left(  {\bf g} - \frac{{\bf g}\cdot{\bf g}}{2}\,{\bf I}\right)
\end{align}
(with ${\bf g} \equiv -\nabla \Phi_{\rm grav}$).

In order to better understand the origin of the gravitational and kinetic stresses in the disk plane in particular, it is also helpful to quantify the degree of non-axisymmetry of the system. Noting that the total surface mass density $\Sigma_{\rm tot}(R,\,\phi)$ within each cylindrical annulus $R$ can be Fourier decomposed into $\Sigma_{\rm tot}(R,\,\phi) = \langle \Sigma_{\rm tot}(R) \rangle \,\left[ 1 + \sum_{m=1}^{\infty}\,a_{m}(R)\,\cos{(m\,[\phi-\phi_{0,\,m}(R)])} \right]$, we extract the coefficients $|a_{m}(R)|$, and plot the first few coefficients $a_{m}$ as a measure of the global asymmetry. The behavior is similar for higher-$m$ modes, but $m=1$ is most relevant for linear global gravitational instabilities interior to the BHROI; see \citealt{hopkins:zoom.sims,hopkins:inflow.analytics,hopkins:cusps.ell,hopkins:cusps.evol}.

The first thing to note is that the torques are large, in a dimensionless sense, $|\boldsymbol{\tau} \cdot \hat{\mathbf{j}}| \sim 0.1\,V_{c}^{2}$, i.e.\ the timescale for angular momentum loss of an initially-circular orbit is just a couple of orbital times ($t_{\rm orbit} = 2\pi/\Omega$). This is expected from the very large $\dot{M}_{\rm in}$ in Fig.~\ref{fig:profile.mass} and shown in Fig.~\ref{fig:profile.mass.timescales.kinematics}: we anticipate $\dot{M}_{\rm in} \sim M_{\rm gas}(r)\, |\boldsymbol{\tau} \cdot \hat{\mathbf{j}}| / (r\,V_{c}) \sim (|\boldsymbol{\tau} \cdot \hat{\mathbf{j}}|/V_{c}^{2})\,\pi\,\Sigma_{\rm gas}\,r\,V_{c} \sim 10-100\,{M_{\odot}\,{\rm yr}^{-1}}$ at these radii (inserting typical values from Fig.~\ref{fig:profile.mass} \&\ Fig.~\ref{fig:profile.torque} in the final evaluation). This means that accretion is fundamentally {\em dynamical} here, occurring on of order the dynamical time, as opposed to a slow, secular, viscous-type process as often assumed for much lower accretion-rate systems.

From a cursory examination of the components of $\boldsymbol{\Pi}^{\ast}$ extracted directly from the simulation, or from the torques in Fig.~\ref{fig:profile.torque} (where various torques fall below the plotted range), or our discussion above of relevant physics and scalings, it is easy to confirm that the physical viscosity ($\boldsymbol{\Pi}_{\rm visc}$), cosmic ray ($\boldsymbol{\Pi}_{\rm cr}$), radiation ($\boldsymbol{\Pi}_{\rm rad}$), and pure-thermal (isotropic by definition $\boldsymbol{\Pi}_{\rm therm}$) terms in the stress tensor contribute negligibly to the torques at essentially all radii modeled here. There are really only three contributions of broad importance: the ``gravitational torque'' (${\bf r} \times {\bf g}$), and the MHD torques arising from a combination of magnetic ($\boldsymbol{\Pi}_{\rm mag}$) and kinetic or Reynolds-like ($\boldsymbol{\Pi}_{\rm kin}$) stresses.

\subsubsection{The Gravitational \&\ ``Stellar Feedback'' Torques}
\label{sec:torques:grav}

On scales $\gtrsim$\,pc, we see that gravitational torques are important (even if not always dominant) for the dynamics of angular momentum exchange, with large-amplitude $|a_{m}| \sim \mathcal{O}(1)$ asymmetries (obvious in the visual morphology of gas and stars) producing strong torques. This is studied on these scales in much greater detail in \citet{daa:20.hyperrefinement.bh.growth}, who show that such torques act most efficiently with gas forced into shocks and dissipation (allowing gas orbits to decay rapidly) via asymmetries in the {\em stellar} distribution which dominates the local mass density (hence the actual masses {\em exerting} the torques in Fig.~\ref{fig:profile.torque}). This does mean that even when gravitational torques are dominant on these scales, the MHD torque is generally order-of-magnitude comparable (the torques induce shocks which have comparable amplitude and [usually] opposite sign, as we see). As shown in \citet{daa:20.hyperrefinement.bh.growth}, phenomena like the sign flip we see in Fig.~\ref{fig:profile.torque} at $\sim 0.5-10\,$kpc are often transient and can flip back-and-forth, with the time-averaged effect of these torques on these scales being to reduce gas angular momentum. This also agrees with the results of previous ``nuclear zoom-in'' simulations \citep{levine2008:nuclear.zoom,prieto:2016.zoomin.sims.to.fewpc.hydro.cosmo.highz,prieto:2017.zoomin.sims.agn.fueling.sne.fb,daa:20.hyperrefinement.bh.growth} and both idealized simulations of small scales in gas+stellar nuclear disks \citep{hopkins:zoom.sims,hopkins:qso.stellar.fb.together,williamson:2022.gizmo.rhd.psph.sims.binary.smbh.torii.radiation.reduces.grav.torques}, 
observations of nearby nuclear disks \citep{lauer:central.minimum.ell,hopkins:m31.disk,querejeta:grav.torque.obs.m51} as well as galaxy-scale simulations of galaxy mergers, strong bars, and large clump-type perturbations, which were the first to describe this gravitational torques process as uniquely efficient in mixed (collisional+collisionless) systems \citep{barnes.hernquist.91,barneshernquist96,hopkins:disk.survival,hopkins:disk.survival.cosmo}.

On galactic scales $\gtrsim$\,kpc, we also see comparable and sometimes dominant MHD torques to gravitational torques, which relate to strong shocks sometimes (as noted above) driven by gravitational motions (e.g.\ infall/accretion, bar or clump-induced shocks) but sometimes also due to e.g.\ strong shocks owing to stellar feedback events (motions that can be traced directly back to e.g.\ superbubbles and outflows). The latter, in particular, is directly related to the large scatter in the magnitude of the MHD torques at $r\gg$\,pc -- cells with $|\boldsymbol{\tau}| \gg V_{c}^{2}$ for example almost entirely owe to material being strongly accelerated by supernovae shocks and superbubbles/winds.\footnote{In \papertwo\ we show more details of the distribution of torques in both space and time, which (unsurprisingly given the large scatter plotted here) exhibits large fluctuations (also found in \citealt{daa:20.hyperrefinement.bh.growth} on $\gg$\,pc scales).} This is again consistent with previous studies and very similar to the results in e.g.\ \citet{prieto:2016.zoomin.sims.to.fewpc.hydro.cosmo.highz,prieto:2017.zoomin.sims.agn.fueling.sne.fb}. If feedback is ``self-regulating'' on these scales (e.g.\ on-average balances gravitational collapse and maintains a super-sonic turbulent $Q_{\rm turb} \sim 1$), then it is (by definition) true that the non-circular (or non-hydrostatic) motions generated by such feedback should be comparable to those generated by gravity (and generally have the opposite sign). This is closely related to the question of ``what powers the turbulence'' in the ISM (gravity or stellar feedback), where self-regulating models necessarily predict that if inflow is balanced by outflow and star formation (as it is here on super-kpc scales; see Figs.~\ref{fig:profile.mass} \&\ \ref{fig:profile.mass.timescales.kinematics}) the two should be comparable \citep{orr:2020.resolved.dispersions.sfrs.correlations}. 

Note that this physical connection, as well as the large scatter, make it difficult to un-ambiguously determine precisely ``how much inflow'' is driven by each mechanism, as transient events like shocks could have large integrated effects on gas orbits. In future work, one can envision following the Lagrangian evolution of individual gas parcels which ultimately accrete onto the SMBH over cosmic time, and asking the question (distinct from what we plot here) of how they lost angular momentum (and whether they preferentially sample lower angular-momentum material in the first place) at different times and locations on their way to being accreted (akin to the studies on larger scales in e.g.\ \citealt{angles.alcazar:particle.tracking.fire.baryon.cycle.intergalactic.transfer,hafen:2018.cgm.fire.origins,hafen:2019.fire.cgm.fates}).

Regardless of this, at smaller radii $\ll$\,pc, with star formation efficiently shut down, we see a transition around $\sim 0.3-0.6\,$pc (Fig.~\ref{fig:profile.mass}) from the local mass density and gravitational field being dominated by stars at larger radii, to entirely gas-dominated at smaller radii. This dramatically reduces the efficacy of gravitational torques and (of course) any torques owing directly or indirectly to stellar feedback. As noted above and shown more formally in \citet{hopkins:inflow.analytics}, the leading-order gravitational torque arises when one has a two-component system with a collisional, dissipative gas component being acted upon by a dominant collisionless component (e.g.\ stars). When the collisionless component becomes small, so the gas disk is effectively ``one-component,'' the strength of the torques (averaged over the orbit of some gas parcel) drops dramatically, until it eventually is reduced to the small and higher-order resonant-only contributions given by \citet{kalnajs:1971}. The analytic prediction from \citet{hopkins:inflow.analytics} is that the torque drops $\propto f_{\ast} \equiv \Sigma_{\ast}/(\Sigma_{\ast} + \Sigma_{\rm gas})$ (in terms of the mean gas and stellar mass densities in an annulus of some radius $r$ from the BH) when $\Sigma_{\ast}$ becomes smaller, which agrees well with the trend we see in the ``transition region'' (comparing Fig.~\ref{fig:profile.torque} \&\ Fig.~\ref{fig:profile.mass}). 

Note that this does not mean that the $m=1$ modes themselves cease; as shown in Fig.~\ref{fig:profile.torque} they propagate towards small $r$, albeit with decreasing amplitude. Indeed \citet{hopkins:zoom.sims,hopkins:m31.disk,hopkins:slow.modes} showed that this lopsided disk mode, if excited at large radius where the disk is marginally self-gravitating, can excite a response at all radii $r\rightarrow 0$. However, the point in \citet{hopkins:inflow.analytics} is that {\em for a given asymmetric mode amplitude} $|a_{m=1}|$, the effective net torque on gas is weaker if $f_{\ast}$ is smaller. Moreover, as shown in \citet{hopkins:slow.modes}, if the mass profile of the collisionless component ceases to rise sufficiently steeply towards $r\rightarrow 0$ (e.g.\ for $\Sigma_{\ast} \propto r^{-\eta}$ with $\eta \lesssim 0.5$), then there is a refraction barrier (akin to an inner Lindblad resonance in many ways) across which the torque switches sign, exactly as we see in Fig.~\ref{fig:profile.torque}.\footnote{As shown in the linear analysis in \citet{hopkins:inflow.analytics} and standard Solar-system texts \citep[e.g.][]{murraydermott}, the exact rate at which the amplitude of the sign-flipped gravitational torque declines as $r\rightarrow 0$ is partly an artifact of our definitions, as it relates to defining ${\bf r}$ relative to the BH position, but this is not important for our analysis here.}

If gravitational torques were the {\em only} mechanism for angular momentum transfer, then as speculated in \citet{hopkins:inflow.analytics}, this barrier would create a trap and ``pileup'' of gas between $\sim 0.1-1\,$pc, until it became so dense it would necessarily fragment and form stars, until sufficient stars formed (assuming a large gas supply continued to flow in) to steepen the profile and reverse the barrier (making the stars-on-gas gravitational torque strong again), moving the barrier gradually inwards and building a steep stellar cusp. And this is what appears to happen in the simulations in \citet{daa:20.hyperrefinement.bh.growth}, as discussed below. 

However, here we see something at first apparently rather remarkable (though perhaps not on further reflection): at the radii where the gravitational torque becomes highly inefficient, the MHD torques ``take over,'' with the same sign and broadly similar magnitude.

\subsubsection{The MHD Torques}
\label{sec:torques:mhd}

In \papertwo, we will study the torques in the inner accretion disk in much greater detail, in order to understand the origins of the strong toroidal field, relative role of the Maxwell versus Reynolds torques, their dominant components and fluctuations in time and space, physical origin and ultimate energy sources, and how they are dynamically maintaining accretion. Here, we simply wish to identify and summarize some basic properties of the ``MHD torques'' across a broad range of radii.

As noted above, we see in Fig.~\ref{fig:profile.torque} that at radii $\sim1-1000$\,pc, the gravitational torques play an important role, and the MHD torques are heavily influenced by stellar feedback, consistent with previous work (\S~\ref{sec:previous} below). There are occasional, usually transient, exceptions, when e.g.\ strong shocks are induced in mergers and the shocks instantaneously dominate the torque -- though as discussed above in such a situation the angular momentum exchange in the shock can be itself ultimately driven/determined by the gravitational forces (or stellar superbubbles and winds), so the ``labeling'' can be somewhat ambiguous \citep{hopkins:inflow.analytics}. 

But to better understand the structure of these torques in either case, we plot the different components (in spherical coordinates centered on the BH) of the magnetic $\boldsymbol{\Pi}^{\rm mag}$ and kinetic $\boldsymbol{\Pi}^{\rm kin}$ stess tensors versus radius in Fig.~\ref{fig:profile.torque}. We normalize the components to the magnitude (Frobenius norm) of the sum internal stress tensor $\| \boldsymbol{\Pi}^{\rm internal} \| \equiv \| \boldsymbol{\Pi}_{\rm kin} + \boldsymbol{\Pi}_{\rm mag} + \boldsymbol{\Pi}_{\rm therm} + \boldsymbol{\Pi}_{\rm visc} + \boldsymbol{\Pi}_{\rm cr} + \boldsymbol{\Pi}_{\rm rad} \|$ (i.e.\ the total stress ignoring the source terms so assuming the tightly-coupling limit for cosmic rays and radiation, and excluding gravitational forces, so representing the internal forces from the gas). At all radii, we see the kinetic components sum close to unity, i.e.\ represent a dominant term in the total.\footnote{Note, as defined here, components can have fractional values $>1$, if there are other components of similar magnitude with opposite sign.} Other than a small range of radii in the CGM, where the thermal pressure contribution to the stress is comparable to the kinetic (as we showed in Fig.~\ref{fig:profile.dynamics}, where the turbulence is trans-sonic), the other (non-kinetic, non-magnetic) terms in the stress are generally fractionally small. At large radii $\gtrsim \,$pc, the kinetic terms are quasi-isotropic (with a mild radial bias from inflow/outflow motion), and dominated by random motions (e.g.\ $| \boldsymbol{\Pi}^{\rm kin}_{rr} | \equiv \langle \rho\,v_{r}^{2} \rangle \gg \langle \rho \rangle \langle v_{r} \rangle^{2}$), with the mixed/off-diagonal terms ($\boldsymbol{\Pi}^{\rm kin}_{r\theta}$, $\boldsymbol{\Pi}^{\rm kin}_{r\phi}$, $\boldsymbol{\Pi}^{\rm kin}_{\theta\phi}$) having essentially random (rapidly alternating) signs as they average out to smaller values than the diagonal terms. All of this is consistent with incoherent (e.g.\ turbulent and/or feedback-dominated) motions at a non-negligible fraction of the circular velocity -- i.e.\ roughly as expected from the virial theorem -- at similar Mach numbers for the components shown in Fig.~\ref{fig:profile.dynamics}. At $\ll  1\,$pc we clearly see the ordered disk form: the azimuthal (rotational) component $\boldsymbol{\Pi}_{\phi\phi}$ dominates the total stress, this component is itself strongly dominated by its mean/coherent component ($\boldsymbol{\Pi}_{\phi\phi} \approx \langle \rho \rangle \langle v_{\phi} \rangle^{2}$), and the radial and azimuthal terms become sub-dominant by a factor of $\sim 100$ (implying turbulent/incoherent velocities more like $\sim 0.1\,V_{c}$). 

We will analyze these terms to study e.g.\ the Reynolds stresses within the accretion disk in \papertwo, but note that what is plotted here, for the sake of comparing the entire stress tensor and radial range, is not the Reynolds stress. Specifically, components like $\boldsymbol{\Pi}^{\rm kin}_{r\phi} \equiv \langle \rho\,v_{r}\,v_{\phi} \rangle$ are defined as the average of the total values of the relevant velocity components like $v_{\phi}$, whereas the Reynolds stress is defined in terms of the incoherent components $\delta v_{\phi} = v_{\phi}-\langle v_{\phi} \rangle$. So the fact that, for example, $\boldsymbol{\Pi}^{\rm kin}_{r\phi}<0$ here at all radii $\lesssim 1\,$pc simply means that there is, at this snapshot in time, net inflow through all radii $\lesssim\,$pc in the disk, because it is dominated by its coherent components $\boldsymbol{\Pi}^{\rm kin}_{r\phi} \sim \langle \rho\rangle \langle v_{r} \rangle \langle v_{\phi} \rangle$ (and $\langle v_{\phi} \rangle >0$ by definition of our coordinate convention for the inner, rotating disk, while $\langle v_{r} \rangle < 0$ denotes inflow). 

For the magnetic stresses $\boldsymbol{\Pi}^{\rm mag}$, we see the corresponding expected behavior: overall $\|\boldsymbol{\Pi}^{\rm mag}\|$ is a fractionally small contribution to $\|\boldsymbol{\Pi}^{\rm internal} \|$ at large radii $\gg\,$pc where magnetic field effects on the dynamics are small and $\beta$ is increasingly large, and at radii $\gtrsim\,$pc the magnetic fields are quasi-isotropic/tangled (with again a mild radial bias), and magnitudes relative to velocity consistent with the \Alf\ Mach numbers in Fig.~\ref{fig:profile.dynamics}. At sub-pc scales, we see the toroidal magnetic field becomes dominant and at $\lesssim 0.01\,$pc begins to contribute at up to an order-unity level to the total stress, though it is still sub-dominant to rotational support of the disk. This component is dominated by the coherent field, while the others remain dominated by largely incoherent fields (more detailed analysis in \papertwo). Here the $r\phi$ component is a more traditional Maxwell stress, though it is dominated by a mix of coherent and incoherent components depending on exactly which radius we analyze; we will study this in detail in \papertwo\ but for our purposes here we can note (a) the sign is positive, which in the convention here means it is transporting angular momentum outwards and therefore promoting inflow; (b) the fractional magnitude (assuming the disk is orbiting at $\sim V_{c}$) is comparable to the values needed to explain the fractional MHD torques and inflow rates in Figs.~\ref{fig:profile.torque} \&\ \ref{fig:profile.mass}; and (c) at the smallest radii, the magnitude $|\boldsymbol{\Pi}^{\rm mag}_{r\phi}|$ is comparable to its kinetic counterpart even with the kinetic term defined in terms of the bulk/coherent components, meaning that the Maxwell stress must be at least comparable to the Reynolds stress (if not larger) at these radii. 

Why do these torques appear to smoothly ``take over'' from the gravitational torques with broadly similar magnitude? This might at first to appear to require some sort of ``conspiracy,'' but closer examination of Fig.~\ref{fig:profile.torque} suggests a more mundane explanation, namely that this is required for continuity/steady-state. First, upon examination of Fig.~\ref{fig:profile.torque} we see that it is {\em not} the case that there is some sort of ``exact'' boundary-condition matching occurring here. Both the gravitational torque and (especially) MHD torque have huge fluctuations in magnitude, so the ``transition'' between one and the other dominating is actually spread out, in a local sense (e.g.\ considering different narrow annuli in solid angle from the SMBH), over at least an order of magnitude in radius (again see \papertwo\ for more details of the distribution of torques and stresses at these radii, in particular). The transition only appears ``narrow'' because we follow and plot so many orders of magnitude in radius. Second, this large scatter also means there are large fluctuations where one or the other dominates outside/inside this radius. Third, we see that even the instantaneous mass-weighed mean specific torque is very clearly not precisely constant as a function of radius, but fluctuates by an order of magnitude: the boundary between the mean torque being MHD-dominated and gravity-dominated is one such example (there is a factor $\sim 10$ fluctuation in their sum between $\sim 0.1-10\,$pc).  It is true that the nearest ``peaks'' in the mean specific torque on either side of $\sim 1$\,pc happen to have remarkably similar amplitude in this particular snapshot, but comparing other snapshots even these peaks are only comparable in an order-of-magnitude sense. With this in mind, it is much easier to understand. Continuity means that if there were a sudden change in the torque efficiency around $\sim 1\,$pc, mass would either ``pile up'' (or be evacuated), which would, for most reasonable models for the origin of the MHD torques (e.g.\ those discussed in \papertwo\ in detail) and for the gravitational torque models lead to a corresponding increase (decrease) in the torques, until $\dot{M}_{\rm in}\sim$\,constant with radius was in approximate steady-state. While it is true in principle that a ``sharp'' discontinuity in the torques could be balanced (for the same $\dot{M}_{\rm in}$) by a similar discontinuity in the gas surface density, the fact that the transition is ``smeared out'' in both space and time by large local fluctuations and turbulence means that this cannot reasonably be self-sustaining (so $\Sigma_{\rm gas}$ must be smooth, hence the specific torques being smooth). In summary, the ``transition'' being continuous is a statistical, order-of-magnitude statement over a fairly wide range of radii.

\begin{figure*}
	\centering\includegraphics[width=0.95\textwidth]{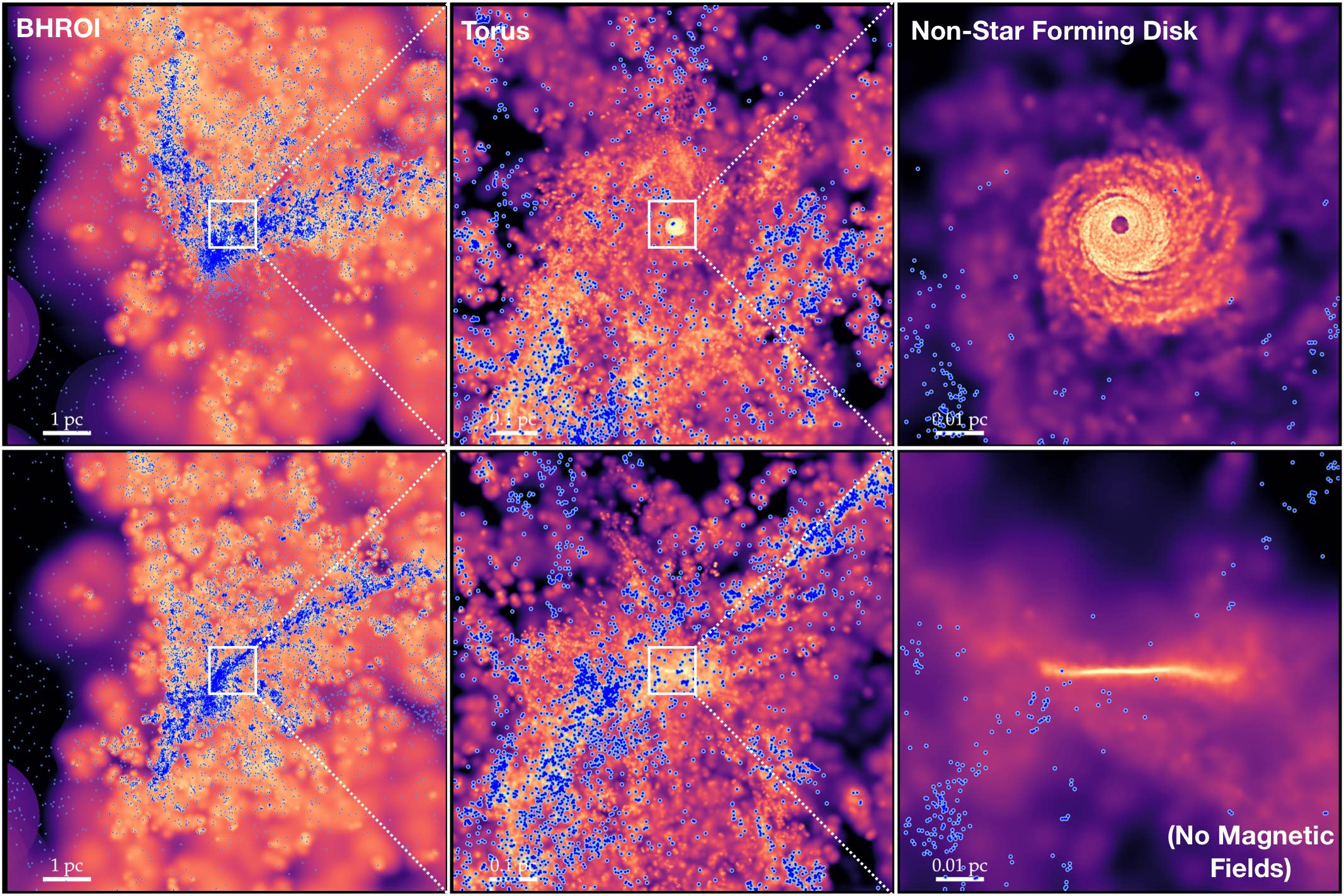}
	\caption{Images of a re-run of our fiducial simulation {\em without} magnetic fields (\S~\ref{sec:no.mhd}). We show gas face-on ({\em top}) and edge-on ({\em bottom}) as Fig.~\ref{fig:images.faceon}-\ref{fig:images.edgeon}. We overlay the star particles which form (blue points) to emphasize that the extreme ``clumpiness'' in the gas is a real effect: these are collapsing dense gas clouds which rapidly form stars and produce runaway star formation on sub-pc scales. A much smaller (spatially and in mass/surface density) inner non-star-forming disk remains, but it is truncated at the radii where the {\em thermal-only} Toomre $Q$ parameter falls below $Q \ll 10$ ($\sim 0.01\,$pc; see Fig.~\ref{fig:profile.dynamics}).
	\label{fig:images.nomhd}}
\end{figure*}
\begin{figure*}
	\centering\includegraphics[width=0.48\textwidth]{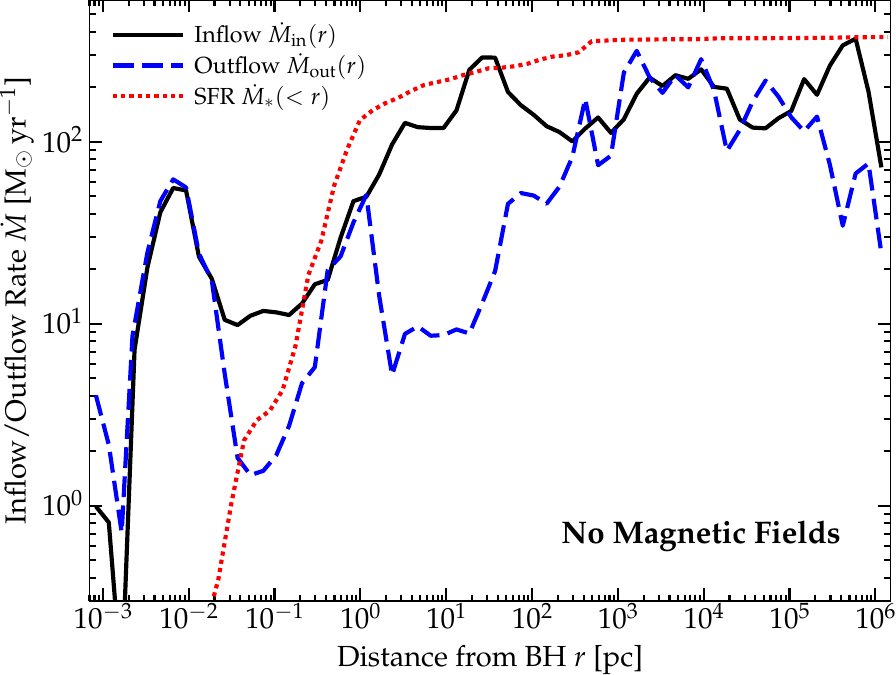}
	\centering\includegraphics[width=0.48\textwidth]{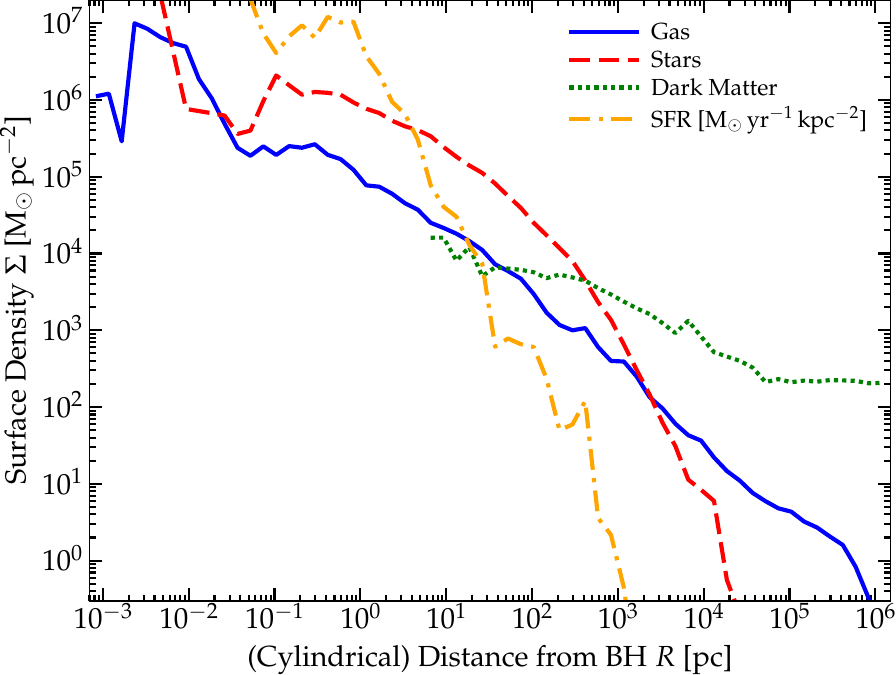} \\
	\centering\includegraphics[width=0.48\textwidth]{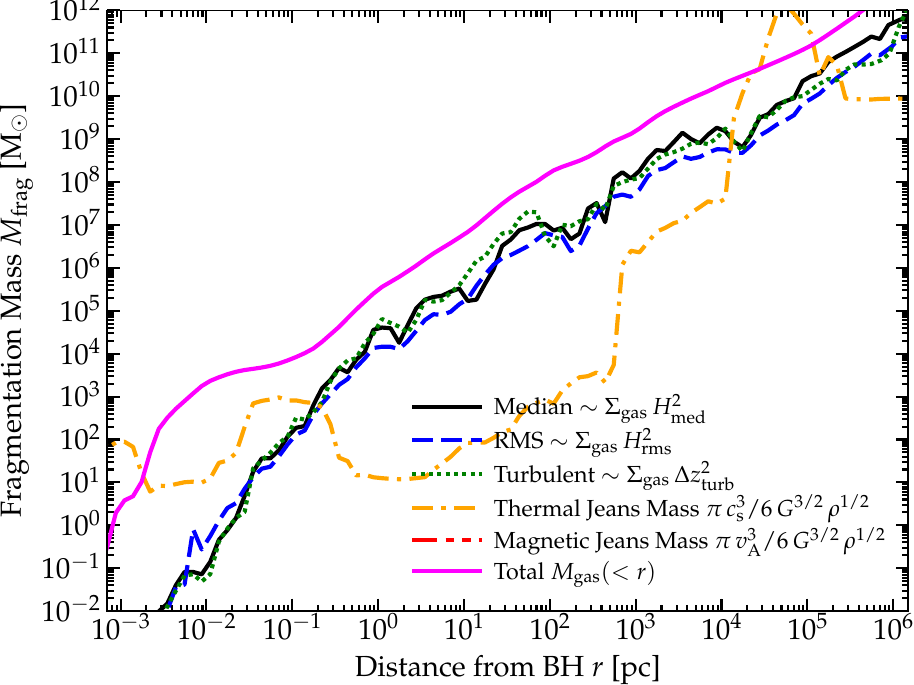}
	\centering\includegraphics[width=0.48\textwidth]{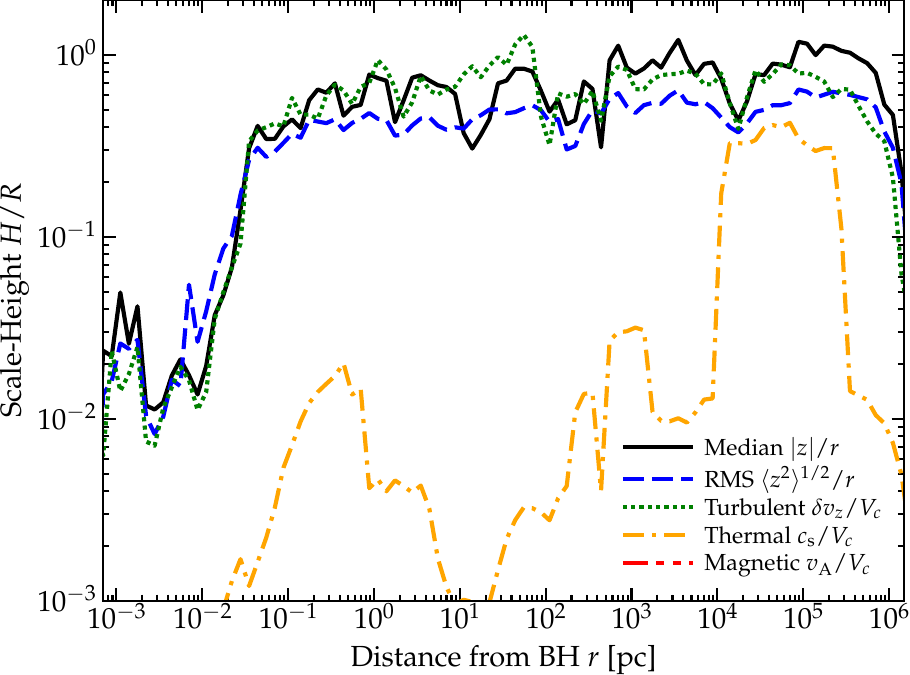} \\
	\centering\includegraphics[width=0.48\textwidth]{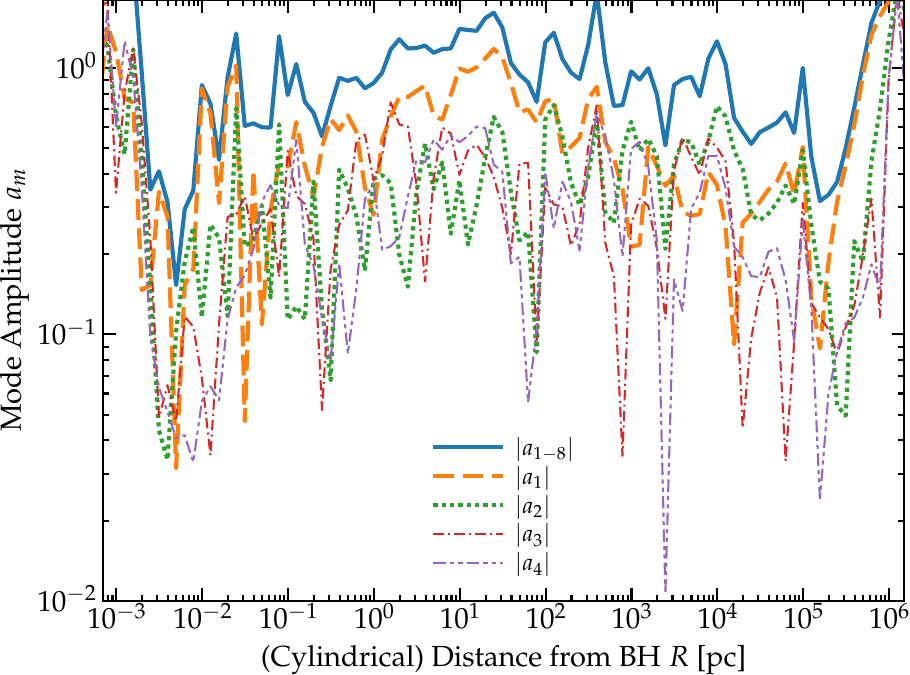} 
	\centering\includegraphics[width=0.48\textwidth]{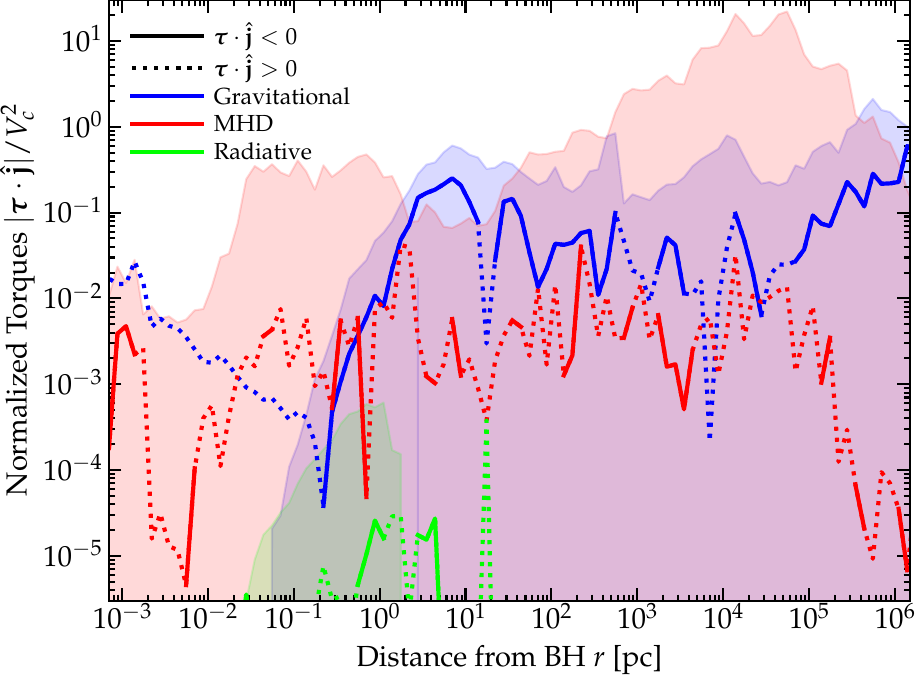} \\
	\caption{Radial profiles as Figs.~\ref{fig:profile.mass}-\ref{fig:profile.torque}, for the resimulation without magnetic fields (\S~\ref{sec:no.mhd}). 
	We specifically compare the inflow/outflow/SF rates ({\em top-left}) and surface densities ({\em top-right}) as Fig.~\ref{fig:profile.mass}; 
	characteristic fragmentation scales ({\em middle-left}) and scale-heights ({\em middle-right}) as Fig.~\ref{fig:profile.dynamics}; 
	and non-axisymmetric mode amplitudes ({\em bottom-left}) and torques ({\em bottom-right}) as Fig.~\ref{fig:profile.torque}. 
	Consistent with the morphology in Fig.~\ref{fig:images.nomhd}, we see that fragmentation and star formation proceeds much more rapidly at sub-pc scales, without magnetic fields to resist gravito-turbulent/Jeans fragmentation nor support a thicker (higher $H/R$), lower-density disk. The SFR density rises monotonically to $r\rightarrow 0$, and the total SFR in the last dynamical time exceeds the inflow rate at all radii $\gtrsim 1\,$pc. Meanwhile the MHD torques are much weaker (and have the opposite sign from that required for accretion) at $\ll1\,$pc, so we actually see net outflow (a decretion disk) with only small episodes of accretion of clumps of gas, in the disk at $\lesssim 0.01\,$pc. Note the ``peak'' in inflow+outflow with the two nearly identical at $\sim 0.01\,$pc arises owing to coherent eccentric motion from the obvious large single-component lopsided disk mode (large $m=1$ mode at small $r$). Together this reduces the total accretion rate (over the duration of this test) into the accretion disk at $R \lesssim 10^{-3}$\,pc by a factor of $\gg 100$, and produces runaway fragmentation and star formation at radii $\lesssim\,$pc. More details of the disk structure are contrasted in \papertwo.
	\label{fig:profile.nomhd}}
\end{figure*}

\section{Simulation without Magnetic Fields}
\label{sec:no.mhd}

In this section, we compare a simulation run without magnetic fields. We begin from the same initial condition/snapshot used for ``zooming in'' as our fiducial simulation, at the same time when we would normally begin our hyper-refinement process. But now we remove magnetic fields. This technically means the ``initial condition'' for the zoom-in is slightly out-of-equilibrium, but recall as shown above (1) it is already a highly non-equilibrium system, (2) on large scales outside of where the iterative hyper-refinement procedure begins, the magnetic fields are not as dynamically important ($\beta$ is larger), and (3) we evolve each hierarchical level of the hyper-refinement several dynamical times before allowing a subsequent level of refinement so that each level can re-equilibrate, so this is given time to occur in these runs. 

Altogether, this experiment appears to support all of the statements above regarding the role of magnetic fields. 

Figs.~\ref{fig:images.nomhd} \&\ \ref{fig:profile.nomhd} repeat some of our earlier comparisons in e.g.\ Fig.~\ref{fig:images.faceon}-\ref{fig:images.star.faceon} and Figs.~\ref{fig:profile.mass}-\ref{fig:profile.torque}, respectively, but for this simulation without magnetic fields. Unsurprisingly, on large scales (again, outside of where the iterative hyper-refinement procedure begins), the change is small. Indeed at all $r \gg 1\,$pc, the morphologies, star formation rates, gas and stellar densities, $m=1$ mode amplitudes, scale-heights, thermochemical properties of the gas (phase distribution, temperatures, ionization states), and strength of gravitational torques are basically the same as in our ``default'' simulation with MHD. However we caution that this simulation is only run a short time, so this could simply arise at the largest scales because the system has no time to come to a new equilibrium -- for galactic scales $\gg 100\,$pc, explicit studies of cosmological simulations with and without magnetic fields are more informative. But at sub-pc scales, we can immediately see some major {\em qualitative} differences appear. 

Visually, we can see much stronger fragmentation setting in on scales $\lesssim 0.1-0.5\,$pc. This is also immediately evident in the surface density of star formation, which in the runs with MHD ``cuts off'' and is strongly suppressed at $\ll 1\,$pc, while without MHD it continues to rise monotonically to small radii. The integrated SFR (averaged over a few dynamical times) inside of $<\,(0.1,\,1,\,10)\,{\rm pc}$ rises from $\sim (0.1,\,10,\,30)\,{\rm M_{\odot}\,yr^{-1}}$ with MHD to $\sim (5,\,120,\,300)\,{\rm M_{\odot}\,yr^{-1}}$ without. The surface density of stars begins to rapidly rise on small scales as gas is depleted: whereas with MHD we saw gas dominate the local density over stars within $r \lesssim 0.1\,$pc, we now see stellar densities increase rapidly and beginning to dominate after a few tens of dynamical times at all radii $r\rightarrow 0$. 

Explaining this rapid enhancement of fragmentation, we see that without magnetic support, the disk $H/R$ becomes smaller/thinner (without MHD) inside of $r\ll$\,pc, by a factor of $\sim 10-30$. There is still highly super-sonic turbulence supporting it, but it is well-established that absent magnetic fields, a disk supported by stronger super-sonic turbulence will actually have more rapid fragmentation \citep{hopkins:excursion.ism}. We see this manifest in much larger $a_{m}$ especially for $m\gg1$ without MHD, as gravo-turbulent fragmentation runs away (boosting the negative gravitational torque at small radii). We also see that without magnetic fields, the hydrodynamic torques on sub-pc scales are much weaker than the magnetized ``MHD torques'' -- moreover the sign of the hydrodynamic torques without MHD is actually opposite (they are net moving material outward). Thus while the stellar densities are still relatively low at $r \ll $\,pc, this creates a ``bottleneck'' or ``pileup'' of material at these radii, which further assists the runaway fragmentation. We also show in \paperthree\ that this leads to an even more top-heavy stellar IMF at these radii. The inefficient hydrodynamic torques without magnetic fields, coupled to efficient fragmentation, mean that the inflow rate to the BH is strongly suppressed at $\ll \,$pc radii (especially at the smallest radii we follow, $\sim 10^{-3}\,$pc). The {\em net} inflow rate into our central resolution element -- i.e.\ total mass growth rate of the sink interior to $<80\,$au is lower by a factor of $\sim 200-300$ on average over the duration of the more limited no-MHD run.\footnote{Note that in Fig.~\ref{fig:profile.nomhd}, it appears as if there is still some modest inflow at $\sim 1\,{\rm M_{\odot}\,yr^{-1}}$ into the smallest radii shown. However, like the inflow ``peak'' at $\sim 0.01\,$pc in this figure, this is dominated by the coherent motion of the eccentric disk, and note that the outflow rate at this annulus is actually {\em larger}. The {\em net} inflow we define by the rate at which the central ``sink'' captures gas which is bound to it with an apocentric orbital radius $<80\,$au. This is reduced to more like $\lesssim 0.05\,{\rm M_{\odot}\,yr^{-1}}$ for the duration of this re-simulation without magnetic fields.} So while still non-zero owing to some transient and non-spherically-homogeneous structure (and indeed still fairly large in an absolute sense), the accretion rates without MHD are dramatically suppressed relative to those with magnetic fields present, at least until a much larger stellar density is able to build up (beyond the duration of our simulation to explore). We will study the more detailed consequences for the disk at $\ll$\,pc scales in this simulation in \papertwo.

\section{Scales Where Different Simulation Physics ``Ingredients'' Become Important}
\label{sec:scales.vs.physics}

From the above, we can refer back to Table~\ref{tbl:physics} and review the major physical ``ingredients'' in these simulations, in order to discuss where each plays a crucial role in determining the {\em dynamics} of the system. We emphasize this because of course, certain physics will always be important by definition if one is interested in them for their own sake, or their influence on certain observations (e.g.\ one could have dynamically negligible magnetic fields but they would still be ``important'' to predict Zeeman observations). This is summarized in Table~\ref{tbl:physics.vs.scale} and Fig.~\ref{fig:scale.fast.vs.slow}.

\subsection{Gravity \&\ Collisional vs.\ Collisionless Dynamics}
\label{sec:scales.vs.physics:gravity}

The importance of gravitational dynamics is self-evident. At radii $\gtrsim 0.01\,$pc, self-gravity is essential to follow the formation of the galaxy, inflows, feedback, and (especially crucial even in idealized simulations of a ``patch'' of this medium) fragmentation to form multi-phase ISM structure and stars. Without this, no meaningful predictions for inflow rates to the SMBH can be made, since this is the primary ``competitor'' with inflow to determine whether or not gas can actually reach the BH (not to mention how it qualitatively changes much of the dynamics). The presence of stars (and at larger radii, dark matter) also means one must be able to integrate collisional+collisionless systems simultaneously.

At smaller radii, where star formation has ceased and the potential is dominated by the SMBH, accurate gravitational orbit integration is obviously necessary: certain numerical methods for example cannot accurately integrate warped or precessing disks, or nearly-Keplerian cold disks, for many orbits before spurious numerical torques or ``grid alignment effects'' (in e.g.\ fixed-mesh codes or many smoothed-particle hydrodynamics methods) will destroy or artificially grid-align the disks \citep[for extensive discussion of this and validation of the methods here in test problems, see][]{gaburov:2011.meshless.dg.particle.method,hopkins:gizmo,zhu:2016.sph.vs.gizmo.cosmo.sims.mw.mass.galaxy,deng:gravito.turb.frag.convergence.gizmo.methods,deng:2020.parametric.instab.free.disks,deng:2021.magnetic.disk.frag.in.gizmo.small.planetesimals,deng:2022.warped.disk.dynamics.need.mfm.mfv,hubber:gandalf.gizmo.methods,fletcher:2019.gizmo.vs.many.codes.comparison.migration.orbital.dynamics.detailed.differences.in.gap.accretion.dynamics.from.sink.and.numerical.viscosities,bonetti:2020.orbit.circularization.gizmo.validation,yamamoto:2021.no.artificial.fragmentation.in.mfm.mfv.sph,franchini:2022.binary.bh.hyperzoom.dynamics,bortolas:2022.gizmo.vs.analytic.bar.satellite.interaction.sinking.dynamics}. Here our high-order Hermite integrator provides the ability to, for example, reasonably integrate a hard stellar binary for $\gg 10^{5}$ orbital times in a strong tidal field \citep{grudic:starforge.methods}, much longer than necessary given the duration that we actually run our simulations to at their highest refinement level. But more importantly, we see that self-gravity is not negligible even at radii $\sim 1000\,R_{\rm g}$. The spiral arms and $m=1$ modes seen plainly in Fig.~\ref{fig:images.faceon} and discussed above can play an important role in the dynamics even for a gaseous disk-to-BH mass $M_{\rm disk}(<r) / M_{\rm BH} \ll 1$.

\subsection{Magnetic Fields}
\label{sec:scales.vs.physics:mhd}

As expected based on most previous studies on galactic scales, at $\gtrsim 100\,$pc magnetic fields play a relatively minimal role in the dynamics or gas thermodynamics \citep{2015ApJ...815...67K,su:2016.weak.mhd.cond.visc.turbdiff.fx,su:fire.feedback.alters.magnetic.amplification.morphology,su:2018.stellar.fb.fails.to.solve.cooling.flow,hopkins:cr.mhd.fire2,ji:fire.cr.cgm,steinwandel:2019.magnetic.bouyancy.galactic.scales,steinwandel:lmc.mass.galaxy.outflows.mfm.validation,martin.alvarez:2021.bfield.amplification,ponnada:fire.magnetic.fields.vs.obs,whitworth:no.mhd.fx.galform.sims}. Even on scales from $\sim 1\,$pc to $\sim 100\,$pc, we see no evidence that the magnetic fields play a major role in the overall gas dynamics, and re-starting our simulation and re-running for $\sim100$ dynamical times on these scales without magnetic fields (without refining down to $\ll 1\,$pc scales) produces no major qualitative differences in our predictions for these scales, despite a plasma $\beta \ll 1$. This is also expected, based on many previous studies of magnetic fields in the cold, neutral ISM on similar scales, where despite $\beta \ll 1$ (because the gas is thermally cold), the magnetic pressure is still sub-dominant to other forms of pressure such as the ``turbulent pressure'' or (in the outer ISM) cosmic ray pressure \citep{federrath:supersonic.turb.dynamo,2018MNRAS.479.3343M,martin.alvarez:2022.cosmological.turb.dynamo,guszejnov:fire.gmc.props.vs.z,guszejnov:environment.feedback.starforge.imf,hopkins:cr.mhd.fire2,hopkins:fire3.methods,grudic:2022.sf.fullstarforge.imf,seta.federrath:2022.turb.dynamo.twophase.medium}. Equivalently, we see above that the turbulence is still super-\Alf{ic} on these scales. In such situations the magnetic fields are largely ``passively'' tracing the local gas dynamics, rather than controlling it. There can of course still be indirect effects via smaller-scale dynamics (e.g.\ magnetic fields modifying the IMF which in turn modifies feedback; see \citealt{guszejnov:2020.mhd.turb.isothermal.imf.cannot.solve,guszejnov:environment.feedback.starforge.imf} and references therein). On smaller scales, however, we clearly see a reversal in this situation: the magnetic field strengths continue to grow, magnetic pressure dominates the vertical disk support and torques (discussed in greater detail in \papertwo), the turbulence becomes trans or sub-\Alf{ic}, and so magnetic fields become essential to the dynamics.

The role of non-ideal effects is more minor. Though formally included, atomic \&\ molecular viscosities and conductivities are everywhere negligible compared to numerical diffusion (and other physical processes), as expected. Anisotropic Braginskii viscosity and conductivity, given their strong temperature dependence, are only expected to be important in the most diffuse, hot phases of the CGM and ISM, and even there have relatively small effects \citep{su:2016.weak.mhd.cond.visc.turbdiff.fx}, so while included here we do not expect it to change any of our conclusions if they were excluded (and we see the viscous stress tensor is almost always relatively small). 

In highly-neutral gas, we have re-run our simulation briefly turning on and off Ohmic resistivity, the Hall effect, and ambipolar diffusion in turn\footnote{The neutral-gas non-ideal MHD terms are parameterized in the usual long-wavelength approximation as $\partial {\bf B}/\partial t = -\nabla \times [ \eta_{\rm O}\,{\bf J} + \eta_{\rm H}\,{\bf J}\times \hat{\bf B} - \eta_{\rm A}\,({\bf J} \times \hat{\bf B}) \times \hat{\bf B}]$ in the induction equation, with the Ohmic/Hall/ambipolar coefficients $\eta_{\rm O}/\eta_{\rm H}/\eta_{\rm A}$ (each a complicated function of plasma parameters including the free electron, ion, and dust charge abundance, see e.g.\ \citealt{keith:2014.planetary.disk.ionization.model}) having the same units of diffusivity here.} to examine their relative importance. A conservative estimate of this comes from comparing the relevant timescale or timestep $\propto \lambda^{2}/\eta_{i}$ for process $i$ on scale $\lambda$ to the other code timescales/timesteps (for e.g.\ other diffusion processes, sound-crossing, \Alf-wave crossing, etc.). We see that Ohmic resistivity is never dominant given the density and ionization fractions we resolve. Ambipolar diffusion can be the most important effect of these three non-ideal terms in the least-dense but still overwhelmingly-neutral phases of gas (e.g.\ ISM-like molecular phases), but we see that including or excluding it has almost no effect on the global dynamics in the simulation, because even in the regions where it dominates over Hall and Ohmic terms, the ambipolar diffusion time is almost always much longer (often by orders of magnitude) than other transport process timescales such as the turbulent dissipation/reconnection timescale ($\mathcal{O}(\lambda/v_{\rm turb})$), as seen in most modern idealized simulations of protostellar core collapse \citep{chen:2014.core.collapse.with.without.ambipolar.similar,wurster:2021.ambipolar.small.fx.starform.hall.fx.larger} and studies of individual GMCs with realistic star formation and feedback \citep{mac-low:2004.turb.sf.review,vazquez.semadeni:2011.weak.fx.ambipolar.diffusion.gmc.sf,sadanari:2022.ambipolar.diffusion.first.stars.fx.small}. The Hall term is most potentially interesting: we do see a regime where the Hall term is dominant among non-ideal MHD effects {\em and} where the relevant timescale is shorter than other resolved timescales. This specifically occurs in extremely-dense gas forming protostellar disks at our highest resolution level ($\sim 10\,$au distance from sink particles) in the star-forming disk (e.g.\ mostly at $\gtrsim 1\,$pc) where the local densities are $\gg 10^{12}\,{\rm cm^{-3}}$ (about a million times higher than the average at those radii; Fig.~\ref{fig:profile.mass}) and the ion fractions can (locally) become extremely small (typically $x_{i} \equiv n_{\rm ion} / n_{\rm neutral} \sim 10^{-17}$). This is not surprising: indeed, studies have shown that Hall effects can be important for the dynamics of proto-stellar disks and planetary disks on these spatial and ionization fraction scales \citep{bai:2017.hall.magnetic.transport.ppds,zhao:2020.hall.fx.ppds,lee:2021.non.ideal.mhd.fx.planet.disks,tsukamoto:hall.mhd.fx.ppd.form}. However, this is clearly not directly important for the global, quasar accretion disk-scale dynamics (the fraction of the total gas mass or volume in such protostellar disks is, as we showed above, negligibly small at these radii). Though the Hall effect could in principle indirectly alter e.g.\ the IMF of stars if it helps regulate accretion through individual protostellar disks onto the proto-stars themselves via Hall MRI, it would require much higher resolution (compared to our simulation) within the individual protostellar disks to resolve this (see \paperthree). Crucially, within the global quasar accretion disk on sub-pc scales, even though the densities are very large, the warmer temperatures mean that the ionization fractions are vastly larger, $\sim 0.01$ as shown in Fig.~\ref{fig:profile.thermochem}. This means that the characteristic timescales for Hall MHD effects within the quasar accretion disk and ISM as a whole are typically $\sim 11-15$ orders of magnitude longer than the disk dynamical time at the radii we model here, so can be safely neglected.\footnote{Given the timestep penalties involved (which come from resolving fast whistler waves in the small number of cells in the $\sim 10\,$au protostellar disks at $\gtrsim 1\,$pc) and related numerical integration challenges \citep{marchand:2018.technical.challenges.hall.mhd.angmom.conservation}, and the fact that the coefficients are extremely uncertain in the regime of greatest interest owing to their sensitivity to the detailed assumptions of the grain chemistry and size distribution \citep[see][for a review]{tsukamoto:2022.dust.size.fx.hall}, we therefore neglect the Hall term in our default simulation and only include it in these tests run for a shorter time period.}

\begin{figure*}
	\centering\includegraphics[width=0.97\textwidth]{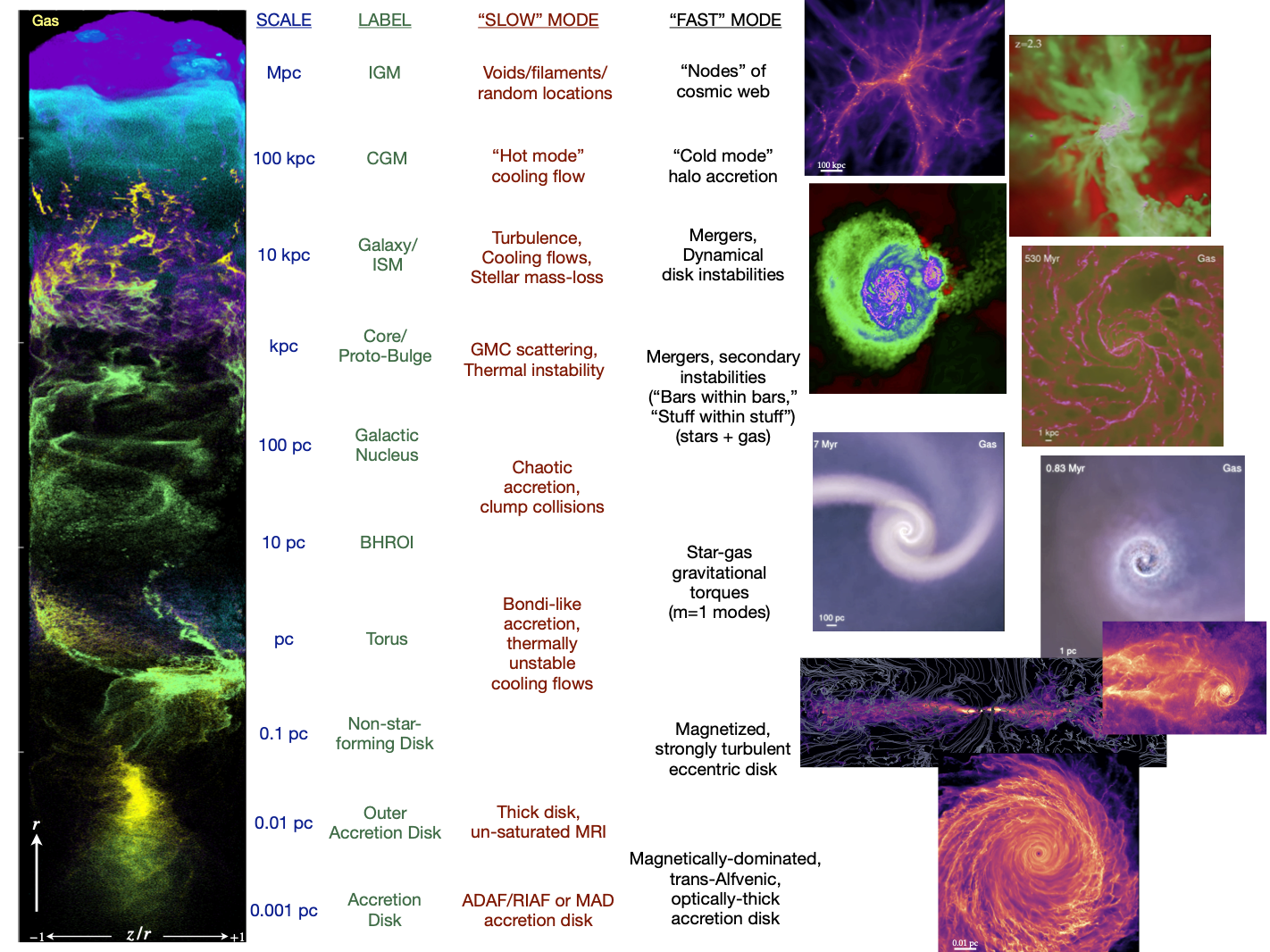}
	\caption{Cartoon illustrating the hierarchy of scales as Fig.~\ref{fig:scale.labels}, with a heuristic description of the process driving fastest angular momentum loss on each scale. We show the image from our simulation, size scale, and descriptor as Fig.~\ref{fig:scale.labels}, together with a list of characteristic processes that drive angular momentum loss on these scales in a ``slow'' or ``secular'' fashion (timescales much longer than the dynamical time) or in a ``fast'' or ``dynamical'' fashion (timescales of order dynamical times). See discussion in \S~\ref{sec:scales.vs.physics}-\ref{sec:previous}. Illustrations of numerical simulations for each ``fast'' scale/process are shown, taken from simulations first presented in \citet{hopkins:zoom.sims,hopkins:2013.fire,torrey.2016:fire.galactic.nuclei.star.formation.instability} for $\gtrsim\,$pc scales and from the simulations here on sub-pc scales. 
	\label{fig:scale.fast.vs.slow}}
\end{figure*}

\subsection{Cosmic Rays}
\label{sec:scales.vs.physics:crs}

We see fairly minor effects turning on/off explicit cosmic ray transport, or switching between the simpler sub-grid model from \citet{hopkins:2022.cr.subgrid.model} and the more detailed and physically-derived explicit cosmic ray dynamics models developed in \citet{hopkins:m1.cr.closure,hopkins:cr.spectra.accurate.integration,hopkins:cr.multibin.mw.comparison,hopkins:2021.sc.et.models.incompatible.obs}, or even simply assuming a uniform cosmic ray background for purposes of ionization rate calculations. This is expected given the detailed studies of cosmic ray dynamics in e.g.\ \citet{su:2018.stellar.fb.fails.to.solve.cooling.flow,su:turb.crs.quench,su:2021.agn.jet.params.vs.quenching,chan:2018.cosmicray.fire.gammaray,chan:2021.cosmic.ray.vertical.balance,hopkins:cr.mhd.fire2,hopkins:2020.cr.outflows.to.mpc.scales,hopkins:2020.cr.transport.model.fx.galform,hopkins:cr.transport.constraints.from.galaxies,ji:fire.cr.cgm,ji:20.virial.shocks.suppressed.cr.dominated.halos,buck:2020.cosmic.ray.low.coeff.high.Egamma,peschken:2022.crs.naab.sims.outflows.similar.to.fire,martin.alvarez:radiation.crs.galsim.similar.conclusions.fire} as well as observational constraints from starburst galaxies \citep{lacki:2011.cosmic.ray.sub.calorimetric,griffin:2016.arp220.detection.gammarays,zhang:2019.new.cosmic.ray.compilation.vs.calorimetry.sub.calor,heesen:2021.radio.cosmic.ray.escape.sub.calorimetric}, which show that for starburst systems and massive high-redshift galaxies the cosmic ray energy is lost via catastrophic and Coulomb+ionization interactions on a timescale short compared to other timescales of interest (i.e.\ the galaxies are approximate proton calorimeters). It is precisely the opposite regime: tenuous CGM/IGM gas around low-redshift dwarf and $\sim L^{\ast}$ galaxies, where CRs are seen in the studies above to have the largest effects (where they are observed to escape from galaxies into the CGM efficiently; see \citealt{lacki:2011.cosmic.ray.sub.calorimetric,rojas.bravo:2016.new.fermi.upper.limits.lots.of.sub.calorimetric.galaxies,lopez:2018.smc.below.calorimetric.crs,persic:2022.cosmic.ray.lmc.smc.not.calorimeters,butsky:2022.cr.kappa.lower.limits.cgm}).

This is not surprising, based both on the more detailed studies above, but also simple analytic considerations. If we assume CRs are injected in the midplane and treat the gas as a uniform slab with tangled magnetic fields, then assuming a typical scattering rate similar to the constraints from the Solar system applies \citep[see][for details]{hopkins:cr.multibin.mw.comparison}, then given the density profiles in Fig.~\ref{fig:profile.mass} at initial injection radii $R_{\rm inj} \lesssim 10\,$kpc catastrophic losses would remove all of the proton energy before propagation to a distance $\lesssim 0.3\,R_{\rm inj}$. If we further assume the injection is proportional to the SNe rate assuming a steady-state SNe rate proportional to the star formation rate, itself scaling as some efficiency $\epsilon_{\rm SF}$ per free-fall time, this would produce a steady-state CR energy density in the ISM (again, taking the profiles from Fig.~\ref{fig:profile.mass}) of $\sim 30\,{\rm eV\,cm^{-3}}\,(\epsilon_{\rm SF}/0.01)\,({\rm kpc}/R_{\rm inj})$ -- reasonably similar to what we measure in the simulation and much less than any of the dominant energy densities that we plot at any radii $r\lesssim 100\,$kpc in Fig.~\ref{fig:profile.thermochem}. And this is essentially an upper limit to the CR energy density, as it ignores other loss terms from trapping, denser sub-structure, advection or streaming.

\subsection{Radiation Transport \&\ Thermo-Chemistry}
\label{sec:scales.vs.physics:rhd}

Clearly {\em some} cooling physics is important on all scales we simulate, but again the nature of that cooling and the role of radiation changes with scale. On large scales $\gtrsim 10\,$pc, the cooling can be well-approximated as optically-thin (the usual approximation in galaxy formation simulations), although the actual chemistry can be enormously complex (as e.g.\ the models here account for a huge range of atomic and molecular and ionization and dust and other processes, with a multi-band radiation background, interactions with cosmic-rays, non-equilibrium photo-chemistry, etc.). On these scales radiation is important for determining self-consistent ionization and photo-heating and radiation pressure dynamics from stars within the galaxy but its global effects can be captured reasonably well by simple approximations such as the LEBRON method \citep[see][]{hopkins:radiation.methods}, and the gas cooling radiation itself can be largely neglected in the dynamics.

On the smallest scales we resolve, cooling is still important -- in fact the disk has a cooling time short compared to its dynamical time even at the smallest scales we resolve (discussed in more detail in \papertwo), so one cannot simply approximate the disk as strictly adiabatic. But the chemistry becomes substantially less complex as dust is sublimated, molecules dissociated, and in general the system becomes more and more locally black-body-like (and eventually at sufficiently small radii in the accretion disk the medium will be largely ionized, with chemistry relevant for second-order [though still potentially important] effects like metal line absorption, see e.g.\ \citealt{proga:disk.winds,jiang:2016.iron.opacity.fx.accretion.disks} and references therein). In this regime radiation is dominated by the cooling radiation itself, although it can increasingly be approximated via simple black-body or gray-body approximations and the photon mean-free paths become short, so methods like flux-limited diffusion or other even simpler analytic radiation treatments may be valid \citep[as in e.g.][]{thompson:rad.pressure,rafikov:2007.convect.cooling.grav.instab.planets,derdzinski:2022.inspirals.agn.disk.model.opacity.models.approx}, and one could even approximate the disk as e.g.\ locally isothermal (or following some effective adiabatic index, with a mean temperature that depends on the distance from the BH, as is common in e.g.\ shearing-box simulations).

The complexity is maximized ``in between,'' here from radii $\sim 0.01-10\,$pc (or more generally given how this should depend on the opacities, from surface densities $\Sigma_{\rm gas} \sim 10^{4} - 10^{7}\,{\rm M_{\odot}\,pc^{-2}}$). Here the system is optically thick to its cooling radiation, but not so optically thick that one can treat the radiation/dust/gas temperatures as in strict local thermodynamic equilibrium (LTE); more complex species such as dust, atoms, and molecules are all present in varying abundances (also not in LTE); and processes such as H$^{-}$ Kramers opacity -- often neglected in {\em both} galaxy-scale simulations and accretion disk simulations -- can dominate the opacities (this is dominant over a significant fraction of the dynamic range at $\ll 1\,$pc, given the high atomic abundance and relatively high free electron fraction producing significant H$^{-}$). Notably, the opacities would be orders-of-magnitude incorrect if we simply ignored dust destruction or neglected H$^{-}$ or detailed gas-phase opacities, and the temperature and cooling rates would also be order-of-magnitude incorrect if we simply assumed LTE and that all (radiation/dust/gas) temperatures were in equilibrium.\footnote{Notably, \citet{derdzinski:2022.inspirals.agn.disk.model.opacity.models.approx} do point out the importance of these opacity terms in quasar accretion disks at broadly similar radii, though the parameter space they explore is rather distinct from that here, and they adopt a simpler analytic disk model with opacity fitting functions calibrated for proto-stellar disks in \citet{lin.papaloizou:1985:opacities,bell:1994.protostellar.disk.model}. However direct comparison with e.g.\ their Table~1 or Figures 1-3 shows that the explicit chemical network and non-equilibrium dynamics evolved in our simulation -- important for capturing conditions in the sub-pc AGN disk that are {\em not} analogous to protoplanetary disks at a similar density and temperature (e.g.\ much shorter dynamical times, stronger radiation and cosmic ray fields, sublimation of dust grains, higher accretion rates) -- can produce order-of-magnitude quantitative differences in the detailed opacities.} Above, we discuss how this produces some substantial differences (most notably, shutting down star formation), compared to previous studies which treated the transition region more simply.

\subsection{Star Formation (Sink Particle and Unresolved) \&\ Stellar Feedback}
\label{sec:scales.vs.physics:sf}

Star formation is clearly important on scales $\sim 0.1 - 10^{4}\,$pc. On much larger CGM/IGM scales we do not expect star formation to occur given the low densities; on much smaller scales we self-consistently see it suppressed so it can be again neglected. On scales $\gg$\,pc, we see that the characteristic ``fragmentation mass'' $\sim \Sigma_{\rm gas}\,H_{\rm gas}^{2} \sim \Sigma_{\rm gas}\,(\delta v/\Omega)^{2}$ expected in any turbulent fragmentation cascade \citep{hopkins:excursion.ism} (akin to the ``Toomre mass'' for a marginally-stable disk) is $\gg 10^{4}\,{\rm M_{\odot}}$, so the stellar IMF should be well-sampled by the typical fragmenting clouds containing most of the mass and most of the star formation \citep{evans:1999.sf.gmc.review,vazquez-semadeni:2003.turb.reg.sfr,hopkins:excursion.imf,hopkins:excursion.clustering,grudic:2019.imf.sampling.fx.on.gmc.destruction}. This means on these scales, approximating the star formation via ``galactic-type'' models wherein one simply seeks to identify fragmenting sub-regions and, from them, statistically samples some IMF, is reasonable. Of course, one could always imagine ``zooming in'' to some sub-region (e.g.\ an individual GMC) on these large scales and studying it with resolved star formation models as in STARFORGE, but this would be more akin to standard studies of star formation in isolated clouds in the typical ISM \citep{guszejnov:2020.starforge.jets,grudic:starforge.methods,guszejnov:2022.starforge.cluster.assembly,guszejnov:environment.feedback.starforge.imf,guszejnov:starforge.environment.multiplicity} and is not strictly necessary for recovery of the large-scale dynamics (though it could of course be indirectly important via calibration of the ``correct'' IMF to use and other related properties of the stars themselves). Our preliminary analysis in \paperthree\ of the resolved IMF at smaller radii also supports the use of a statically-sampled universal IMF on these larger scales.

``Resolved'' star formation physics is therefore strictly necessary over a relatively narrow range of intermediate radii, here primarily from $\sim 0.1-1\,$pc. Still it plays a crucial role in the simulation here of actually allowing us to {\em validate} that SF should indeed cease at $\ll 0.1\,$pc. The galactic-type models cannot truly self-consistently predict this, since the SFR for a small ``patch'' of the ISM with certain properties is assumed, not self-consistently resolved. On these scales the characteristic upper limit of the fragmentation mass is still relatively large, $\gtrsim 100\,M_{\odot}$, but not so massive that a well-sampled IMF can be assumed. But the thermal conditions of the gas and its increasing warmth and magnetic support mean that the sites of individual star formation are highly constrained, as we discuss above and in more detail in \paperthree.

\section{Comparison to Previous Results}
\label{sec:previous}

\subsection{Galactic Scales ($\gtrsim 100\,$pc)}
\label{sec:previous:galactic}

As noted in \S~\ref{sec:scales.vs.physics}, on relatively large scales $\gtrsim 100\,$pc, which one could reasonably call ``galactic,'' our results are broadly consistent with previous FIRE studies of massive, high-redshift galaxies. But it is worth reiterating some basic conclusions (some of which are summarized in Fig.~\ref{fig:scale.fast.vs.slow}): (1) the galaxies are very much not in steady-state or equilibrium, with large clumps, mergers, and feedback-driven perturbations to the potential of order unity \citep{ma:2015.fire.escape.fractions,ma:fire.reionization.epoch.galaxies.clumpiness.luminosities,oklopcic:clumpy.highz.gals.fire.case.study.clumps.not.long.lived,price:FIRE.size.mass.recovery.z2}; (2) they feature prominent ``cold flows'' in the halo contributing substantial ``cold mode'' accretion onto the galaxy \citep{feldmann.2016:quiescent.massive.highz.galaxies.fire,fg15,faucher.2016:high.mass.qso.halo.covering.fraction.neutral.gas.fire,sravan.2015:metal.emission.line.cgm.fire,chen:2020.kbss.vs.fire.mocks}; (3) ``gravitational torques'' play a key role in the dynamics of angular momentum exchange, with the gas predominantly being forced into shocks and dissipation by asymmetries in the stars \citep{daa:BHs.on.FIRE,ma:2021.seed.sink.inefficient.fire,trapp2022}; (4) the galactic ISM is highly multi-phase and unstable, with relatively short-lived structures \citep{orr:stacked.vs.bursty.sf.fire,kim:gc.form.FIRE,ma:2020.no.missing.photons.for.reion.supershells,smith:lyalpha.rt.fire.sim.escape.fraction}; (5) the plasma $\beta \gg 1$ except in the cold-phase ISM, where magnetic pressure is larger than thermal but still order-of-magnitude sub-dominant to turbulent energy densities (so magnetic fields are not dynamically dominant; \citealt{su:2016.weak.mhd.cond.visc.turbdiff.fx,su:fire.feedback.alters.magnetic.amplification.morphology,su:2018.stellar.fb.fails.to.solve.cooling.flow,guszejnov:imf.var.mw,guszejnov:2019.imf.variation.vs.galaxy.props.not.variable,guszejnov:fire.gmc.props.vs.z,hopkins:cr.mhd.fire2}); (6) stellar feedback rapidly becomes less efficient above a critical acceleration scale $\sim \langle p_{\ast}/m_{\ast} \rangle \sim 10^{-8}\,{\rm cm^{2}\,s^{-1}}$ (corresponding to a total-enclosed-mass effective surface density $M_{\rm enc}/\pi\,r^{2} \gtrsim 10^{3}\,{\rm M_{\odot}\,pc^{-2}}$; \citealt{grudic:sfe.cluster.form.surface.density,grudic:max.surface.density,grudic:balls.of.fire.starcluster.from.galaxy.sims,ma:2020.globular.form.highz.sims,shi:2020.imbh.budget.star.clusters, byrne2023}); (7) star formation is rapid but inflows are {\em dynamical}, with $\dot{M} \sim M_{\rm gas}\,\Omega$, and co-exist with outflows (as there is no spherical symmetry) so can still out-pace star formation \citep{sparre.2015:bursty.star.formation.main.sequence.fire,orr:2021.fire.cmz.analog,2021MNRAS.501.4812F,ma:fire2.reion.gal.lfs}.

There are novel advantages of the simulation here. For one, it includes a variety of physics not included in all previous FIRE studies: e.g.\ magnetic fields with non-ideal MHD, detailed thermochemical treatments of non-equilibrium chemistry and opacities for the highly optically-thick and/or dust-free regimes, explicit M1 radiation hydrodynamics (as compared to simpler RHD treatments). For another, it reaches significantly higher resolution than some previous FIRE studies cited above, with mass resolution $\sim 10^{3}-10^{4}\,{\rm M_{\odot}}$ throughout the galaxy. This allows us to confirm that, at least in the simulation run up to the time of ``hyper-refinement'' in the nucleus, these additional physics and numerics improvements do not appear to have a major qualitative effect on any of the conclusions of those previous papers regarding global galaxy properties. The weak effects of these physics on gross properties at large scales had been noted before \citep{hopkins:fire2.methods,hopkins:radiation.methods,hopkins:cr.mhd.fire2,hopkins:fire3.methods,su:2016.weak.mhd.cond.visc.turbdiff.fx,su:2018.stellar.fb.fails.to.solve.cooling.flow,wheeler:ultra.highres.dwarfs}, but those studies focused on lower-mass systems, so we extend them here. 

However, the simulation here also has a serious and obvious disadvantage compared to previous studies: we only simulate one case study, and once we turn on hyper-refinement, only simulate it for a very short cosmic time. So studies of galaxy-scale properties are generally better-served by dedicated simulations without hyper-refinement (using e.g.\ sub-grid models for BH accretion and feedback) that can evolve for longer times. The primary purpose of our including these scales in our simulation here is to generate self-consistent initial and boundary conditions for smaller scales, and of course to inform and refine such sub-grid models for future studies.

\subsection{Galactic Nuclei Scales ($\sim 1-100\,$pc)}
\label{sec:previous:nuclear}

On scales $\sim 1-100\,$pc, our conclusions are largely similar to those in the dedicated hyper-refinement experiments in \citet{daa:20.hyperrefinement.bh.growth} (themselves in key respects similar to previous studies such as \citealt{levine2008:nuclear.zoom,levine:sim.mdot.pwrspectrum,wada:torus.mol.gas.hydro.sims,hopkins:zoom.sims,hopkins:qso.stellar.fb.together,torrey.2016:fire.galactic.nuclei.star.formation.instability,prieto:2016.zoomin.sims.to.fewpc.hydro.cosmo.highz,prieto:2017.zoomin.sims.agn.fueling.sne.fb} or other more idealized nuclear simulations in e.g.\ \citealt{emsellem:nuclear.fueling.sims,beckmann:cosmological.sims.super.refinement.for.time.variability,sivasankaran:2022.iso.galaxy.sims.refined.center.bhgrowth.smuggle}; see \citealt{daa:20.hyperrefinement.bh.growth} \S~5.1 for a summary). For example (see also Fig.~\ref{fig:scale.fast.vs.slow}): (1) gravitational torques between gas and stars (largely stars at similar radii to the gas) again dominate the accretion physics (even more strongly than on galactic scales); (2) angular momentum support is the key ``barrier'' to inflows and accretion; the accretion is qualitatively distinct from a radial or Bondi or turbulent accretion problem, and application of Bondi-Hoyle-Lyttleton-type accretion-rate estimators based on the gas properties (as opposed to those which account for effects like gravitational torques, star formation, and stellar feedback; see \citealt{wada:torus.mol.gas.hydro.sims,hopkins:inflow.analytics,hopkins:2021.bhs.bulges.from.sigma.sfr}) at these scales gives an accretion rate which is typically incorrect by $\sim 4-8$ orders of magnitude; (3) the ISM is highly multi-phase and unstable and rapidly star-forming, with most of the gas mass at these radii in cold/warm neutral phases, but (4) accretion is dynamical owing to said gravitational torques.

As noted in \S~\ref{sec:mdot}, stellar-feedback driven wind/outflow rates decline interior to $\lesssim 100\,$pc, We note that this is evident even before our refinement begins (so is a largely resolution-independent statement) and appears both in our simulations with and without magnetic fields. The same effect is seen in previous zoom-in simulations \citep{levine2008:nuclear.zoom,daa:20.hyperrefinement.bh.growth} as well as more idealized simulations of galactic nuclei \citep{wada:torus.mol.gas.hydro.sims,hopkins:qso.stellar.fb.together,hopkins:2021.bhs.bulges.from.sigma.sfr} and dense molecular clouds \citep{2017MNRAS.471.4844G,2018ApJ...859...68K,grudic:sfe.cluster.form.surface.density}. As predicted analytically \citet{fall:2010.sf.eff.vs.surfacedensity,grudic:max.surface.density,grudic:mond.accel.scale.from.stellar.fb}, this occurs when the gravitational acceleration scale $|{\bf a}| = V_{c}^{2}/r$ exceeds the IMF-averaged momentum flux per unit mass of a young stellar population ($\sim 10^{-7}\,{\rm cm\,s^{-2}}$), so stellar feedback becomes inefficient at wind-launching.

The much more detailed physics (MHD both ideal and non-ideal, multi-band explicit RHD, expanded opacities, individual star formation/evolution) and higher resolution ($\sim 3$ orders-of-magnitude improved) here do lead to some differences of potential importance for observational diagnostics on these scales, but do not change the key qualitative physics of accretion, outflows, and star formation above. Given the cold ISM prominence, we see $\beta \ll 1$, but magnetic fields do not dominate the torques and are sub-dominant to turbulent and ``gravitational'' pressure and do not strongly alter the dynamics on these scales (much like in GMCs and HI filaments in the ``normal'' ISM of $z\sim 0$ galaxies; see \citealt{su:2016.weak.mhd.cond.visc.turbdiff.fx,2018MNRAS.479.3343M,martin.alvarez:2021.bfield.amplification,martin.alvarez:2022.cosmological.turb.dynamo,guszejnov:fire.gmc.props.vs.z,benincasa:2020.gmc.lifetimes.fire}). The inner galaxy begins to become optically thick to its own cooling radiation as we reach $\Sigma_{\rm gas} \gtrsim 10^{3}-10^{4}\,{\rm M_{\odot}}$ inside of $\sim 10-100\,$pc, and this does modify the phase structure: there is a large amount of warm molecular gas at $\sim 10^{3}\,$K, the dust temperature rises to $\sim 100\,$K (well above the CMB temperature at this redshift),  and the ISRF becomes strongly dominated by the re-radiated/IR radiation. These are notably similar to observed properties of gas at similar densities in the nuclei of local starburst galaxies such as Arp 220, NGC 6240 and others \citep{tacconi:ngc6240.gasdynamics,lonsdale:vlbi.agn.cores,evans:agn.host.sfr,iono:ngc6240.nuclear.gas.huge.turbulence,mason:ngc1068.torus.obs,greve:2009.sb.molgas.props,2011ApJ...742...95O,scoville:arp220.nuclear.disk.super.dense.alma.obs} as well as inferences in high-redshift quasar hosts \citep{casey:highz.ulirg.pops,riechers:2009.qso.edd.lim.starbursts,younger:warm.ulirg.evol,wang:highz.qso.ir,izumi:nuclear.disk.plus.outflow.equals.mass.acc,lelli:2022.qso.co.similar.to.superzoom.warm.co.high.inflow.smbh.mass.similar}. In future work, we will investigate their consequences for observables as well as whether they have any impact on the stellar IMF, but for now, they do not appear to qualitatively change the most important conclusions from \citet{daa:20.hyperrefinement.bh.growth} regarding accretion.

Given this, the primary purpose of our extended physics and resolution on these scales is to (1) test and validate the conclusions of \citet{daa:20.hyperrefinement.bh.growth} with more detailed simulations accounting for a range of physics neglected therein; (2) make more accurate predictions for observables and future sub-grid models on these scales; (3) enable exploration of more detailed quantities like the IMF; and (4) to provide self-consistent initial and boundary conditions for even smaller scales.

\subsection{Approaching The Accretion Disk ($\sim 0.01-1$\,pc)}
\label{sec:previous:torus}

On scales $\ll 1\,$pc, however, we see significant deviations from the behavior seen in \citet{daa:20.hyperrefinement.bh.growth} and other similar studies described above (and see also \citealt{kawakatu:disk.bhar.model,hopkins:m31.disk,hopkins:zoom.sims,hopkins:qso.stellar.fb.together,wada:torus.mol.gas.hydro.sims,schartmann:2010.1068.star.cluster.fueling,hobbs:turbulence.agn.feeding,izumi:nuclear.disk.plus.outflow.equals.mass.acc,2020ApJ...897...26W,kawakatu:2020.obscuration.torus.from.stellar.fb.in.torus}). There are several closely-related key qualitative differences. Perhaps most importantly, we see star formation shut down, as the magnetic $Q_{\rm mag} \gg 1$ and thermal $Q_{\rm thermal} \gtrsim 1$ as the system becomes more optically-thick and magnetically-dominated. This obviously involves both the MHD and RHD physics (coupled explicitly to the thermo-chemistry) here, as well as a resolved individual star model which can detect and mass-resolve individual stellar-mass patches that might be collapsing (or not). In contrast, in simulations like \citet{daa:20.hyperrefinement.bh.growth} and almost all the other simulation examples in \S~\ref{sec:previous:nuclear} above, a simple ``galaxy-scale'' sub-grid star formation prescription was adopted at all scales and magnetic fields were neglected, which meant star formation could not ``cease'' in this manner, and as a result, continued to be efficient (even growing in efficiency at smaller and smaller scales) at all resolved scales therein.

This immediately produces important consequences. With star formation suppressed, we see a transition at $\lesssim 0.5$\,pc where the local mass density becomes gas-dominated, and gravitational torques (whose efficiency depends strongly on there being a dominant {\em collisionless} component of the local mass density to drive shocks in the gas) become less efficient (and even reverse sign). However, Maxwell and Reynolds torques from the strongly-magnetized, gravito-turbulent disk take over and continue efficient inflow (Fig.~\ref{fig:scale.fast.vs.slow}). The detailed structure of the disk, its turbulence and magnetic fields, their origins, and how they drive accretion, will be the subject of detailed study in \papertwo, but depend directly on magnetic fields. Strong $m=1$ modes persist and a lopsided disk with clear spiral structure forms, and some star formation does occur, which will be studied in \paperthree, but the SFR inside these radii is small compared to inflow rates. 

As noted above, there is a body of work with overlapping physics and results here in historical idealized simulations of nuclear ``torus'' scales around AGN, such as those in \citet{kawakatu:disk.bhar.model,hopkins:m31.disk,hopkins:zoom.sims,hopkins:qso.stellar.fb.together,wada:torus.mol.gas.hydro.sims,schartmann:2010.1068.star.cluster.fueling,hobbs:turbulence.agn.feeding,izumi:nuclear.disk.plus.outflow.equals.mass.acc,2020ApJ...897...26W,kawakatu:2020.obscuration.torus.from.stellar.fb.in.torus}. These studies generally were more akin to \citet{daa:20.hyperrefinement.bh.growth} in that they included a more limited range of physics (often, but not always, neglecting MHD and RHD, and in all cases using a much simpler thermo-chemical network compared to that here). But most crucially, these were like \citet{daa:20.hyperrefinement.bh.growth} in using idealized star formation prescriptions with some statistically-averaged star formation rate per free-fall time above some density put in ``by hand'' as opposed to explicitly resolving individual stars and star formation. As such, their conclusions, like \citet{daa:20.hyperrefinement.bh.growth} as summarized above, have a great deal in common with ours on larger scales but diverge from ours when star formation shuts down. 

There has however been some work specifically using simulations designed to resolve individual star formation to predict e.g.\ the IMF of stars forming in circum-nuclear disks \citep{nayakshin:2005.top.heavy.imf.mw.center,nayakshin:sfr.in.clumps.vs.coolingrate,klessen:2007.imf.from.turbulence,hopkins:excursion.imf.variation,bonnell.rice:2008.gmc.infall.smbh.sf.sims,alexander:2008.mw.nucdisk.sims,hobbs:2009.mw.nucdisk.sim,2019MNRAS.489...52F}, to which we will compare in more detail in \paperthree. These studies have found some similar conclusions to those here, e.g.\ that coherent lopsided modes in disks are ubiquitous, as expected from analytic considerations as discussed above (and other conclusions specific to the IMF, like its being somewhat top-heavy, see \paperthree). But again these usually neglected physics such as magnetic fields and self-consistent radiation hydrodynamics tied to the thermo-chemistry of molecular/atomic/neutral phases -- all crucial, as we argued above, to follow the self-consistent suppression of star formation and transition in structure on these scales. Even more importantly, these simulations in the past have been much more limited in the dynamic range of scales around the SMBH which they could probe, so needed to adopt somewhat ad-hoc initial and outer boundary conditions at $\sim$\,pc, and therefore could not self-consistently predict the transition between star-forming and accretion disk. Of equal importance, all of those IMF studies referenced above explored parameter space orders-of-magnitude distinct from that here, with much lower gas masses and densities (in most cases because they were designed to specifically understand the sub-pc stellar disk around Sgr A$^{\ast}$, rather than the most luminous quasar environments like our study here). 

Alternatively, some recent studies have extended accretion disk models ``outwards'' to these scales \citep{2016MNRAS.460..980N,2023ApJ...948..120C} to explore fragmentation. But again, these simulations necessarily focused on a small dynamic range with specific initial/boundary conditions and physics, so were not attempting to link different scales in the same way as we do here. 

As such, we stress that these different types of intermediate-scale simulations are highly complementary to studies like those here. Our hope is that a study like this provides additional motivation and improved understanding of the necessary initial and boundary conditions and choice of ``physics included'' in these sorts of idealized, more-restricted-in-scale nuclear simulations, in the future.

\subsection{Within the Accretion Disk ($\lesssim 0.01\,$pc)}
\label{sec:previous:disk}

By the time we get to the more traditional ``accretion disk'' scales at $\lesssim 0.01\,$pc, star formation is inefficient, so the dominant {\em physical ingredients} are broadly similar to those traditionally invoked in AGN accretion disk simulations: ideal MHD and radiation-hydrodynamics in a nearly-Keplerian potential. 

Still, we see a several key {\em qualitative} differences between our predictions and the assumptions of the vast majority of existing quasar accretion disk simulation literature (although some recent work shows striking similarity, see e.g.\ \citealt{kudoh:2020.strong.b.field.agn.acc.disk.sims.compare}), which will be studied in detail in \papertwo\ so we only briefly review them here. Most of these have to do with the initial/boundary conditions. A strong $m=1$ coherent eccentric gas disk mode persists, induced (and propagating inwards) by the asymmetry from large radii \citep{hopkins:slow.modes}. Self-gravity is not completely negligible and some non-zero star formation persists (\paperthree), along with some gravito-turbulence (\papertwo). The disk is still predominantly neutral even at these outer radii (though this will change when the disk temperature rises to $\gg 10^{4}\,$K at smaller radii) and can still efficiently cool, with a cooling time still less than its dynamical time, so the gas cannot be treated as adiabatic, and the opacities include important contributions from mostly-neutral gas contributors like H$^{-}$, usually ignored in accretion disk simulations (which generally, if they follow explicit RHD, assume just some combination of free-free/Compton and metal line opacities). As a result, the turbulence is vigorous: trans-\Alf{ic} and highly super-sonic. And perhaps most importantly, the disk is strongly magnetized as a result of flux-freezing from the magnetic flux being fed to it from the ISM (as detailed in \papertwo), sustaining a ``flux-frozen'' or ``flux-fed'' disk with plasma $\beta \ll 1$, even in the midplane. Again, the hope is that the simulations here will provide additional motivation for new generations of accretion-disk simulations exploring these rather distinct portions of parameter space.

\section{Conclusions}
\label{sec:conclusions}

We present novel simulations which utilize the galaxy-scale cosmological physics of the FIRE simulations to inform the small-scale physics of individual star formation and stellar evolution of STARFORGE. This allows us to run a cosmological simulation employing a super-Lagrangian refinement technique to reach $\sim 10^{-4}\,$pc resolution in a $\sim (100\,{\rm cMpc})^{3}$ box, i.e.\ the equivalent of a $\gtrsim (10^{12})^{3}$ uniform-resolution simulation. More importantly, we incorporate a range of physics including (non-ideal and kinetic) magneto-hydrodynamics; self-gravity, star formation, stellar evolution, and (proto)stellar feedback (including jets, main-sequence mass-loss, multi-band radiation, core-collapse and Ia supernovae); explicit multi-band radiation-MHD (with separately evolved dust, gas, and radiation field temperatures/bands); and detailed thermo-chemistry accounting for a huge range of processes and opacities including dust-gas coupling, sublimation, non-equilibrium atomic and molecular chemistry, metal lines, H$^{-}$ and others directly coupled to the RMHD solver. This allows us, for the first time, to self-consistently treat {\em both} the limits of ``traditional'' accretion disk simulations and traditional ``ISM-scale'' simulations and the transition in between, all in the same simulation.

Our most notable conclusions from this particular study include:
\begin{itemize}[labelindent=0pt,labelwidth=10pt,labelsep*=0pt,leftmargin=!,align=parleft]
\item{Magnetic fields play a key role}: We show that magnetic fields are critical for a wide range of effects on sub-pc scales within the accretion disk, ranging from maintaining efficient torques and high inflow rates, explaining the scale heights and vertical profiles of the disk structure, the outer size/boundary of the accretion disk, and perhaps most importantly the suppression of star formation at sub-pc scales. Without magnetic fields, a disk still forms, but it is an order of magnitude or more smaller in spatial scale and mass, and produces factor of $\gtrsim 100$ times lower accretion rates into the SMBH, with runaway fragmentation and orders-of-magnitude larger nuclear SFRs on $\sim 0.1-10$\,pc scales. The accretion disk that forms also has qualitatively different structures as a result of magnetic flux-freezing and flux-feeding from the ISM, to be studied in detail in \papertwo.

\item{Quasar-Level Inflow Rates are Plausible and Can be Maintained:} Extending previous studies like those in \citet{daa:20.hyperrefinement.bh.growth} with both much higher resolution and more detailed micro-physics relevant on small scales, we confirm that strong torques on sub-kpc scales can maintain inflow rates as large as $\gtrsim 10\,{\rm M_{\odot}\,yr^{-1}}$ into a QSO accretion disk at $<80\,$au, for extended periods of time (hundreds of thousands of dynamical times at the smallest radii simulated here). On scales $\sim$\,pc-kpc these are dominated by ``gravitational torques'' in a multi-component (gas+stellar) disk, inducing strong shocks and inflow. On sub-pc scales where star formation becomes inefficient these become weak but strong MHD torques (Maxwell+Reynolds stress) in a turbulent, strongly-magnetized outer flux-fed accretion disk take over and are able to sustain such large inflow rates down to the smallest resolved radii here, well within the ``traditional accretion disk'' range of scales. 

\item{Suppression of star formation:} On sub-pc scales, star formation is strongly suppressed. Models which simply assume some fixed star formation efficiency per free-fall time in sufficiently dense and/or self-gravitating gas (the standard on galactic scales and in our FIRE simulations) would not necessarily fully capture or be able to predict this effect {\em a priori}, but on these scales the simulations here explicitly resolve individual proto-stellar cores (at $<0.01\,M_{\odot}$ mass resolution) with models for resolved single-star formation from STARFORGE that would, if star formation were ``missed'' incorrectly, allow the gas to collapse to infinitely high densities. Just as important, for the first time this prediction is made using physics and numerical methods which have been explicitly shown to reproduce reasonable observed star formation efficiencies, stellar masses/IMFs, and stellar multiplicity distributions under typical Solar-neighborhood ISM/GMC conditions. With these physics, we show that a combination of increasing optical depths producing warmer gas in the galactic nucleus, plus (crucially) strong toroidal magnetic fields raising the magnetic critical mass to be larger than the disk mass and strongly suppressing gravito-turbulent fragmentation, leads to a dramatic, almost complete suppression of star formation at distances $\ll$\,pc from the SMBH. 
\end{itemize}

There are many properties which could and should be studied in more detail in the simulations here. In \papertwo, we explore the structure of the strongly-magnetized and flux-frozen/fed accretion disk, nature of the MHD (Maxwell/Reynolds) torques, origin and dynamics of these strong fields on $\ll$\,pc scales, and their consequences for accretion disk theory. In \paperthree\ we explore the detailed predictions for the dynamics and process of star formation and the resulting IMF of stars in the hyper-resolved inner region around the QSO accretion disk (the ``circum-quasar medium''). There are obvious extensions of the work here and therein modeling many different observables, contrasting how, for example, the much more detailed radiation-magneto-thermochemistry models produce different predictions for dust and atomic and molecular gas properties (and hence observables) from galactic nuclei and the quasar ``torus'' region, compared to previous-generation simulations with more simplified physics. 

In future work, there are also multiple ways one might extend the actual simulation work here. The most obvious is to consider other initial conditions of different galaxies at different times, to explore other regimes. Perhaps the biggest caveat here is that (owing to the computational expense of these simulations) we have studied just one case, so it is not obvious how much of our conclusions can be generalized to very different conditions with e.g.\ much lower accretion rates, let alone extremely low-accretion rate systems like M87 or Sgr A$^{\ast}$. Another obvious limitation of the work here is that we do not include any ``outward'' fluxes from the un-resolved accretion disk at $<80\,$au, e.g.\ radiation or jets from the inner disk. Our hope is that, given the unexpected properties of the disks which form here, this work will first motivate smaller-scale simulations of accretion disks (going down to the ISCO) with outer boundary conditions broadly similar to our inner boundary conditions, and those can provide some motivation for including such ``inner disk feedback'' prescriptions in a subsequent generation of simulations.

In principle, one could also attempt to improve the simulations here even further in resolution or run-time. However the simulations here already push the boundaries of what is possible, and it is not obvious such a strategy would be most efficient. Instead, we argued that conclusions on large (galactic) scales (where one would ideally like to run the simulations for much longer) are better studied in simulations without ``hyper-refinement'' (but using simulations like those here to inform the sub-grid models for BH accretion and feedback). Meanwhile one can much more efficiently explore the physics and parameter space of accretion disk physics with classical, dedicated accretion disk simulations. But in those simulations, the initial and boundary conditions are largely arbitrary, so our goal here is to provide predictive values for those, motivating qualitatively new parameter space to be studied in future work.

\begin{acknowledgements}
Support for PFH was provided by NSF Research Grants 1911233, 20009234, 2108318, NSF CAREER grant 1455342, NASA grants 80NSSC18K0562, HST-AR-15800. 
DAA acknowledges support by NSF grants AST-2009687 and AST-2108944, CXO grant TM2-23006X, Simons Foundation Award CCA-1018464, and Cottrell Scholar Award CS-CSA-2023-028 by the Research Corporation for Science Advancement.
CAFG was supported by NSF through grants AST-2108230  and CAREER award AST-1652522; by NASA through grants 17-ATP17-0067 and 21-ATP21-0036; by STScI through grant HST-GO-16730.016-A; and by CXO through grant TM2-23005X. Support for MYG was provided by NASA through the NASA Hubble Fellowship grant \#HST-HF2-51479 awarded  by  the  Space  Telescope  Science  Institute,  which  is  operated  by  the   Association  of  Universities  for  Research  in  Astronomy,  Inc.,  for  NASA,  under  contract NAS5-26555. 
Numerical calculations were run on the Caltech compute cluster ``Wheeler,'' allocations AST21010 and AST20016 supported by the NSF and TACC, and NASA HEC SMD-16-7592.  This research is part of the Frontera computing project at the Texas Advanced Computing Center. Frontera is made possible by National Science Foundation award OAC-1818253.
\end{acknowledgements}


\bibliographystyle{mn2e}
\bibliography{ms_extracted}

\end{document}